\documentclass[sn-basic]{sn-jnl}%  Reference Style
\jyear{2023}%

\theoremstyle{thmstyleone}%
\theoremstyle{thmstyletwo}%

\theoremstyle{thmstylethree}%

\begin{document}

\title [Accepted in Journal of Big Data (Q1, IF:8.1, SCIE) on Jan 19, 2024] {Machine learning-based network intrusion detection for big and imbalanced data using oversampling, stacking feature embedding and feature extraction}

\author*[1,2]{Md. Alamin Talukder} \email{alamintalukder.cse.jnu@gmail.com}
\author[2]{Md. Manowarul Islam} \email{manowar@cse.jnu.ac.bd}
\author[3]{Md Ashraf Uddin} \email{ashraf.uddin@deakin.edu.au}
\author[4]{Khondokar Fida Hasan} \email{fida.hasan@unsw.edu.au}
\author[2]{Selina Sharmin} \email{selina@cse.jnu.ac.bd}
\author[5]{Salem A. Alyami} \email{saalyami@imamu.edu.sa}
\author*[6]{Mohammad Ali Moni} \email{m.moni@uq.edu.au}

\affil[1]{Department of Computer Science and Engineering, International University of Business Agriculture and Technology, Dhaka, Bangladesh}
\affil[2]{Department of Computer Science and Engineering, Jagannath University, Dhaka, Bangladesh}

\affil[3]{School of Information Technology, Deakin University, Waurn Ponds Campus, Geelong, Australia}
\affil[4]{School of Professional Studies, University of New South Wales (UNSW), 37 Constitution Avenue, ACT 2601, Australia}
\affil[5]{Department of Mathematics and Statistics, Faculty of Science, Imam Mohammad Ibn Saud Islamic University (IMSIU), 11432, Saudi Arabia} 
\affil[6]{Artificial Intelligence \& Data Science, School of Health and Rehabilitation Sciences, Faculty of Health and Behavioural Sciences, The University of Queensland St Lucia, QLD 4072, Australia.}

\abstract{

Cybersecurity has emerged as a critical global concern. Intrusion Detection Systems (IDS) play a critical role in protecting interconnected networks by detecting malicious actors and activities. Machine Learning (ML)-based behavior analysis within the IDS has considerable potential for detecting dynamic cyber threats, identifying abnormalities, and identifying malicious conduct within the network. However, as the number of data grows, dimension reduction becomes an increasingly difficult task when training ML models.
Addressing this, our paper introduces a novel ML-based network intrusion detection model that uses Random Oversampling (RO) to address data imbalance and Stacking Feature Embedding based on clustering results, as well as Principal Component Analysis (PCA) for dimension reduction and is specifically designed for large and imbalanced datasets. This model's performance is carefully evaluated using three cutting-edge benchmark datasets: UNSW-NB15, CIC-IDS-2017, and CIC-IDS-2018.
On the UNSW-NB15 dataset, our trials show that the RF and ET models achieve accuracy rates of 99.59\% and 99.95\%, respectively. Furthermore, using the CIC-IDS2017 dataset, DT, RF, and ET models reach 99.99\% accuracy, while DT and RF models obtain 99.94\% accuracy on CIC-IDS2018. These performance results continuously outperform the state-of-art, indicating significant progress in the field of network intrusion detection.
This achievement demonstrates the efficacy of the suggested methodology, which can be used practically to accurately monitor and identify network traffic intrusions, thereby blocking possible threats.

}


\maketitle

\section{Introduction}
\label{intro}

% Cybersecurity has become one of the prime concerns worldwide. In practice, cybersecurity is a set of technological safeguards that ensure that a system operates without interruption, completing tasks while avoiding unwanted, unauthorized and unanticipated intervention. Intrusion Detection Systems (IDS) is considered one of the fundamental tools to enforce perimeter security \citep{sarker2020cybersecurity, hasan2021security}. Traditional signature-based intrusion detection mechanism seems insufficient to tackle the growing attack, where Machine Learning (ML) enabled behaviour analysis-based detection systems can be promising \citep{singh2021approach}. ML-enabled IDS can offer better accuracy and faster detection speed. It has the ability to dramatically alter the privacy scene, which is a privacy protection concern for IDS and data science is at the forefront of an innovative materialistic worldview \citep{mohammadi2019cyber, hey2009fourth}. It is a comprehensive approach to extracting pattern information from distinct inputs without focusing on an algorithmic process to produce intelligent outcomes \citep{mandal2020improved, talukder2022machine, talukder2023efficient}. 

Cybersecurity, in the current era, has emerged as an international imperative, driven by the critical need to protect systems from unwanted, unauthorized, and unforeseen interference \citep{mueller2021facing}. These interferences can range from data breaches and information theft to threats that undermine the integrity and functionality of systems. Safeguarding against such threats is paramount in ensuring the smooth operation of systems, protecting sensitive data, and preserving user trust \citep{marwala2023cybersecurity}. Intrusion Detection Systems (IDS) have traditionally served as a cornerstone of perimeter security \citep{george2023digitally, nguyen2023strengthening}. 

These systems are crafted with the purpose of uncovering and responding to suspicious or malicious activities within a network or system. Nevertheless, the conventional signature-based intrusion detection methods, reliant on established attack patterns and signatures, have been found lacking in the face of ever-evolving and sophisticated cyber threats. These solutions are engineered to identify and react to questionable or potentially harmful actions occurring within a network or system. Nevertheless, conventional intrusion detection methods, which hinge on established attack patterns and signatures, have demonstrated their inadequacy when confronted with ever-changing and increasingly complex cyber threats \citep{khan2022deep, talukder2023dependable}.

To address the shortcomings of traditional IDS, the cybersecurity community has turned its attention to Machine Learning (ML) as a promising solution. ML-enabled IDS leverages behavior analysis to detect anomalies and threats, offering the potential for significantly higher accuracy and faster detection times \citep{schmitt2023securing, preuveneers2021sharing}. This paradigm shift in intrusion detection holds the promise of not only bolstering security but also reshaping the privacy landscape. This shift towards ML-enabled intrusion detection has sparked concerns regarding both privacy and the field of data science \citep{singh2023artificial, mohammadi2019cyber}. ML algorithms, while effective at identifying threats, often require access to sensitive data. Balancing the need for security with privacy concerns is a challenge that demands innovative and ethical solutions \citep{allahrakha2023balancing}.

ML in cybersecurity serves as a powerful tool to augment the ability of systems to understand diverse patterns and forecast potential data threats \citep{sarker2020cybersecurity}. It optimizes processing and training procedures to construct models that can effectively safeguard systems against dubious and spyware activities \citep{hussain2020machine, talukder2023dependable}. It is a transformative technology that empowers systems to learn and adapt from data, making intelligent decisions without being explicitly programmed \citep{mishra2022role}. In the context of IDS, ML algorithms utilize historical and real-time data to identify patterns of normal behavior and anomalies that may indicate security threats. By training on diverse datasets, these algorithms become proficient at recognizing new and emerging attack vectors. ML enhances IDS systems by providing faster and more accurate threat detection, reducing false positives, and adapting to evolving threats \citep{jayalaxmi2022machine}. It empowers security systems to efficiently safeguard networks and data against unauthorized access and malicious activity \citep{kafi2023securing}.

In today's landscape, the optimization of processing and training procedures is imperative for constructing models that can effectively safeguard systems against dubious and spyware activities \citep{sarker2020cybersecurity}. However, it's worth noting that many contemporary ML-IDS solutions tend to be limited by their reliance on small, outdated and balanced datasets for model development \citep{istiaque2021performance, cholakoska2021analysis, narayanasami2021biological}. The focus on these smaller, often outdated datasets, coupled with imbalances in the data distribution, while facilitating preprocessing and training with diverse ML algorithms, raises questions regarding the practical applicability of these models in real-world scenarios, specifically when dealing with big data. The achievable accuracy of such models often hinges on the intricacies of dataset preprocessing and the selection of suitable algorithms, adding an additional layer of complexity to their effectiveness \citep{norwahidayah2021performances, bhati2021intrusion}.

Therefore, we need to develop and validate ML-based intrusion detection for large, imbalanced datasets where all potential attack scenarios are encompassed. To address this gap, Our research places a significant emphasis on constructing a robust and well-structured framework that accommodates the detection of network intrusion in a more efficient manner on substantial datasets. We employ data preprocessing techniques, including data normalization, feature resampling, Stacking Feature Embedded and dimension reduction techniques, to address the unique challenges posed by big datasets. The Key techniques of our approach are as follows:
\begin{itemize}
    \item Stacking Feature Embedded (SFE): This technique enhances detection accuracy by introducing meta-features, providing a deeper insight into data patterns and anomalies.

    \item Random Oversampling (RO): By mitigating class imbalance issues, RO ensures equitable consideration of minority classes, resulting in a more balanced and reliable intrusion detection system.

    \item Principal Component Analysis (PCA): PCA optimizes the feature space, reducing dimensionality while preserving vital information, thus enhancing the efficiency and effectiveness of our ML models.
\end{itemize}
This comprehensive approach seeks to bridge the gap in intrusion detection, accommodating the intricacies of large, imbalanced datasets, and improving the robustness of security measures in the face of evolving threats.

% The performance of our proposed model is evaluated using a variety of ML classifiers, such as Decision Tree (DT), Random Forest (RF), Extra Tree (ET) and Extreme Gradient Boosting (XGB), which all are used to train our model with a smaller number of features. We have evaluated a variety of performance indicators in our investigation, including Accuracy, Precision and Recall, as well as F1-score, RMSE, Confusion Matrix and the ROC Curve. The classifier algorithms in the proposed framework can detect attacks with an accuracy rate of above 99.9\%. 

Our proposed model's performance is rigorously evaluated across a spectrum of ML classifiers, including Decision Tree (DT), Random Forest (RF), Extra Tree (ET), and Extreme Gradient Boosting (XGB). These classifiers are trained using a reduced feature set. We assess our model using a comprehensive set of performance indicators, encompassing precision, recall, f1-score, confusion matrix, accuracy and the roc curve. The ML algorithms integrated into our framework demonstrate an exceptional ability to detect attacks, consistently achieving accuracy rates exceeding 99.9\%. This thorough performance evaluation ensures the robustness and reliability of our intrusion detection system, highlighting its effectiveness in identifying and countering potential threats.
In summary, this paper's contribution can be encapsulated as follows:
\begin{itemize}
    \item We proposed a novel intrusion detection approach using efficient preprocessing, oversampling management, stacking feature embedding, and dimensionality reduction to enhance intrusion detection performance.
    \item We addressed the issue of imbalanced data by implementing a random oversampling strategy to ensure balanced consideration of minority and majority classes, leading to more robust intrusion detection.
    \item The introduction of Stacking Feature Embedded (SFE) enhances detection accuracy by introducing meta-features, providing a comprehensive understanding of data patterns and anomalies.
    \item Our utilization of Principal Component Analysis (PCA) for feature extraction optimizes intrusion detection performance while minimizing the dimensionality of the original dataset.
    \item Evaluation of our approach with various ML algorithms on popular large and imbalanced datasets demonstrates its effectiveness in significantly improving intrusion detection accuracy and robustness.
\end{itemize}

The subsequent sections of this paper offer a concise overview of our proposed work. In Section \ref{sec:Related}, we delve into the literature review, providing insights into the existing body of knowledge. Section \ref{sec:Method} is dedicated to the detailed description of our proposed methodology. Section \ref{sec:experimental} outlines the experimental setup and evaluation procedures. Finally, in Section \ref{sec:conclusion}, we draw the conclusions from our findings and explore avenues for future research.

\section{Literature Review}
\label{sec:Related}

ML strategies have been widely utilized in cybersecurity over the last several decades due to their capacity to retrieve hidden patterns on the variations between malevolent and legitimate patterns \citep{zhang2020effective, das2023integration}. ML is an effective research tool for detecting any anomalies in the network stream of traffic \citep{bhavani2020network}. As a result, previous researchers explored a variety of algorithms based on ML as well as hybrid and DL models in IDS.

\subsection{ML Approaches}

\cite{moualla2021improving} proposed a revolutionary network IDS model that plays a crucial role in network security and combats existing cyberattacks on networks utilizing the UNSW-NB15 data as a baseline. It was a dynamically scalable multiclass ML-based network with several phases. The imbalance was handled by SMOTE technique, after that based on the Gini Impurity criterion, it employed the ET Classifier and finally, a pre-trained Extreme Learning Machine (ELM) was utilized to classify each of the attacks using binary. Using the outputs of the ELM classifier as inputs to a fully connected layer, a logistic regression layer was employed to produce soft judgments for all classes and attained 98.43\% accuracy.

\cite{kasongo2020performance} presented a filter-based feature-dropping technique on the UNSW-NB15 IDS dataset, employing the XGB algorithm, and assessed its performance using various predictive algorithms, including Decision Tree (DT), Artificial Neural Network (ANN), Logistic Regression (LR), K-Nearest Neighbor (KNN), and Support Vector Machine (SVM). They demonstrated that their approach led to a significant enhancement in binary accuracy, increasing it from 88.13\% to 90.85\%. The overall accuracy rates for binary were 90.85\% for DT, 84.4\% for ANN, 77.64\% for LR, 84.46\% for KNN, and 60.89\% for SVM. For multiclass, the accuracy rates were 67.57\% for DT, 77.51\% for ANN, 65.29\% for LR, 72.30\% for KNN, and 53.95\% for SVM, all of which were evaluated using the 19 optimal selected features.

\cite{nimbalkar2021feature} offered a feature selection method for IDS using Information Grain (IG) and Grain Ratio (GR) where they selected 50\% of the top most features to build their model detecting Denial of Service (DoS) and Distributed Denial of Service (DDoS) attacks. The studies were carried out using well-known datasets, such as KDDCUP'99 and BOT-IOT. They selected 16 and 19 features for BOT-IOT and KDDCUP'99 datasets, respectively and then trained the model using the JRip classifier to reach the desired performance. They achieved 99.99\% and 99.57\% accuracy for BOT-IOT and KDDCUP'99 datasets, respectively.

\cite{kumar2020statistical} presented IDS model that detects intrusions based on misuse to protect networks from modern threats, such as DoS attacks or exploits, probes, generics and so on. Intrusion detection rate (IDR) and false alarm rate (FAR) were determined using the UNSW-NB15 dataset. The IG and C5 classifier were utilized where, IG was used to pick 13 out of 47 features for feature selection and C5 gave 99.37\% accuracy rate.

\cite{ahmad2021intrusion} proposed a feature clustering-based ML model where clusters were applied for Flow, Message Queuing Telemetry Transport (MQTT) and Transmission Control Protocol (TCP). The clustering process eliminated the overfitting that was the curse of dimensionality and data-set inequity. Various supervised ML methods were utilized on the clusters, including RF and SVM. They employed the UNSW-NB15 dataset to train and test the model and found that RF produced 98.67\% accuracy in binary and 97.37\% accuracy in multiclass.

\cite{ahmad2021intrusion} introduced an innovative ML model based on feature clustering, with distinct clusters created for Flow, Message Queuing Telemetry Transport (MQTT), and Transmission Control Protocol (TCP). This clustering approach effectively addressed the challenges of overfitting arising from high dimensionality and dataset imbalances. They applied a range of supervised machine learning methods, including RF and SVM, to these clusters. Using the UNSW-NB15 dataset for model training and evaluation, their results showed that RF achieved impressive accuracy, reaching 98.67\% in binary classification and 97.37\% in multi-class classification.

\cite{kshirsagar2021efficient} presented a filter-based feature selection technique that leveraged IG Ratio (IGR), Correlation Ratio (CR), and ReliefF (ReF). This method generated a feature subset by considering the average weight of each classifier, complemented by a Subset Combination Strategy (SCS). For CICIDS 2017 dataset, the number of features was lowered from 77 to 24 and for KDDCUP'99, it was cut from 41 to 12. For CIC-IDS2017, it achieved a 99.95\% accuracy rate in 133.66 sec using PART and for the KDDCUP'99 dataset, it achieved a 99.32\% accuracy rate and took 11.22 sec.

In order to minimize the volume of the dataset, \citep{mugabo2021intrusion} employed evolutionary approach-based feature selection (EFS) and a concurrent MapReduce technique was applied to partition the input data into the most crucial characteristics. After that, their model was classified for either normal or attack using the RF classifier. They used the popular KDDCUPP'99 dataset to evaluate performance and classify normal and deviant behavior. They selected 15 features to assess their model's performance and its accuracy was found almost 93.9\%.

\cite{talita2021naive} developed an innovative approach that integrates Particle Swarm Optimization (PSO) for feature selection with the Naive Bayes (NB) classification algorithm, which was applied to the KDDCUP'99 dataset. This dataset comprised of over 400 thousand records and featured more than 40 characteristics. To optimize computational resources and memory usage, PSO was employed to select the most relevant 38 features from the original set of over 40. The outcome of this method yielded an impressive accuracy rate of 99.12\%, demonstrating superior efficiency in terms of both computation time and classification accuracy when compared to other feature selection techniques.

\cite{seth2021novel} developed an IDS model that reduces prediction delay by minimizing the model's sophistication using a hybrid feature selection (HFS) strategy. A quick gradient boosting technique called Light Gradient Boosting Machine (LightGBM) was used to build the model. This approach cut prediction latency by 44.52\% to 2.25\% and model development time by 52.68\% to 17.94\% using the CIC-IDS2018 dataset. It can achieve excellent accuracy by combining attribute choosing and LightGBM. The developed model achieved 97.73\%, 96\%,  99.3\%, accuracy, sensitivity, precision rate respectively and a relatively low prediction latency.

\cite{hammad2021t} presented a model using t-SNERF to identify the cyber-attacks where t-SNERF was used for feature correlations, data reduction and trained the model using RF. To evaluate the model UNSW-NB15, CIC-IDS2017 and Phishing were employed. The offered innovative methodology outperformed current methods. The accuracy rate was 100\% for UNSW-NB15, 99.70\% for Phishing and 99.78\% for CIC-IDS2017.

\cite{guezzaz2021reliable} focused on enhancing the reliability of Network Intrusion Detection (NID) through the utilization of the DT (Decision Tree) algorithm. Their approach entailed two key steps: data quality improvement and feature selection based on entropy decision, followed by the construction of a dependable NID system using the DT classifier. This proposed model underwent evaluation using two well-known datasets, NSL-KDD and CIC-IDS2017, and yielded impressive results. Specifically, the model achieved a remarkable accuracy of 99.42\% on the NSL-KDD dataset and 98.80\% on the CIC-IDS2017 dataset.

\cite{stiawan2020cicids} offered a strategy for analyzing integral and essential aspects of massive network data, increasing traffic anomaly detection accuracy and speed. They used the CIC-IDS2017 dataset and picked important and significant features using IG, as well as sorting and grouping features according to their minimal weight values and then train the dataset using various ML classifiers. The number of relevant and meaningful attributes generated by IG has a considerable impact on accuracy and execution time. A number of ML algorithms, such as Random Tree (RT), RF, NB, Bayes Net (BN) and J48 were used to train the model but RF provided the best accuracy. In the RF classifier, they used 22 relevant selected features, had the best accuracy of 99.86\%, whereas the J48 classifier algorithm, which used 52 relevant selected features but took longer to execute, had the highest accuracy of 99.87\%. 

\subsection{Deep Learning Approaches}

\cite{aleesa2021deep} explored the application of DL models for binary and multiclass classification in the context of IDS. They conducted their evaluations using the UNSW-NB15 dataset. Specifically, the study assessed the effectiveness of three distinct types of neural network models: Deep Neural Network (DNN), Recurrent Neural Network (RNN), and ANN. They used data cleaning techniques, such as handling missing values and categorical values, followed by min-max normalization in order to make it more accurate. The efficiency was evaluated on accuracy; where, the ANN, RNN and DNN provided 99.26\%, 85.42\%, 99.22\% for binary classification and 97.89\%, 85.4\% and 95.9\% for multilabel classification respectively.

% \cite{choudhary2020analysis} proposed an IDS method using DNN to recognize the attacks in IoT. The performance of DNN to detect the attacks had been assessed using the three most commonly used datasets such as KDDCUP'99, NSL-KDD and UNSW-NB15. The proposed method using DNN showed that the accuracy rate was only 91.50\% with each dataset. 

\cite{choudhary2020analysis} introduced an IDS based on DNN for identifying IoT-related attacks. They evaluated the effectiveness of the DNN-based approach by testing it on three widely employed datasets: UNSW-NB15, KDDCUP'99 and NSL-KDD. The outcomes indicated that the DNN-based method achieved an accuracy rate of 91.50\% when applied to each of these datasets.

In the study by \cite{al2021stl}, a more effective IDS model was introduced, utilizing a combination of Long Short-Term Memory (LSTM) and Convolutional Neural Network (CNN) architectures. To address the issue of imbalanced datasets, they applied the Synthetic Minority Oversampling Technique (SMOTE) in conjunction with Tomek-Link, referred to as the STL method. The research implementation was carried out using PySpark, and two distinct datasets, namely CICIDS-001 for multiclass and UNSW-NB15 for binary classification, were used to evaluate the model's performance. The proposed method was benchmarked against various popular ML and DL algorithms. In multilabel classification, the proposed approach performed an outstanding accuracy of 99.83\%, while in binary classification, it demonstrated a high accuracy of 99.17\%.
% \color{blue}

\cite{adeyemo2019ensemble} explored the effectiveness of network IDS using DL and ensemble techniques. They applied an LSTM model, a homogeneous approach with an optimized bagged RF and a heterogeneous approach with four standard classifiers. The evaluation was conducted on the UNSW-NB15 dataset, split into two configurations: two and multi-attack datasets. The results showed that the LSTM achieved an 80\% detection accuracy for the binary dataset and 72\% for the multi-attack dataset. The homogeneous ensemble method reached impressive accuracy rates of 98\% for the binary dataset and 87.4\% for the multi-attack dataset. Similarly, the heterogeneous ensemble method also performed well, with a detection accuracy of 97\% for the binary and 85.23\% for the multi-attack dataset. This research highlights the promising performance of ensemble methods in the context of Intrusion Detection Systems.

\cite{kim2020cnn} developed a DL model primarily focusing on detecting denial of service (DoS) attacks. The KDDCUP'99 and CSE-CIC-IDS2018 were employed to assess the model's performance. Notably, the latter dataset contained more sophisticated DoS attacks than the former. The study concentrated on the DoS category within both datasets and harnessed a CNN for model development. A comparative analysis was conducted between the CNN and RNN. In the case of KDDCUP'99, the CNN demonstrated impressive accuracy rates, surpassing 99\% accuracy in binary and multiclass, while the RNN achieved 99\% accuracy in binary and 93\% accuracy in multiclass. For the CSE-CIC-IDS2018 dataset, the CNN model exhibited an average accuracy of 91.5\%, whereas the RNN model averaged 65\% accuracy. The CNN model consistently outperformed the RNN model in identifying specific DoS attacks characterized by similar attributes.

% \color{black}

\subsection{Hybrid Approaches}
\cite{bhardwaj2021hybrid} presented a hybrid strategy that combines a DNN model with Ant Colony Optimization (ACO) for learning premium hyperparameters for successful DNN classification in a cloud setting. DNN detects attacks more accurately by the usage of ideal settings. They used the CIC-IDS2017 dataset and got well performance. The detection and accuracy performance are both superior to state-of-the-art approaches, at 95.74\% and 98.25\%, respectively.

\cite{khan2021hcrnnids} introduced a novel approach known as the Hybrid Convolutional Recurrent Neural Network for IDS (HCRNNIDS). This approach represents a hybrid IDS paradigm that leverages DL techniques to predict and classify malicious intrusions on the internet. In this model, the CNN was responsible for gathering local information through convolution, while the RNN was employed to capture temporal features, thereby enhancing the effectiveness and predictive capabilities of the ID system. To assess the model's performance, the researchers conducted simulations using publicly available ID data, focusing on the contemporary and reputable CSE-CIC-DS2018 dataset. Through the application of a 10-fold cross-validation methodology, the proposed hybrid model exhibited significant improvements over traditional ID approaches. It achieved a remarkable level of accuracy in fraudulent detection and prevention, reaching up to 97.75\% for the CIC-IDS2018 dataset.

\cite{kasongo2020deep} presented a Wrapper-based Feature Extraction Unit (WFEU) that leverages the Extra Trees technique to create a reduced optimum feature vector for a Feed-Forward Deep Neural Network (FFDNN) wireless IDS system. The proficiency was studied using the UNSW-NB15 and AWID ID datasets. Several ML algorithms, such as RF and SVM are compared to WFEU-FFDNN as well as NB and DT. Using the UNSW-NB15, the WFEU produced an ideal feature vector consisting of 22 features and achieved an overall accuracy of 87.10\% for binary and 77.16\% for multiclass. On the other hand, for AWID, 26 features were selected and revealed an accuracy of 99.66\% for binary and 99.77\% for multiclass.

\cite{kasongo2020deep} introduced a Wrapper-based Feature Extraction Unit (WFEU), which harnessed the power of the Extra Trees technique to construct an optimized and reduced feature vector tailored for a Feed-Forward Deep Neural Network (FFDNN) wireless IDS. To evaluate its efficiency, they conducted experiments on the UNSW-NB15 and AWID intrusion detection datasets. Comparative analyses were performed against various ML algorithms, including RF and SVM, along with conventional approaches such as NB and DT. On the UNSW-NB15 dataset, the WFEU method yielded an optimal feature vector consisting of 22 features, achieving an impressive overall accuracy of 87.10\% for binary and 77.16\% for multiclass. For the AWID dataset, a feature set of 26 features was selected, resulting in remarkable accuracy rates of 99.66\% for binary and 99.77\% for multiclass.

\cite{zhang2020effective} proposed a novel strategy for addressing imbalanced intrusion detection, surpassing existing intrusion detection algorithms. They introduced the SGM-CNN model, which combined Synthetic Minority Over-sampling Technique (SMOTE) with a Gaussian Mixture Model (GMM). The model was validated using the UNSW-NB15 and CIC-IDS2017 datasets. On the UNSW-NB15 dataset, the model exhibited remarkable accuracy rates of 99.74\% for binary and 96.54\% for multiclass. Furthermore, for the CIC-IDS2017 dataset, achieved an impressive detection rate of 99.85\%.

\cite{hassan2020hybrid} introduced a hybrid DL model for efficient network intrusion detection, combining CNN and Weight-Dropped Long Short-Term Memory (WDLSTM). The CNN was employed to extract crucial features from IDS, while the WDLSTM was utilized to capture long-term dependencies and mitigate gradient vanishing issues. Their experiments focused on the UNSW-NB15 dataset. The CNN-WDLSTM model exhibited an overall accuracy rate of 97.17\% for binary and 98.43\% for multiclass.

The summary of related papers can be found in Table \ref{tab:related_summary}

\begin{table}[!h]
\resizebox{\textwidth}{!}{
\begin{tabular}{lclllcl}
\hline
\multirow{2}{*}{SL. NO} & \multicolumn{1}{l}{\multirow{2}{*}{ML Technique}} & \multirow{2}{*}{Algorithm} & \multirow{2}{*}{Author} & \multirow{2}{*}{Dataset} & \multicolumn{2}{l}{Accuracy (In \%)} \\ \cline{6-7} 
 & \multicolumn{1}{l}{} &  &  &  & \multicolumn{1}{l}{Binary} & Multi-class \\ \hline
1 & \multirow{22}{*}{ML} & RF & \cite{ahmad2021intrusion} & \multirow{8}{*}{UNSW-NB15} & \multicolumn{1}{c}{98.67} & 97.37 \\ \cline{1-1} \cline{3-4} \cline{6-7} 
2 &  & ELM & \cite{moualla2021improving} &  & \multicolumn{1}{c}{} & 98.43 \\ \cline{1-1} \cline{3-4} \cline{6-7} 
\multirow{5}{*}{3} &  & DT & \multirow{5}{*}{\cite{kasongo2020performance}} &  & \multicolumn{1}{c}{90.85} & 67.57 \\ \cline{3-3} \cline{6-7} 
 &  & ANN &  &  & \multicolumn{1}{c}{84.4} & 77.51 \\ \cline{3-3} \cline{6-7} 
 &  & LR &  &  & \multicolumn{1}{c}{77.64} & 65.29 \\ \cline{3-3} \cline{6-7} 
 &  & KNN &  &  & \multicolumn{1}{c}{84.46} & 72.30 \\ \cline{3-3} \cline{6-7} 
 &  & SVM &  &  & \multicolumn{1}{c}{60.89} & 53.95 \\ \cline{1-1} \cline{3-4} \cline{6-7} 
4 &  & C5 & \cite{kumar2020statistical} &  & \multicolumn{1}{c}{} & 99.3 \\ \cline{1-1} \cline{3-7} 
\multirow{2}{*}{5} &  & \multirow{2}{*}{PART} & \multirow{2}{*}{\cite{kshirsagar2021efficient}} & CIC-IDS2017 & \multicolumn{1}{c}{} & 99.95 \\ \cline{5-7} 
 &  &  &  & \multirow{3}{*}{KDDCUP’99} & \multicolumn{1}{c}{} & 99.32 \\ \cline{1-1} \cline{3-4} \cline{6-7} 
6 &  & \multicolumn{1}{c}{MapReduce+RF} & \cite{mugabo2021intrusion} &  & \multicolumn{1}{c}{} & 93.9 \\ \cline{1-1} \cline{3-4} \cline{6-7} 
7 &  & PSO+NB & \cite{talita2021naive} &  & \multicolumn{1}{c}{} & 99.12 \\ \cline{1-1} \cline{3-7} 
8 &  & HFS+LightGBM & \cite{seth2021novel} & CIC-IDS2018 & \multicolumn{1}{c}{} & 97.73 \\ \cline{1-1} \cline{3-7} 
\multirow{2}{*}{9} &  & \multirow{2}{*}{IG+GR+JRip} & \multirow{2}{*}{\cite{nimbalkar2021feature}} & KDDCUP’99 & \multicolumn{1}{c}{} & 99.57 \\ \cline{5-7} 
 &  &  &  & BOT-IOT & \multicolumn{1}{c}{} & 99.99 \\ \cline{1-1} \cline{3-7} 
\multirow{3}{*}{10} &  & \multirow{3}{*}{t-SNERF} & \multirow{3}{*}{\cite{hammad2021t}} & UNSW-NB15 & \multicolumn{1}{c}{} & 100 \\ \cline{5-7} 
 &  &  &  & CIC-IDS2017 & \multicolumn{1}{c}{} & 99.78 \\ \cline{5-7} 
 &  &  &  & Phishing & \multicolumn{1}{c}{} & 99.70 \\ \cline{1-1} \cline{3-7} 
\multirow{2}{*}{11} &  & \multirow{2}{*}{DT} & \multirow{2}{*}{\cite{guezzaz2021reliable}} & NSL-KDD & \multicolumn{2}{c}{99.42} \\ \cline{5-7} 
 &  &  &  & \multirow{3}{*}{CIC-IDS2017} & \multicolumn{2}{c}{98.80} \\ 
\multirow{2}{*}{12} &  & IG+RF & \multirow{2}{*}{\cite{stiawan2020cicids}} &  & \multicolumn{2}{c}{99.86} \\ \cline{3-3} \cline{6-7} 
 &  & IG+J48 &  &  & \multicolumn{1}{c}{99.87} &  \\ \hline
\multirow{3}{*}{13} & \multicolumn{1}{l}{\multirow{10}{*}{Deep Learning}} & DNN & \multirow{3}{*}{\cite{aleesa2021deep}} & \multirow{3}{*}{UNSW-NB15} & \multicolumn{1}{c}{99.92} & 95.9 \\ \cline{3-3} \cline{6-7} 
 & \multicolumn{1}{l}{} & RNN &  &  & \multicolumn{1}{c}{85.42} & 85.4 \\ \cline{3-3} \cline{6-7} 
 & \multicolumn{1}{l}{} & ANN &  &  & \multicolumn{1}{c}{99.26} & 97.89 \\ \cline{1-1} \cline{3-7} 
\multirow{3}{*}{14} & \multicolumn{1}{l}{} & \multirow{3}{*}{DNN} & \multirow{3}{*}{\cite{choudhary2020analysis}} & KDDCUP'99 & \multicolumn{2}{c}{\multirow{3}{*}{91.50}} \\ \cline{5-5}
 & \multicolumn{1}{l}{} &  &  & NSL-KDD & \multicolumn{2}{l}{} \\ \cline{5-5}
 & \multicolumn{1}{l}{} &  &  & UNSW-NB15 & \multicolumn{2}{l}{} \\ \cline{1-1} \cline{3-4} \cline{5-5} \cline{6-7} 
\multirow{2}{*}{15} & \multicolumn{1}{l}{} & \multirow{2}{*}{LSTM+CNN} & \multirow{2}{*} {\cite{al2021stl}} & CICIDS-001 & \multicolumn{1}{c}{} & 99.83 \\ \cline{5-7} 
 & \multicolumn{1}{l}{} &  &  & UNSW-NB15 & \multicolumn{1}{c}{99.17} &  \\  \cline{1-1} \cline{3-7} 
\multirow{2}{*}{16} &  & \multirow{2}{*}{CNN} & \multirow{2}{*}{\cite{hassan2020hybrid}} & KDDCUP’99 & \multicolumn{1}{c}{} & 99 \\ \cline{5-7} 
 &  &  &  & CIC-IDS2018 & \multicolumn{1}{c}{} & 91.50 \\ \hline
 
17 & \multirow{10}{*}{Hybrid Learning} & DNN+ACO & \cite{bhardwaj2021hybrid} & CIC-IDS2017 & \multicolumn{1}{c}{} & 98.25 \\ \cline{1-1} \cline{3-7} 
18 &  & CNN+RNN & \cite{khan2021hcrnnids} & CIC-IDS2018 &  & 97.75
 \\ \cline{1-1} \cline{3-7} 
\multirow{2}{*}{19} &  & \multirow{2}{*}{WFEU+FFDN} & \multirow{2}{*}{\cite{kasongo2020deep}} & AWID & \multicolumn{1}{c}{99.66} & 99.77 \\ \cline{5-7} 
 &  &  &  & \multirow{2}{*}{UNSW-NB15} & \multicolumn{1}{c}{87.10} & 77.16 \\ \cline{1-1} \cline{3-4} \cline{6-7} 
\multirow{2}{*}{20} &  & \multirow{2}{*}{SGM+CNN} & \multirow{2}{*}{\cite{zhang2020effective}} &  & \multicolumn{1}{c}{99.74} & 96.54 \\ \cline{5-7} 
 &  &  &  & CIC-IDS2017 & \multicolumn{1}{c}{} & 99.85 \\ \cline{1-1} \cline{3-4} \cline{5-5} \cline{6-7} 
21 &  & WDLSTM+CNN & \cite{hassan2020hybrid} & UNSW-NB15 & \multicolumn{1}{l}{97.17} & 98.43 \\ \hline
\end{tabular}
}
\caption{Related work summary of various ML Techniques }
\label{tab:related_summary}
\end{table}

\subsection{Limitations of the Existing Works}

One notable limitation in the existing works is their reliance on older datasets such as KDDCUP'99 and NSLKDD, which lack recent attack scenarios featured in datasets like CIC-IDS2018. Consequently, these works may not effectively address contemporary and evolving cyber threats. Furthermore, many of these works overlook the importance of data balancing techniques. This omission leads to variations in their model's performance, particularly in terms of false positive and negative as well as true positive and negative rates. Moreover, the performance scores of existing works are inadequate for efficient network attack detection. As a result, their models may not be suitable for real-world scenarios where data imbalance is a common issue. Additionally, most of these papers employ the use of the complete feature set for conducting their experiments. This approach can be time-consuming and computationally intensive, making it less practical for real-time anomaly detection in large-scale network environments. Another notable limitation is the absence of time complexity analysis in the majority of these papers. Understanding the computational demands of their algorithms is crucial for assessing their feasibility in real-world applications. Lastly, No existing works explore the Stacking Feature Embedded approach, a method that can potentially enhance intrusion detection accuracy by incorporating meta-features. The omission of this approach limits the comprehensiveness and effectiveness of their intrusion detection models.

\section{Methodology}
\label{sec:Method}

\begin{figure*}[!htbp]
\centering
  \includegraphics[width=0.85\textwidth]{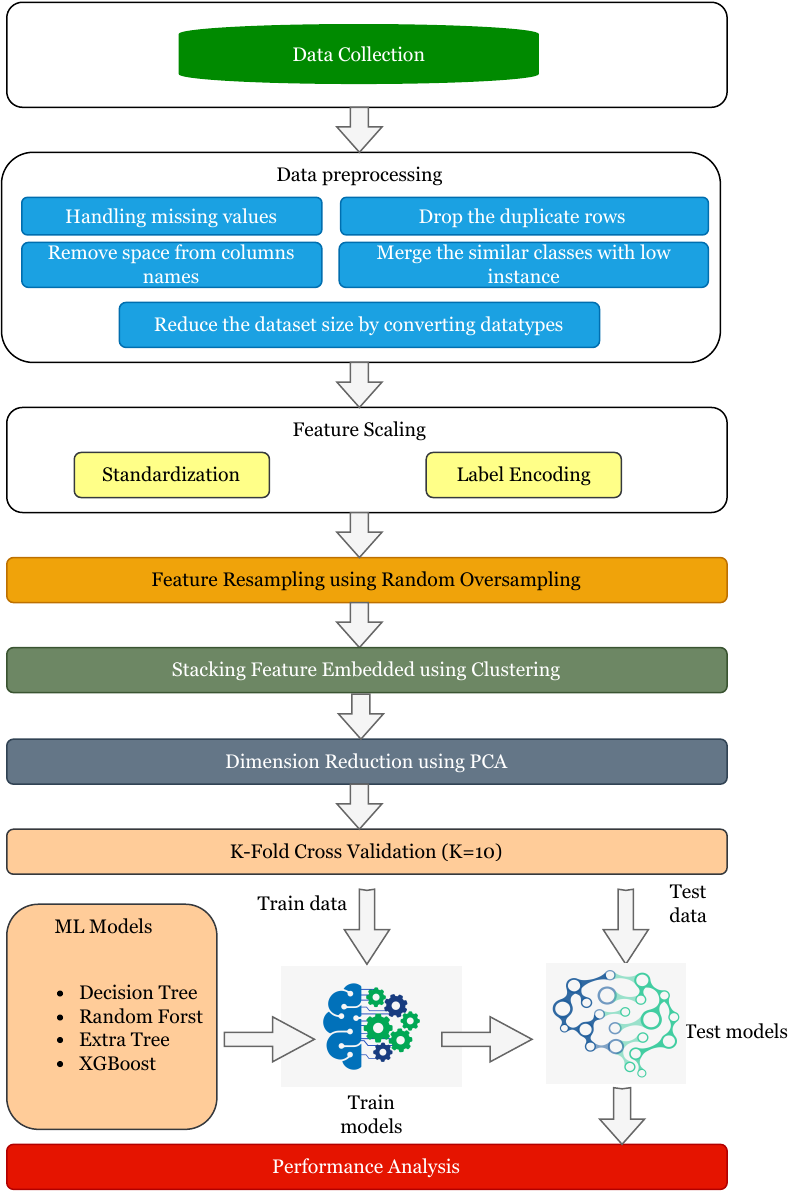}
\caption{Proposed framework for stacking feature embedded with PCA for intrusion detection.}
\label{fig:proposal}
\end{figure*}

This section describes our proposed framework and data preprocessing techniques, including feature resampling and scaling, stacking feature embedded and feature extraction. We also provide a brief recap of the ML algorithms employed for intrusion detection. The proposed framework for stacking feature embedded and dimension reduction for intrusion detection on big and imbalanced datasets can ensure a reliable secure network by detecting the receiving packets as normal or attack packets. The schematic block diagram of our proposed paradigm is illustrated in Figure \ref{fig:proposal}. The approach is structured into the following phases:
\begin{itemize}
    
    \item {Step-1: } Initially, preprocessing is accomplished by handling the missing value, removing space from column names, dropping the duplicate rows, merging the similar classes with low instances and reducing the size of the dataset by converting data types int64 to int32 and float64 to float 32. This step is crucial for data quality improvement. It addresses missing values, data type conversions, and other data cleaning tasks to prepare the dataset for analysis.

    \item {Step-2: } In the feature scaling step, we use standardization for input features and label encoding for output features. Standardization of input features and label encoding of output features ensure that the data is on the same scale, which is essential for ML algorithms.
    
    \item {Step-3: } In the feature resemble step, we use Random Oversampling (RO) by adding random samples to the minority class to solve the problem of the imbalanced dataset and make it a balanced dataset. To address the class imbalance using RO in the dataset, making it more suitable for training ML models.
    
    \item {Step-4: } In the stacking feature embedded step, we utilize the clustering results as meta-features, enhancing the original features within the IDS dataset. Using clustering results as meta-features enriches the dataset with additional information derived from underlying patterns and structures, which can enhance model performance.

    \item {Step-5: } In the feature extraction step, we reduce the dimensionality of the dataset by extracting the feature using Principal Component Analysis (PCA). Dimensionality reduction using PCA can help improve model efficiency by reducing the number of features while retaining essential information.
    
    \item {Step-6: } In this phase, we perform data splitting to create separate training and testing subsets from the pre-processed dataset. We employ a k-fold cross-validation technique with k set to 10. This step is crucial to enhance accuracy and evaluate the model's performance effectively. It ensures that the model is rigorously tested on different subsets of the data.

    \item {Step-7: } In this step, we evaluate the model's performance using four established ML algorithms: DT, RF, ET, and XGB Classification. We employ k-fold cross-validation for these models, aiming to identify the most suitable ML models for IDS.

    \item {Step-8: } In the final phase, we assess the model's performance using various performance metrics, including Precision, Recall, Confusion Matrix, Accuracy, F1-score, and ROC Curve. These metrics serve as benchmarks for comparing our model's performance with existing models, enabling us to select the best-performing model for the IDS task.

\end{itemize}

\subsection{Data Collection}

We have utilized two benchmark big datasets for our research schemes, namely UNSW-NB15 \citep{moustafa2015unsw} and CIC-IDS2017 \citep{sharafaldin2018toward}. Both datasets are realistic to the IDS environments and have up-to-date attack categories to detect attacks. The following sections give a closer look at each of the datasets.

\subsubsection{UNSW-NB15}
\begin{figure*}[!htbptbp]
	\centering
	\subfloat[before pre-process]{\includegraphics[scale=.43]{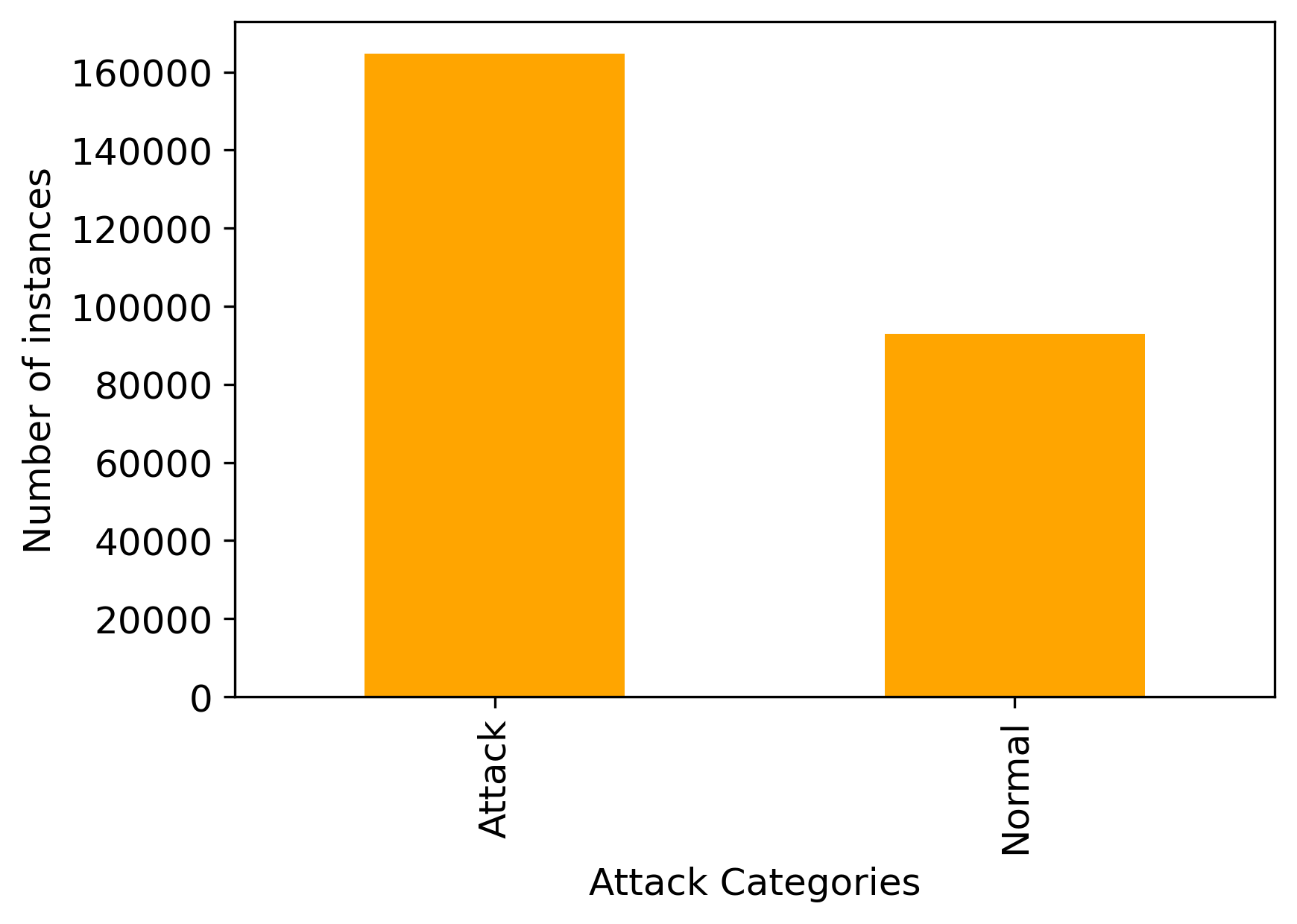}}
	\subfloat[after pre-process]{\includegraphics[scale=.43]{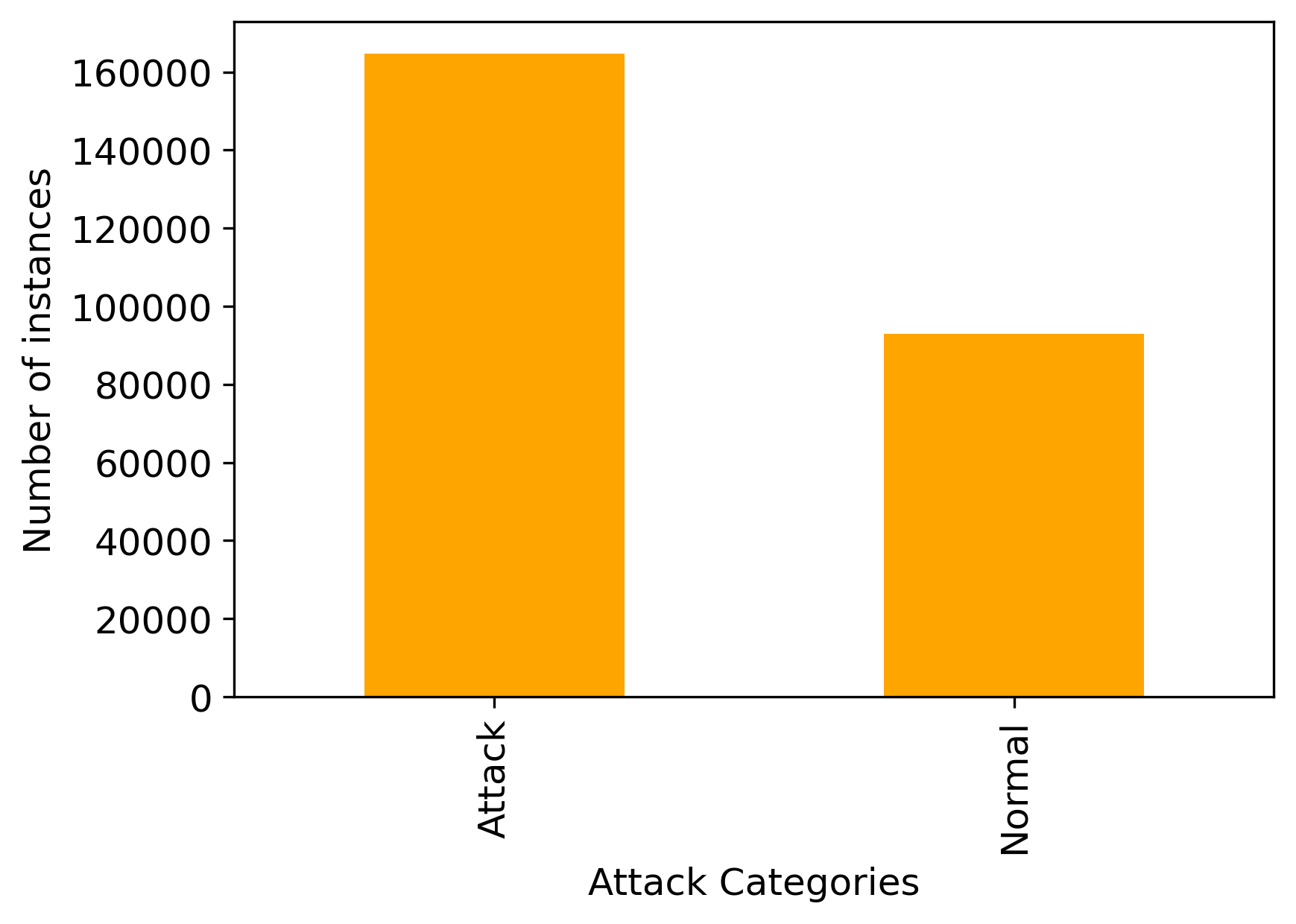}}
 
	\subfloat[after oversampling]{\includegraphics[scale=.5]{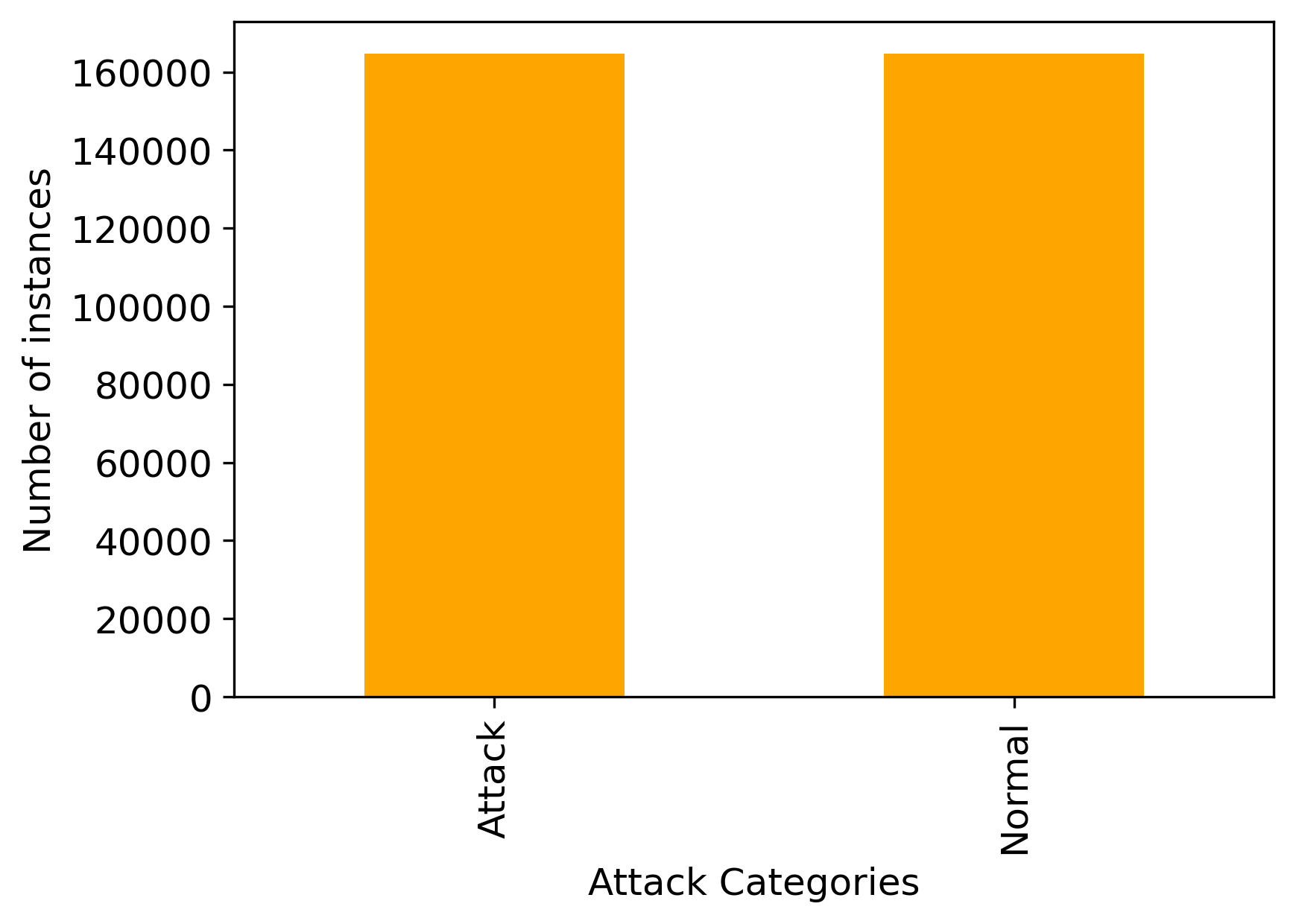}} 
 
	\caption{Binary frequency distribution of UNSW-NB15 dataset}
	\label{fig:bfdis_unswnb}
\end{figure*}

\begin{figure*}[!htbptbp]
	\centering
	\subfloat[before pre-process]{\includegraphics[scale=.43]{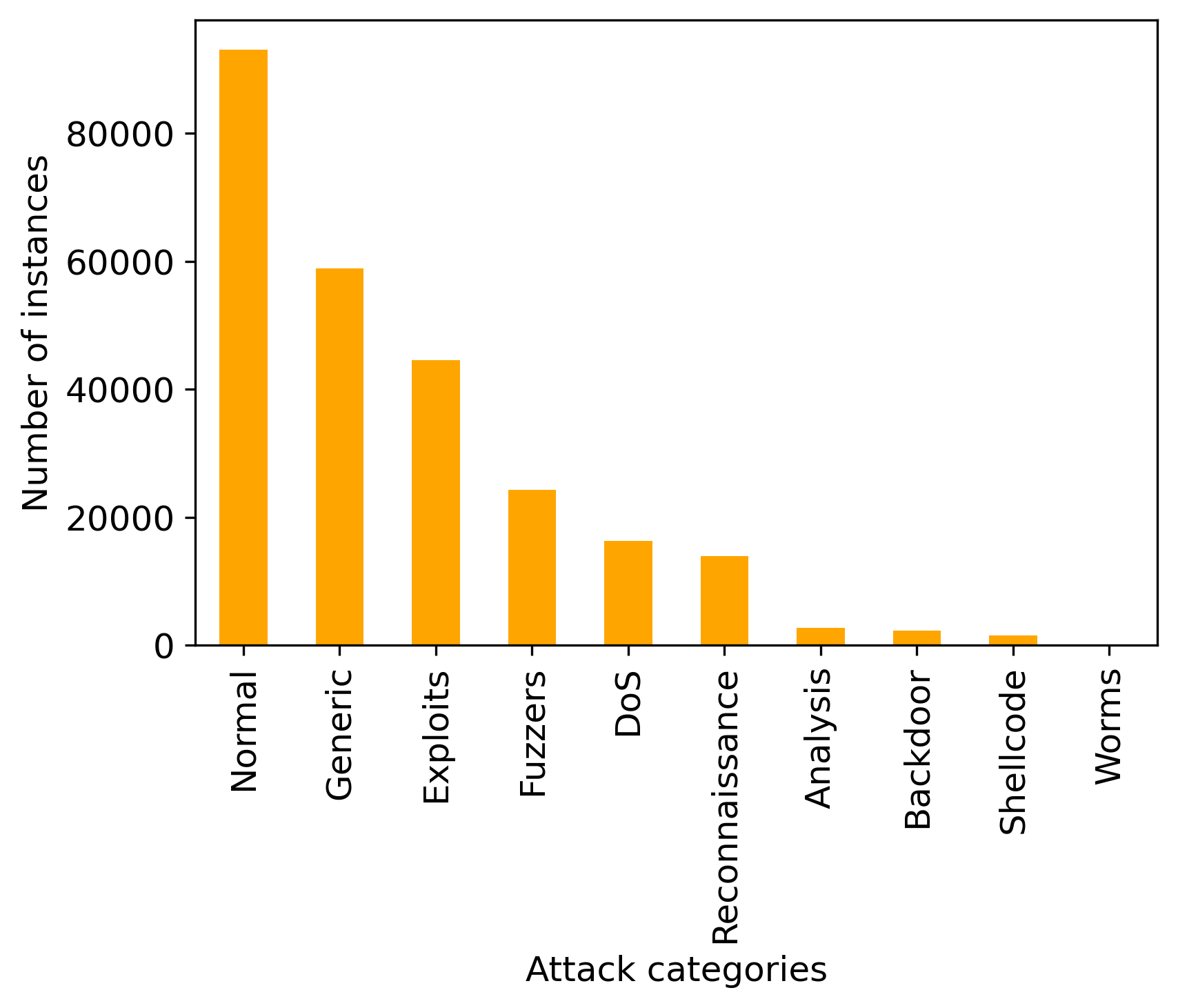}}
	\subfloat[after pre-process]{\includegraphics[scale=.43]{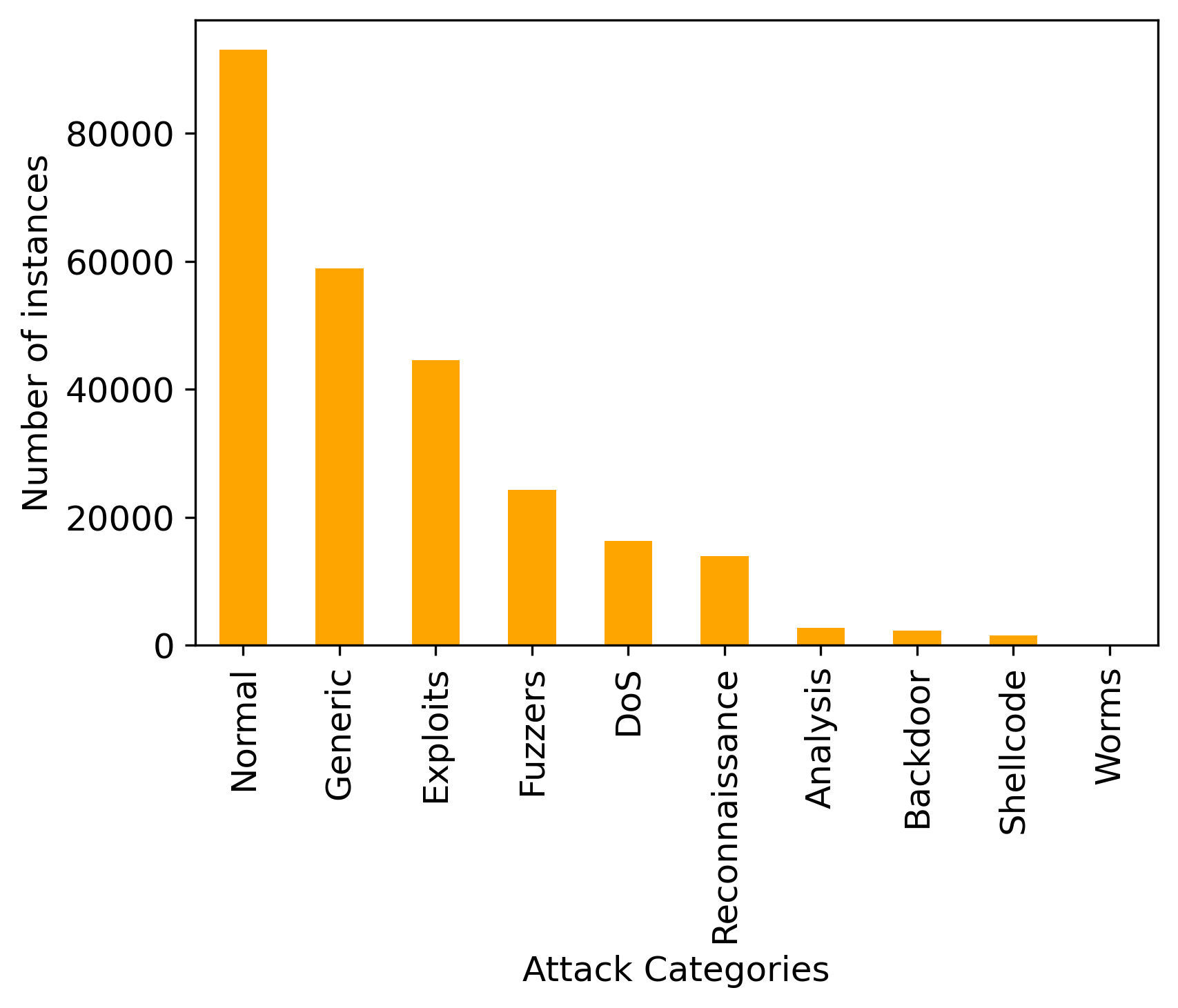}}
 
	\subfloat[after oversampling]{\includegraphics[scale=.5]{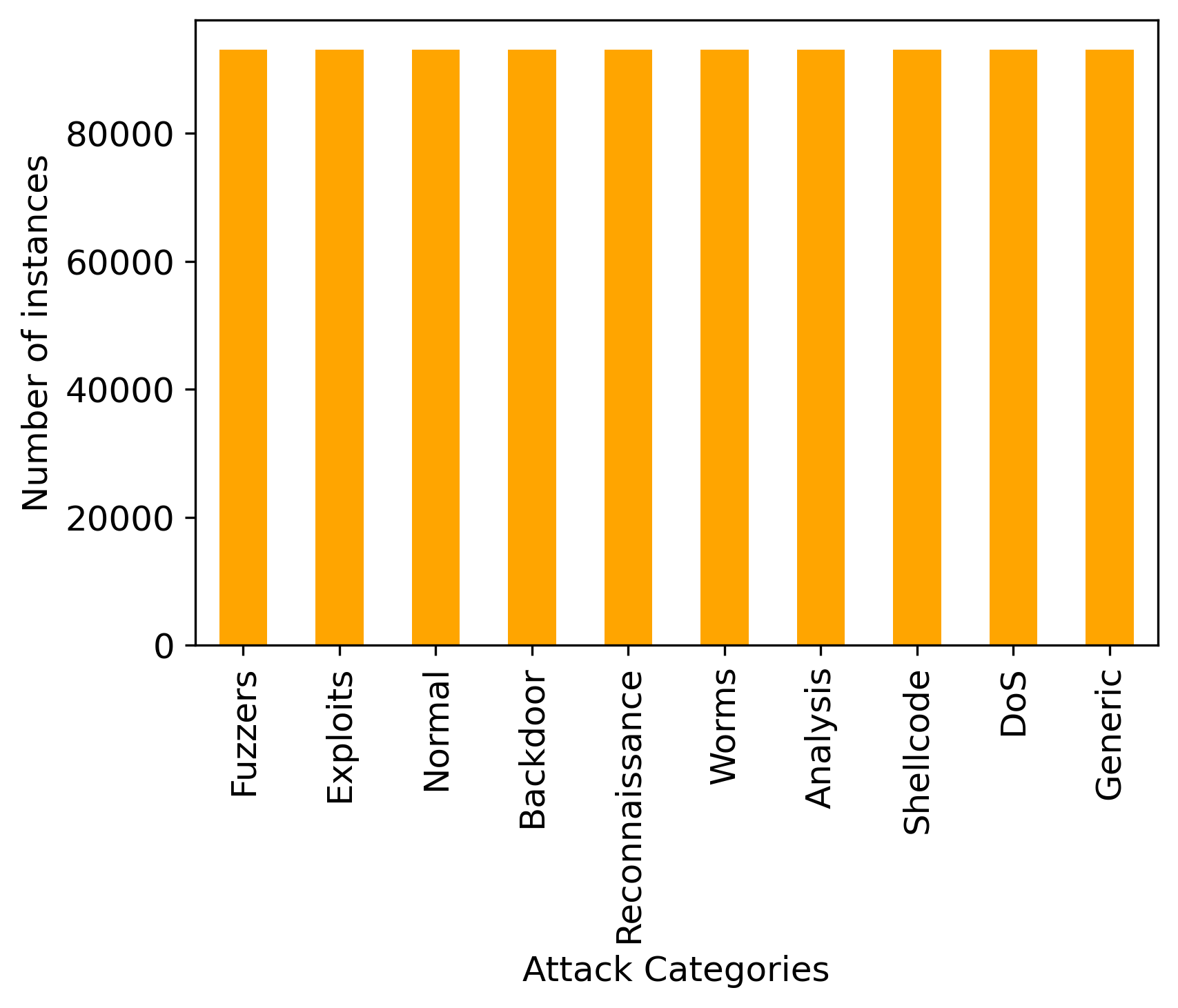}} 
	\caption{multilabel frequency distribution of UNSW-NB15 dataset}
	\label{fig:mfdis_unswnb}
\end{figure*}

UNSW-NB 15 is a fairly recent dataset that comprises a vast quantity of internet traffic patterns with 9 types of malicious activities, as opposed to KDD’98, NSL-KDD, KDDCUP' 99, CAIDA, Kyoto 2006 + and ISCX dataset \citep{protic2023numerical}. It includes current minimal imprint assaults as well as contemporary Netflow for both regular and unusual situations. The IXIA perforectStrom apparatus was utilized within the Cyber Run Lab of the ACCS (Australian Center for Cyber Security) to make synthetic modern assault and genuine advanced ordinary behaviors for creating the crude network packets of the UNSW-NB15 dataset \citep{moustafa2015unsw}. By catching100 GB of the crude activity utilizing the Tcpdump instrument. To create completely 49 features with the class, 12 algorithms are created by employing Argus and Bro-IDS devices. Within the four CSV records, two million and 540,044 records are put away. A setup is conducted on the dataset by dividing it into a training set and a testing set. It has a total of 257673 entries were 175341 entries in the training set and 82332 entries in the testing set. With their respective class labels, the dataset includes both real-world modern typical behavior and staged attack actions from the present day. It incorporates 9 distinct contemporary new attacks as well as a large range of real-world activities \citep{moustafa2016evaluation}. Among the sorts of attacks are fuzzier, backdoor, analysis, reconnaissance, exploit, generic, DoS, shellcode and worm attacks. The data are very unbalanced in this dataset. Figures \ref{fig:bfdis_unswnb} and \ref{fig:mfdis_unswnb} demonstrate the distribution of benign and attack data for binary and multi labels before preprocessing, after preprocessing and after oversampling, respectively, while Table \ref{tab:describe_unswnb15} briefly describes all of the attack classes in this dataset. In Table \ref{tab:unswnb_attacks} shows the frequency distribution of attack categories.

\begin{table}[]
\centering
\resizebox{\textwidth}{!}{

\begin{tabular}{|p{3cm}|p{10cm}|}
\hline
Attack Categories & Brief description \\ \hline
Fuzzers & By supplying a vast volume of random data, the insider tries to crash a software,   operating system, or network. \\ \hline
Backdoor & Cyber attackers can get illegal access to websites using this form of software. By focusing on vulnerable entry points, the intruders were able to disseminate malware throughout the system. \\\hline
Analysis & Pay special attention to malware attacks and computer intrusions in which attackers gain permissions by utilizing their technological capabilities. \\\hline
Reconnaissance & Gathers data on system flaws that can be used to gain control of the system. \\\hline
Exploit & A piece of software that exploits security flaws and vulnerabilities. An attacker can gain unrestricted access with this attack. \\\hline
Generic & Has the ability to decrypt all block ciphers without having to know the cipher's structure. \\ \hline
DoS & User access to machines and network resources can be suspended by an attacker. By delivering too much confusing traffic, the attacker overwhelms the network. \\ \hline
Shellcode & It is a   sequence of instructions that executes software commands to harm a machine. \\ \hline
Worm & It includes security flaws that attack the host machine and spread throughout the network. It is capable of exploiting many applications' security flaws.\\\hline
\end{tabular}
}
\caption{Briefly describe all of the attack classes in the UNSW-NB15 dataset}
\label{tab:describe_unswnb15}
\end{table}

\begin{table}[]
\centering
\begin{tabular}{|l|l|l|}
\hline
Attack   Categories & Count & \% (percentage) \\ \hline
Normal & 93000 & 36.09 \\
Generic & 58871 & 22.85 \\
Exploits & 44525 & 17.28 \\
Fuzzers & 24246 & 9.41 \\
DoS & 16353 & 6.35 \\
Reconnaissance & 13987 & 5.43 \\
Analysis & 2677 & 1.04 \\
Backdoor & 2329 & 0.90 \\
Shellcode & 1511 & 0.59 \\
Worms & 174 & 0.07 \\ \hline
Total & 257673 & 100\\ \hline
\end{tabular}
\caption{The frequency distribution of attack categories of the UNSW-NB15 dataset}
\label{tab:unswnb_attacks}
\end{table}

\subsubsection{CIC-IDS2017}
% Intrusion Detection Systems (IDSs) and Intrusion Prevention Systems (IPSs) are the most important defense tools against sophisticated and ever-growing network attacks. Due to the lack of reliable test and validation datasets, anomaly-based intrusion detection approaches are suffering from consistent and accurate performance evolutions \citep{sharafaldin2018toward}. The CIC-IDS2017 dataset includes benign and current frequent assaults, closely resembling real-world data (PCAPs). It also reports the summary of traffic monitoring using CICFlowMeter, which comprises labelled flows regarding the time stamp, source and destination IPs, source and destination ports, protocols and attack vectors (CSV files) \citep{ sharafaldin2018intrusion} are among the assaults that have been employed. 

\begin{figure*}[!htbptbp]
	\centering
	\subfloat[before pre-process]{\includegraphics[scale=.43]{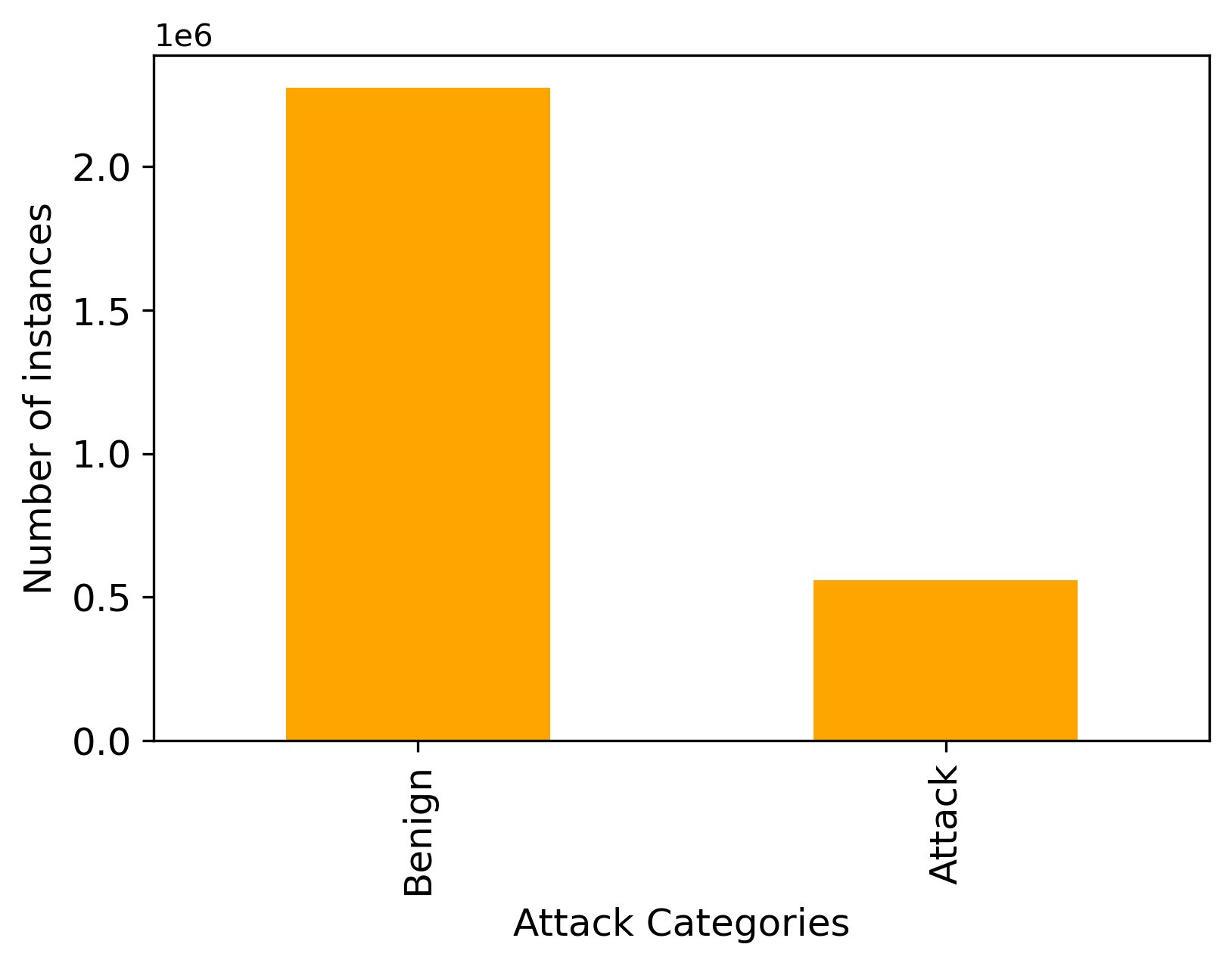}}
	\subfloat[after pre-process]{\includegraphics[scale=.43]{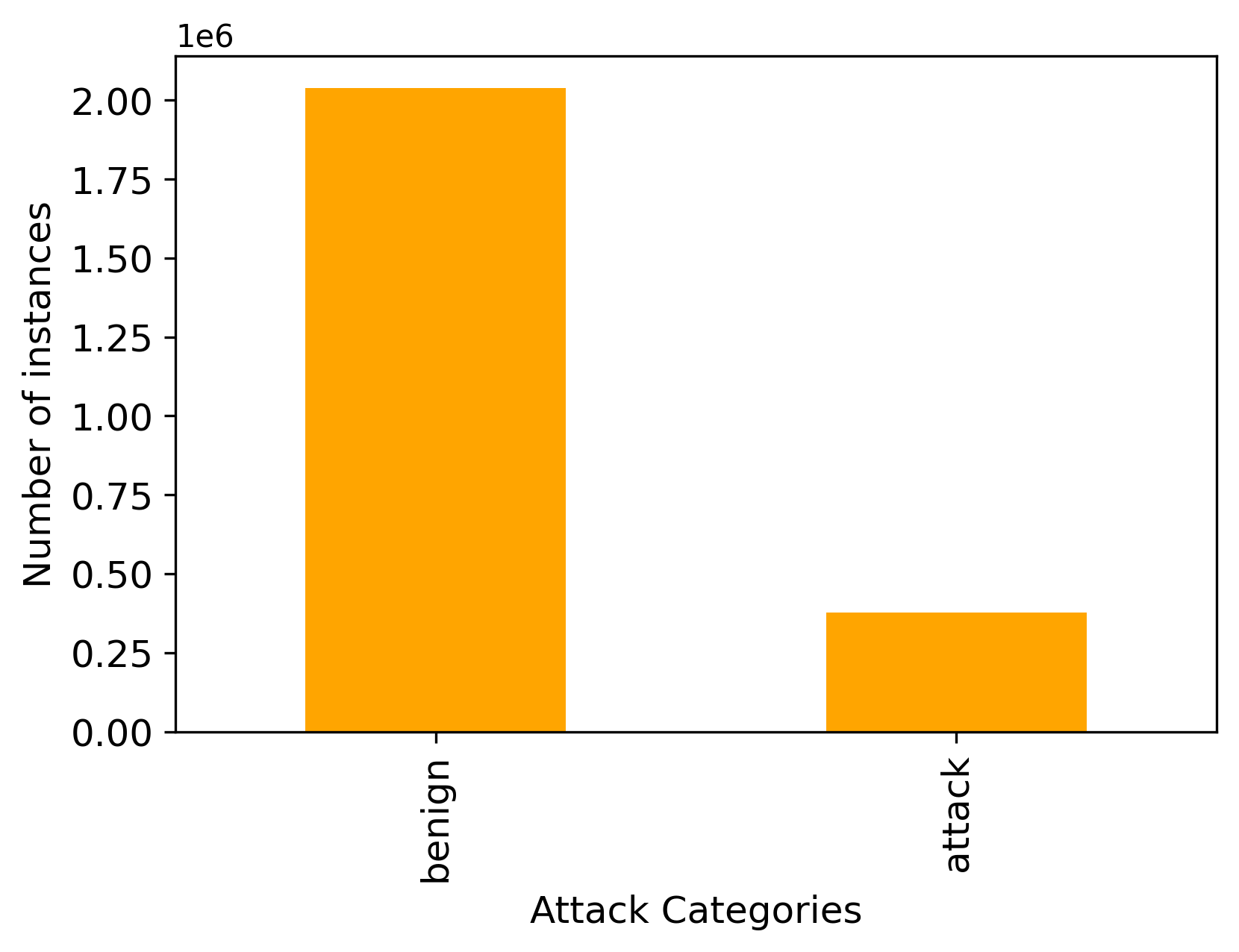}}
 
	\subfloat[after oversampling]{\includegraphics[scale=.5]{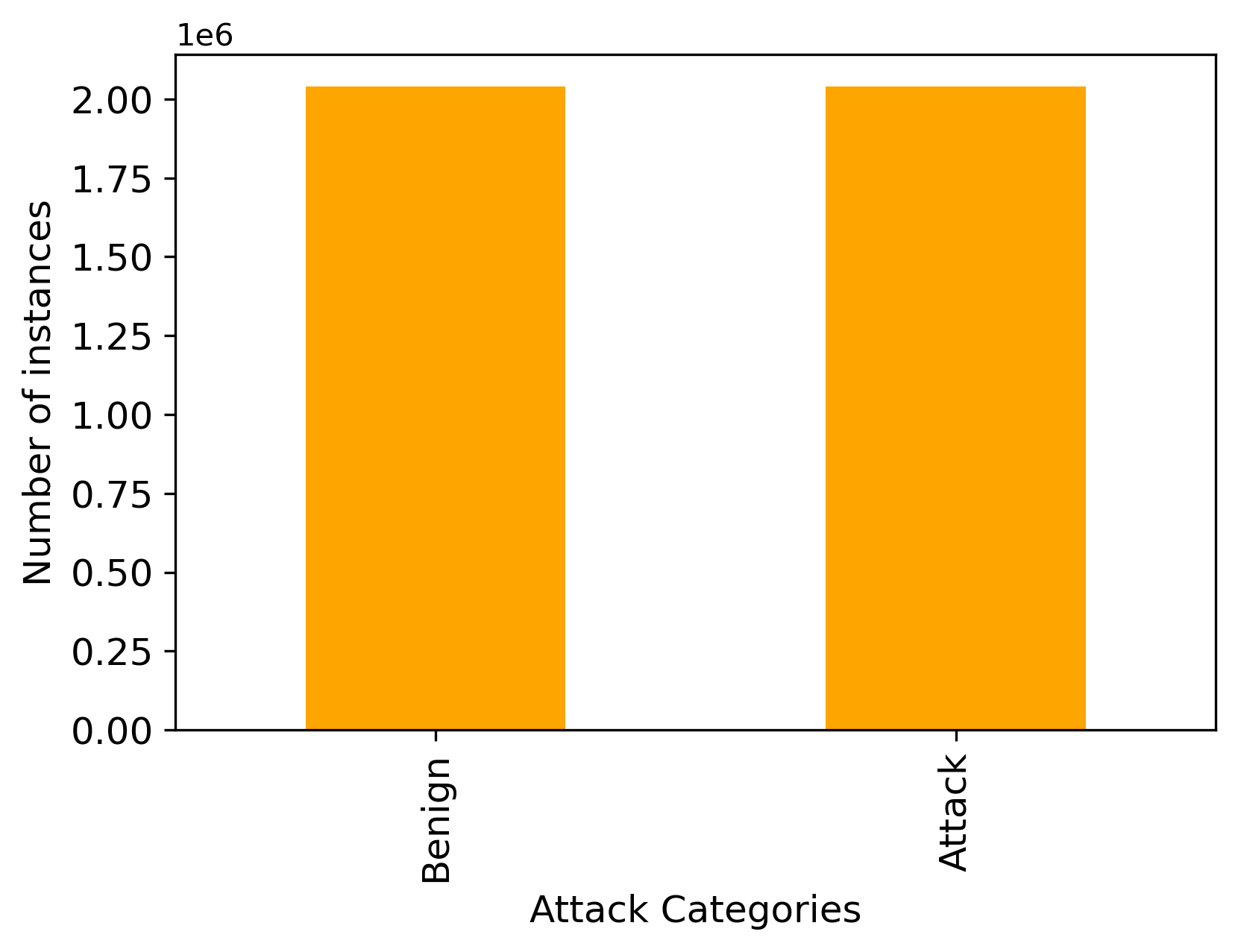}} 
 
	\caption{Binary frequency distribution of CIC-IDS2017 dataset}
	\label{fig:bfdis_cicids}
\end{figure*}

\begin{figure*}[!htbptbp]
	\centering
	\subfloat[before pre-process]{\includegraphics[scale=.43]{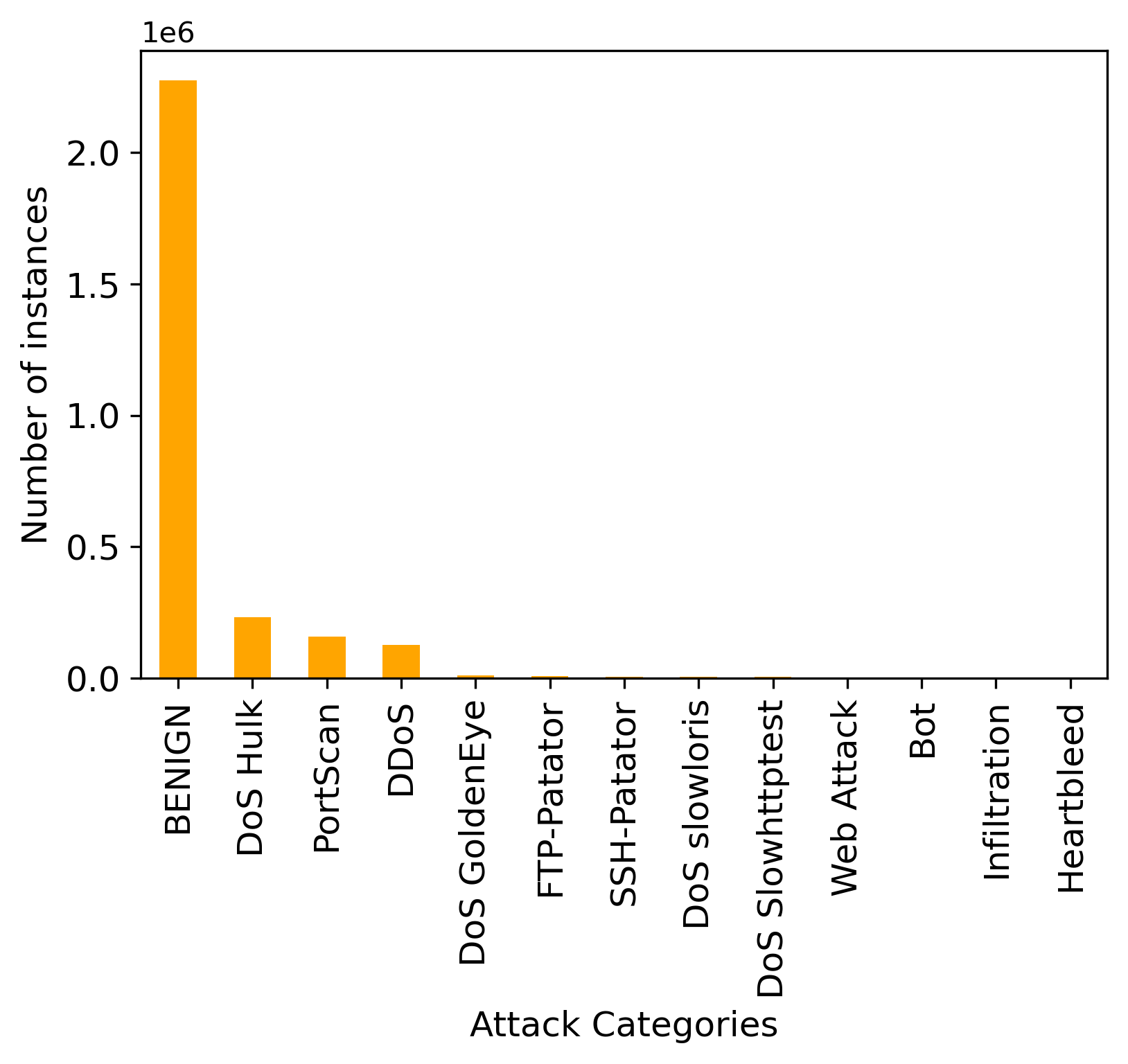}}
	\subfloat[after pre-process]{\includegraphics[scale=.43]{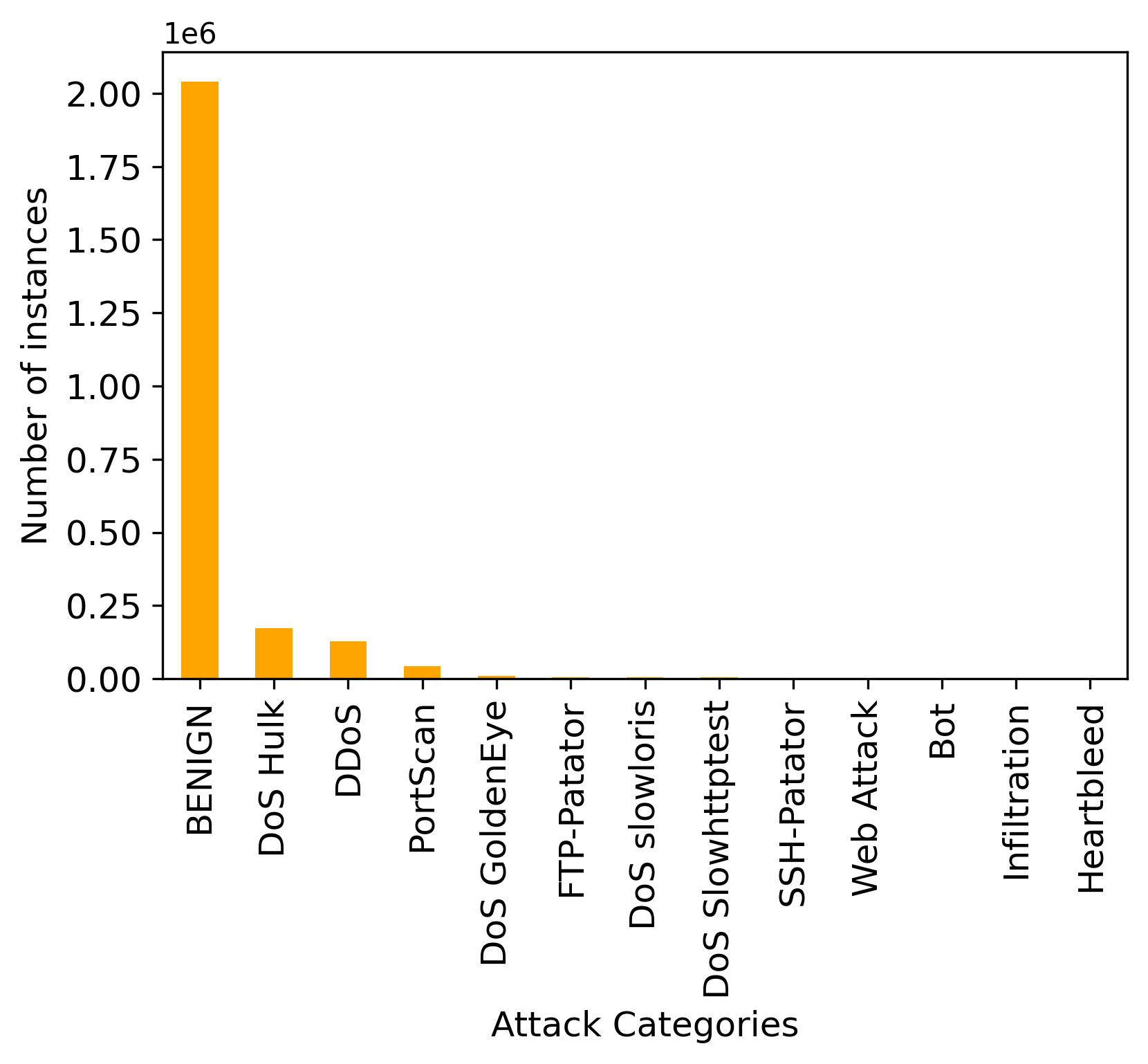}}
 
	\subfloat[after oversampling]{\includegraphics[scale=.5]{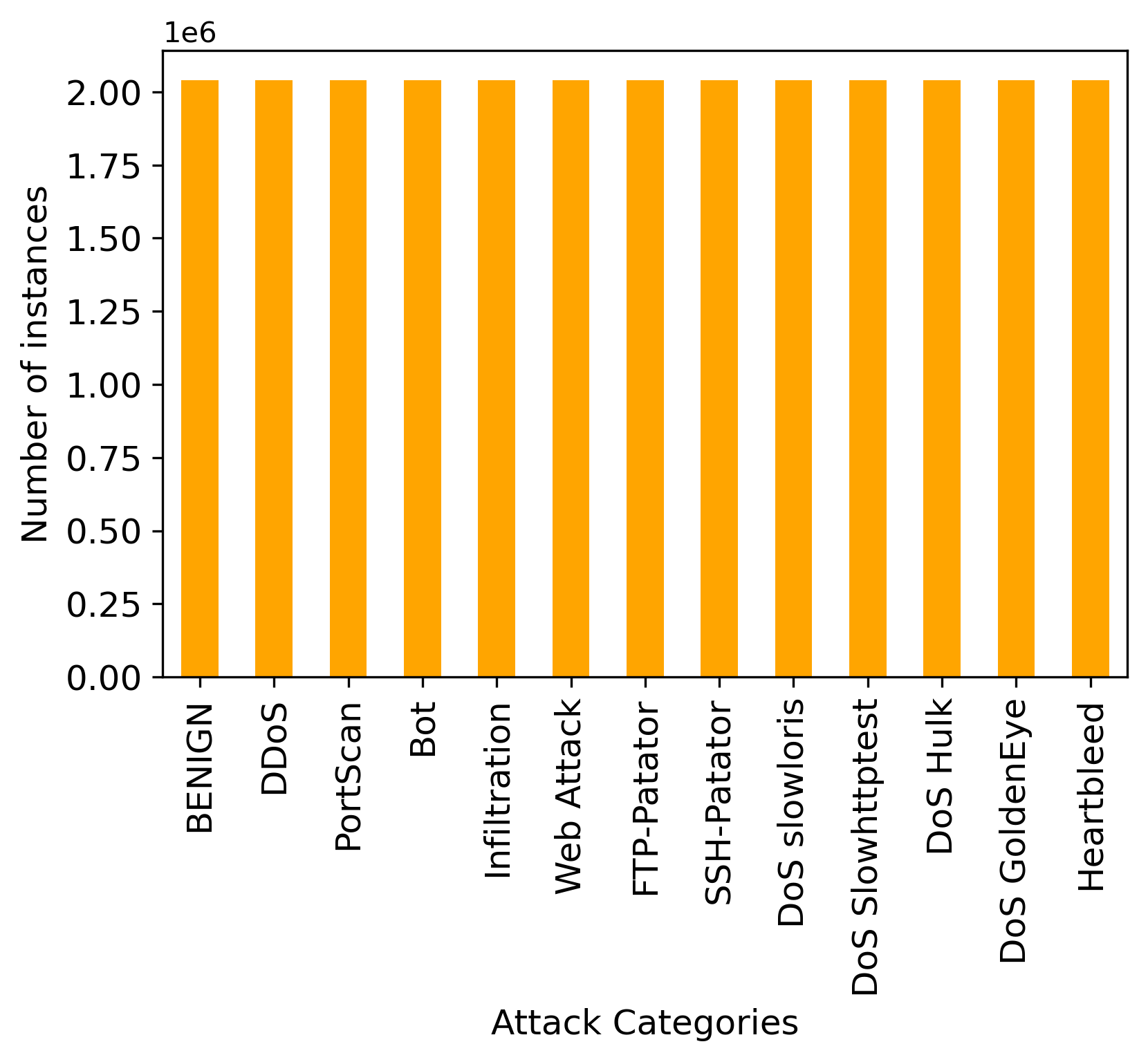}} 
 
	\caption{multilabel frequency distribution of the CIC-IDS2017 dataset}
	\label{fig:mfdis_cicids}
\end{figure*}

Intrusion Detection Systems (IDSs) and Intrusion Prevention Systems (IPSs) serve as vital defenses against the relentless and increasingly sophisticated landscape of network attacks. An ongoing challenge in the field is the scarcity of dependable test and validation datasets, which hinders the consistent and accurate evaluation of anomaly-based intrusion detection methods, as noted by Sharafaldin et al. in 2018 \citep{sharafaldin2018toward}. One promising solution to this issue lies in the CIC-IDS2017 dataset, which features a comprehensive collection of benign network traffic and a diverse range of contemporary, frequently encountered cyber attacks. This dataset closely mirrors real-world scenarios by utilizing PCAPs. Moreover, it provides a detailed summary of traffic monitoring through the utilization of CICFlowMeter, offering labeled network flows with key information, including timestamps, source and destination IP addresses, source and destination ports, protocols, and delineated attack vectors, all meticulously documented in CSV files, as described by Sharafaldin et al. in their 2018 work on intrusion detection \citep{sharafaldin2018intrusion}.
To create various kinds of assaults such as Brute Force FTP, Brute Force SSH, DoS, Heartbleed, Web Attack, Infiltration, Botnet and DDoS presented the B-Profile technology, which profiles the conceptual activity of individual contacts and produces lifelike benign baseline flow. According to the dataset's appraisal methodology which was proposed \citep{gharib2016evaluation}, there are 11 requirements that must be met in order to create a trustworthy benchmark dataset. only this dataset meets all of the requirements and hardly any of the prior IDS datasets were capable of covering those 11 requirements. The distribution of binary and multi categories of attack labels before preprocessing, after preprocessing and after oversampling is depicted in Figure\ref{fig:bfdis_cicids} and Figure\ref{fig:mfdis_cicids} and attack categories are shown in Table \ref{tab:cicids_attacks}.

\begin{table}[!htbptbp]
\centering
\begin{tabular}{|l|l|l|}
\hline
Attack Categories & Count & \% (percentage) \\
\hline
BENIGN & 2273097 & 80.3 \\
DoS Hulk & 231073 & 8.16 \\
PortScan & 158930 & 5.61 \\
DDoS & 128027 & 4.52 \\
DoS GoldenEye & 10293 & 0.36 \\
FTP-Patator & 7938 & 0.28 \\
SSH-Patator & 5897 & 0.21 \\
DoS slowloris & 5796 & 0.2 \\
DoS Slowhttptest & 5499 & 0.19 \\
Web Attack & 2180 & 0.08 \\
Bot & 1966 & 0.07 \\
Infiltration & 36 & 0.01 \\
Heartbleed & 11 & 0.01 \\ \hline
Total & 2830743 & 100\\ \hline
\end{tabular}
\caption{The frequency distribution of attack categories of the CIC-IDS2017 dataset}
\label{tab:cicids_attacks}
\end{table}

% \begin{figure*}[]
% \centering
%   \includegraphics[width=0.65\textwidth]{binary_cicids.png}
% \caption{Binary frequency distribution of CIC-IDS2017 dataset}
% \label{fig:bfdcicids}
% \end{figure*}
% \begin{figure*}[]
% \centering
%   \includegraphics[width=0.65\textwidth]{multi_cicids.png}
% \caption{Multi frequency distribution of CIC-IDS2017 dataset}
% \label{fig:bfdcicids}
% \end{figure*}

\subsubsection{CIC-IDS2018}

\begin{figure}[!htbptbp]
	\centering
	\subfloat[before pre-process]{\includegraphics[scale=.350]{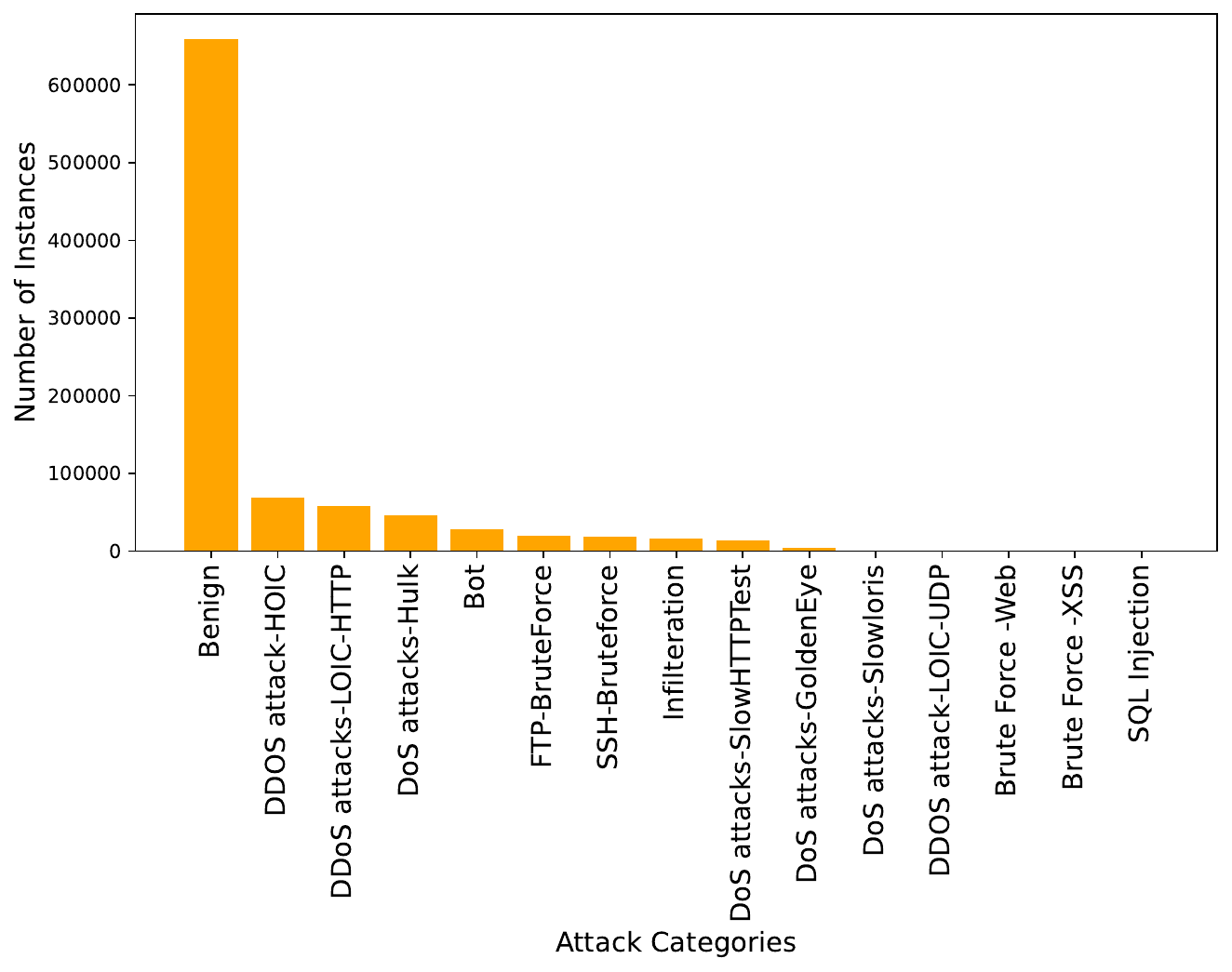}}
 
	\subfloat[after pre-process]{\includegraphics[scale=.350]{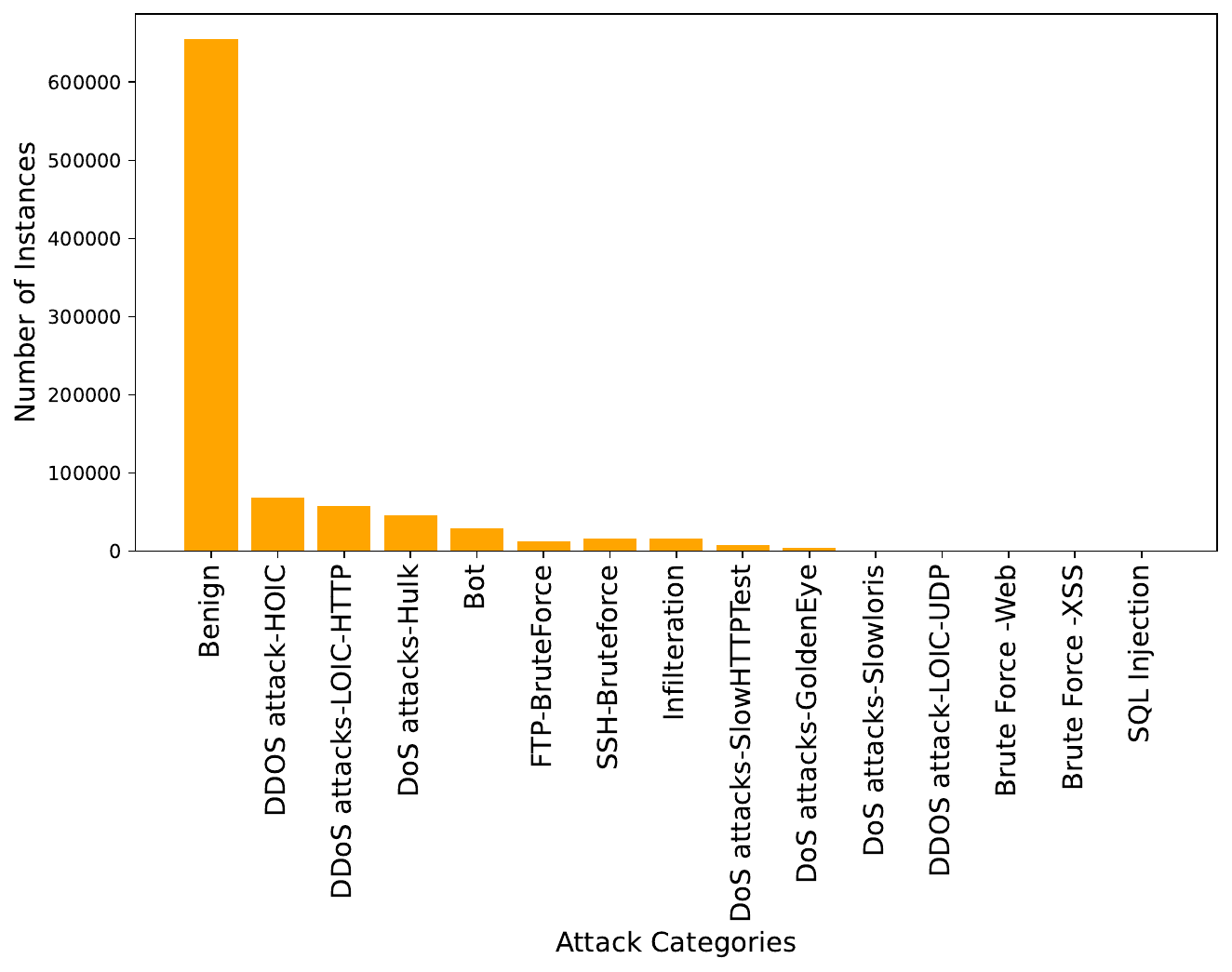}}
 
	\subfloat[after oversampling]{\includegraphics[scale=.350]{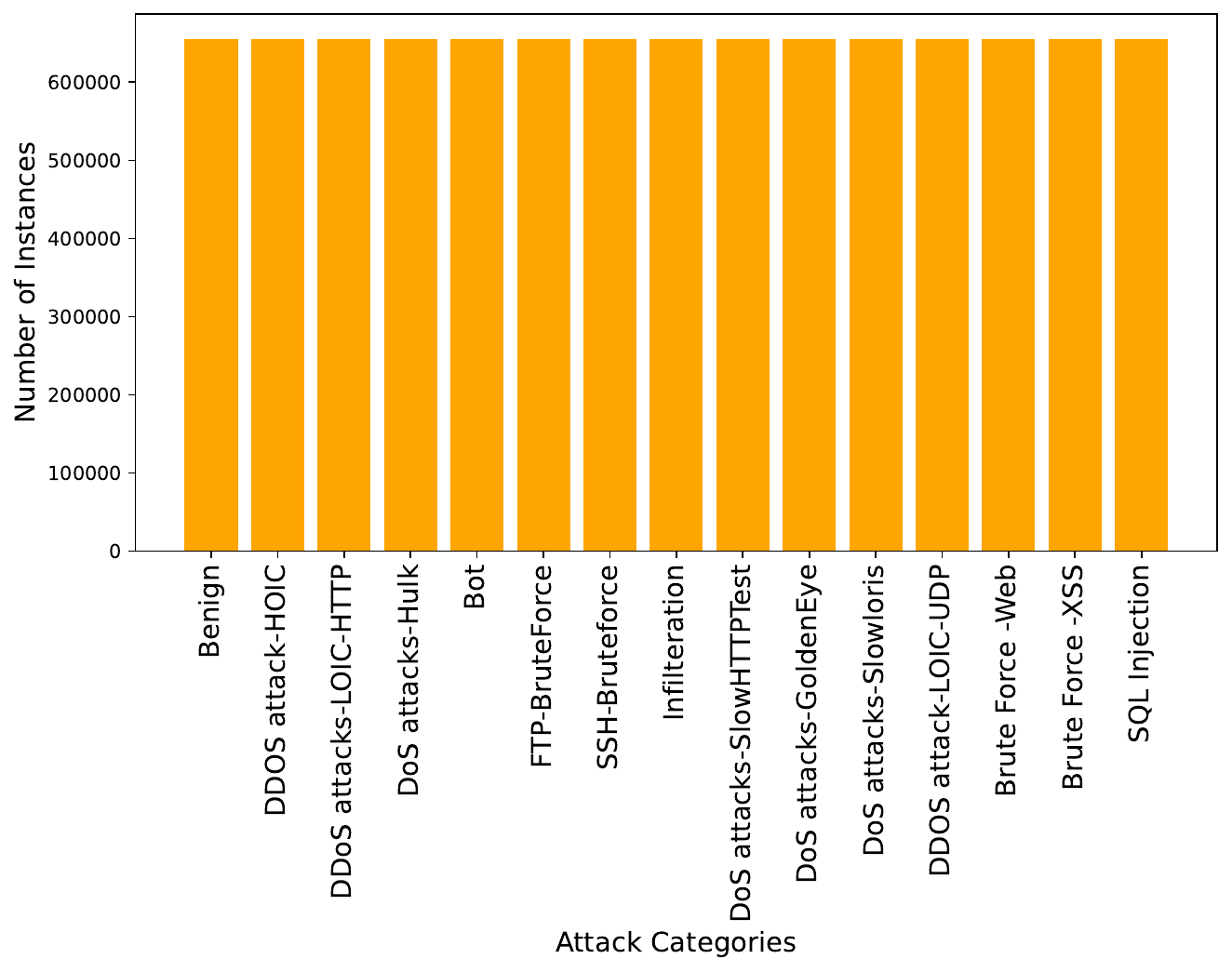}} 
	\caption{The frequency distribution of CIC-IDS2018 Dataset}
	\label{fig:mfdis_cicids18}
\end{figure}

The CSE-CIC-IDS2018 dataset \citep{sharafaldin2018toward}, a collaborative initiative by the Communications Security Establishment (CSE) and the Canadian Institute for Cybersecurity (CIC), endeavors to meet the pressing need for comprehensive datasets suitable for rigorously testing intrusion detection systems, with a specific focus on network-based anomaly detection. Anomaly detection holds significant promise for identifying emerging threats, but its practical implementation has been hindered by inherent complexities, demanding extensive testing and evaluation. Conventional datasets used for these purposes have shown limitations, stemming from privacy constraints, excessive anonymization, and a lack of representation of contemporary threat trends. This project seeks to overcome these limitations by introducing a structured approach for crafting benchmark datasets. This approach revolves around the creation of user profiles that offer abstract representations of network events and behaviors. These profiles are thoughtfully aggregated to construct datasets that exhibit distinctive features, encompassing a wide range of evaluation scenarios. The final dataset encompasses seven distinct attack scenarios, namely Brute-force, Heartbleed, Botnet, Denial of Service (DoS), Distributed Denial of Service (DDoS), Web attacks, and network infiltration. The attack infrastructure comprises 50 machines, while the target organization consists of 5 departments, incorporating 420 machines and 30 servers. The dataset includes meticulously collected network traffic and system logs from each machine, along with the extraction of 80 features through the application of CICFlowMeter-V3. This dataset constitutes an invaluable resource for the systematic evaluation of intrusion detection systems and offers a response to the growing demand for dynamic, adaptable, and comprehensive datasets within the domain of cybersecurity. It holds substantial promise for contributing to the advancement of intrusion detection research and its practical implementation in real-world security scenarios.In our research, we have sampled 10\% of the dataset from each class to accommodate computational resource constraints. Our experimental dataset comprises 933,277 data points with 80 distinct features and encompasses 15 attack classes. The frequency distribution of attack categories for the CIC-IDS2018 dataset is detailed in Table \ref{tab:cicids18_attacks}. The distribution of attack categories in the bar chart, both before preprocessing and after preprocessing, as well as after applying oversampling techniques, is illustrated in Figure \ref{fig:mfdis_cicids18}.

\begin{table}[]
\centering
\begin{tabular}{|l|l|l|}
\hline
Attack Categories        & Count  & (\%) Percentage \\ \hline
Benign                   & 658454 & 70.553         \\ \hline
DDOS attack-HOIC         & 68601  & 7.351          \\ \hline
DDoS attacks-LOIC-HTTP   & 57619  & 6.174          \\ \hline
DoS attacks-Hulk         & 46191  & 4.949          \\ \hline
Bot                      & 28619  & 3.067          \\ \hline
FTP-BruteForce           & 19336  & 2.072          \\ \hline
SSH-Bruteforce           & 18759  & 2.01           \\ \hline
Infilteration            & 16193  & 1.735          \\ \hline
DoS attacks-SlowHTTPTest & 13989  & 1.499          \\ \hline
DoS attacks-GoldenEye    & 4151   & 0.445          \\ \hline
DoS attacks-Slowloris    & 1099   & 0.118          \\ \hline
DDOS attack-LOIC-UDP     & 173    & 0.019          \\ \hline
Brute Force -Web         & 61     & 0.007          \\ \hline
Brute Force -XSS         & 23     & 0.002          \\ \hline
SQL Injection            & 9      & 0.001          \\ \hline
Total                    & 933277 & 100            \\ \hline
\end{tabular}
\caption{The frequency distribution of attack categories of the CIC-IDS2018 Dataset}
\label{tab:cicids18_attacks}
\end{table}

\subsection{Data preprocessing}
Data preprocessing is a crucial part of any ML model. Models without preprocessing can create problems with invalid, overfitting, generating error models, providing low accuracy and much more. So, preprocessing is a very significant part of an ML model. To analyze our model, we have used some preprocessing techniques such as: handling the missing value by eradicating rows containing null, -inf and inf values, removing space from columns names to work with columns smoothly and dropping the duplicate rows by keeping the first one and delete the rest from the dataset, merge the similar classes with low instance from output columns and finally, reduce the dataset size by converting data types from int64 to int32 and float64 to float32 to train models with less dataset size but same dataset entries.

\subsection{Feature Scaling to normalize the features}
Feature scaling is a crucial preprocessing step aimed at normalizing the values of features within a consistent range. In our approach, we have employed both standardization and label encoding techniques to achieve this objective.

\subsubsection{Standardization}
Standardization, also known as z-score normalization, is a pivotal method for feature scaling. It involves subtracting the mean from each feature's value and dividing it by the standard deviation. This technique is especially effective when there is a substantial disparity in feature values within the input data. Post-standardization, all features share a common scale, boasting a mean ($\mu$) of zero and a standard deviation ($\sigma$) of one. This process significantly enhances the accuracy of our predictive models. Equation \ref{equ:std} presents the mathematical representation of the Z-score normalization.

\begin{equation}
{x_{new}} = \frac{{x - \mu }}{\sigma }
\label{equ:std}
\end{equation}

In this equation, $x$ represents the original feature value, ${x_{new}}$ signifies the standardized value, $\mu$ corresponds to the mean of the original feature, and $\sigma$ denotes the standard deviation of the original feature.

\subsubsection{Label Encoding}
Label encoding is the practice of converting categorical data into numerical values, facilitating their utilization in machine learning algorithms. To train a machine learning model, we must transform categorical values into numerical representations to facilitate the model-building process during the training phase. This is achieved by replacing categorical values with integers ranging from 0 to (n-1), where 'n' represents the total number of unique classes. For instance, if there are 11 different categorical classes, we assign integers from 0 to 10 in place of these classes. Table \ref{tab:label_encode} exemplifies the label encoding process.

\begin{table}[!htbp]
\centering
\begin{tabular}{|l|l|}
\hline
Attack types & Label Encoding \\
\hline
Normal & 0 \\ \hline
Generic & 1 \\ \hline
Exploits & 2 \\ \hline
Fuzzers & 3 \\ \hline
DoS & 4 \\ \hline
Reconnaissance & 5 \\ \hline
Analysis & 6 \\ \hline
Backdoor & 7 \\ \hline
SSH-Patator & 8 \\ \hline
Shellcode & 9 \\ \hline
Worms & 10\\ \hline
\end{tabular}
\caption{Label encoding process}
\label{tab:label_encode}
\end{table}

\subsection{Feature Resampling using Random Oversampling (RO)}
\begin{figure*}[!htbptbp]
\centering
  \includegraphics[width=0.55\textwidth]{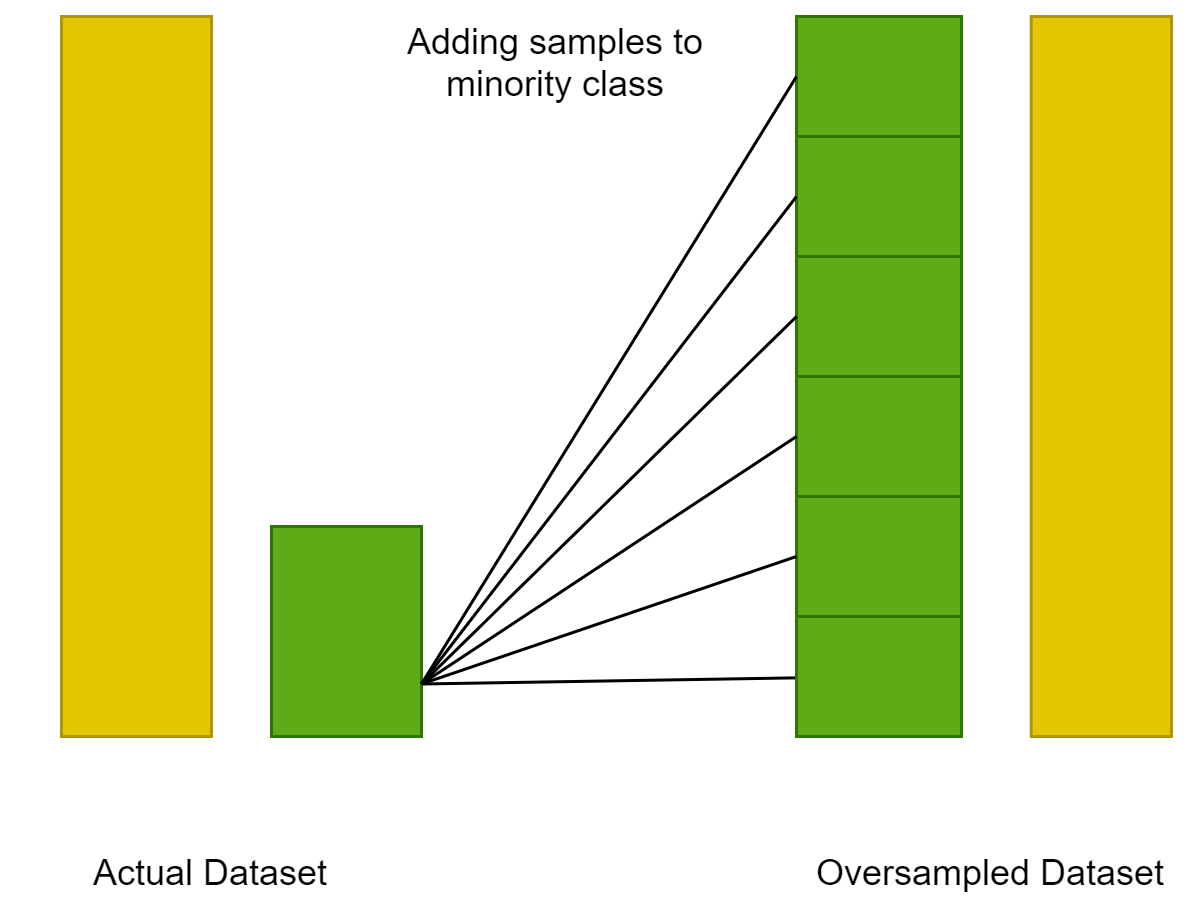}
\caption{Random oversampling process}
\label{fig:randomsample}
\end{figure*}

Feature resampling is a process to rebalance the feature from the imbalanced features in a dataset. The Random oversampling (RO) delivers a naive method to rebalance the class spreading for an imbalanced dataset. It performs an arbitrarily replicating instances from the minority group and incorporating them into the training part. For instance, if the ratio of the dataset’s class is 20:80, then 20 belongs to the minority and 80 belongs to the majority class. It is efficient for skewed distribution algorithms and for a class that can stimulate to fit for the model by replicating instances. In this proposed framework we considered big imbalanced datasets where RO is very crucial to balance the dataset for improving the performance without occurring overfitting problem. The RO process is depicted in Figure\ref{fig:randomsample}.

% SFE-PCA
\subsection{Stacking Feature Embedded using Clustering with PCA}
Within our experimental framework, we introduce a novel methodology known as Stacking Feature Embedded with PCA (SFE-PCA). This approach combines clustering and dimensionality reduction techniques to improve the performance of our ML models.

In the "Stacking Feature Embedded" phase, we first employ clustering methods, such as K-Means and Gaussian Mixture (GM) Clustering, to group data points based on their intrinsic patterns. The clustered results are then embedded as meta-dataset points into the original feature space. This augmentation adds a layer of complexity to our dataset, capturing finer details that might be missed by conventional approaches. Subsequently, Principal Component Analysis (PCA) is applied to this enriched feature set. PCA allows us to reduce the dimensionality while retaining the most informative features. This step ensures that we maintain a set of highly relevant and discriminative features, optimizing the input for our machine learning models.

The integration of clustering and PCA into the SFE-PCA approach aims to strike a balance between detailed feature representation and computational efficiency. By capturing essential information through clustering and refining it with PCA, we aim to empower our ML models with a more focused and effective feature set, ultimately contributing to improved performance and precision in our experimental results.

\subsubsection{Stacking Feature Embedded using Clustering}

The proposed SFE methodology serves as a fundamental component of our experimental framework. It is designed to address the challenges posed by big and imbalanced datasets, particularly in the context of machine learning-based network intrusion detection. This approach combines the strengths of clustering techniques and feature embedding to improve detection accuracy. The SFE process is illustrated in Figure \ref{fig:sfe}. The following are the working principles of this approach:

\begin{figure*}[!htbptbp]
\centering
  \includegraphics[width=0.65\textwidth]{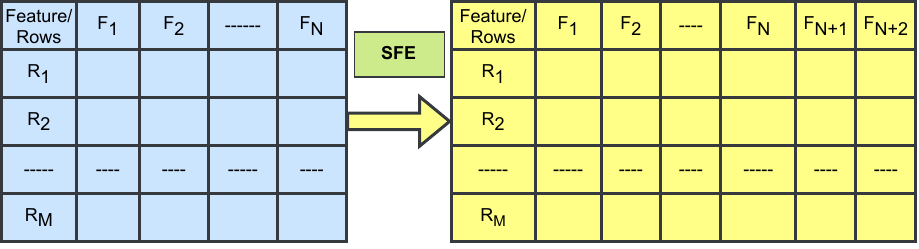}
\caption{SFE process}
\label{fig:sfe}
\end{figure*}

\begin{itemize}
    \item Cluster Formation: The process begins with the application of two clustering methods: K-Means and Gaussian Mixture Clustering. These techniques group data points into coherent clusters based on shared characteristics and patterns. Clustering reveals the underlying structure in the data, enabling a more comprehensive understanding.

    \item Feature Embedding: The output generated by the clustering phase is then embedded into the original feature space. This integration creates a set of additional features, often referred to as meta-dataset points. These new features capture nuanced information that enriches the overall dataset.

    \item Enhanced Data Representation: The dataset now includes the original features alongside the newly embedded meta-dataset points. This augmented representation offers a more comprehensive view of the data, enabling the detection of subtle anomalies and patterns.
\end{itemize}

The approach was adopted to address the limitations of traditional intrusion detection methods when dealing with big and imbalanced data. By integrating clustering techniques and feature embedding, our objectives encompass several key aspects. Firstly, we seek to enrich the reliability and accuracy of intrusion detection, providing a more robust defense against cyber threats. Additionally, our approach enables the capture of fine-grained details within network traffic data, improving our ability to discern subtle anomalies. Moreover, it facilitates the detection of previously undetected threats, contributing to a more comprehensive security posture. Lastly, by incorporating these techniques, we seek to improve the overall performance and precision of our ML models, making them more effective in safeguarding network environments.

This approach represents a crucial advancement in the field, promising to contribute to the development of more robust and effective intrusion detection systems for real-world network security challenges.

% KMeans
% GMC
% added as  a features 

\subsubsection{Feature Extraction using PCA}
The curse of a high dimensional dataset makes a model more complex and leads to overfitting that fallout an ill performance. It is essential to reduce dimension for getting reduced dataset, less computation time, quickly visualize the data and remove redundant features from the dataset.

Feature reduction in a dataset involves the generation of new features from existing ones, with the aim of preserving the essential information present in the original features. PCA is a statistical technique that employs an orthogonal transformation to convert a set of correlated variables into a set of uncorrelated variables. In both exploratory data analysis and the development of predictive machine learning models, PCA stands as a fundamental and widely employed tool. Additionally, It serves as an invaluable unsupervised statistical method for exploring the relationships between a set of variables. It differs from regression in that it seeks to create a line of best fit, which is often referred to as a form of generic factor analysis.
To reduce the dimension of the features from n to k, the following steps should be preceded:\\
1. Equation \ref{equ:normalize} is used to equalize the data's initial attribute values by the mean and variance.
\begin{equation}
\mu  = \frac{1}{n}\sum\limits_{i = 1}^n {{x_i}} 
\label{equ:normalize}    
\end{equation}
Here $n$ represents the instances number and  ${x_i}$ represents the data points.
\\
2. Substitute ${x_i}$ by ${x_i} - \mu $ \\
3. Each vector  ${x_{j(i)}}$  should be rescaled to have unit variance.
\begin{equation}
{\sigma _j}^2 = \frac{1}{n}\sum\limits_{i = 1}^n {{{({x_{j(i)}})}^2}}
\end{equation}
4. Substitute ${x_{j(i)}}$ by ${{{x_{j(i)}}} \over \sigma }$  \\
5. The Covariance Matrix $C_M$ should be calculated as follows:

 \begin{equation}
     C{_M} = \frac{1}{n}\sum\limits_{i = 1}^n {{x_i}.{{({x_i})}^T}}
 \end{equation}
\\
6. Determine Eigen-vectors and their related Eigen-values of $C_M$.\\
7. To generate $w$, sort the Eigen-vectors by decreasing their Eigen-values and choose $k$ Eigen-vectors with the largest Eigen-values.\\
8. Equation \ref{equ:teq} is used to convert the data onto the new subspace using $w$.

\begin{equation}
    y = {w^T}*x\
    \label{equ:teq}
\end{equation}

where $x$ represents one sample as a $d \times 1$ dimensional vector and $y$ represents the converted $k \times 1$ dimensional vector in the resulting subspace.\\
The number of features $D{_p}$ that each data point represents determines the computational complexity of running the designed PCA \citep{zou2006sparse}.
 \begin{equation}
     O({D{_p}^3})
 \end{equation}

The reduction ratio (RR) is the number of output dimensions divided by the number of input dimensions. \citep{vasan2016dimensionality}. The efficiency of PCA is inverse to RR. The lower the RR value, the higher the PCA's efficiency.

\begin{figure*}[!htbptbp]
\centering
  \includegraphics[width=0.75\textwidth]{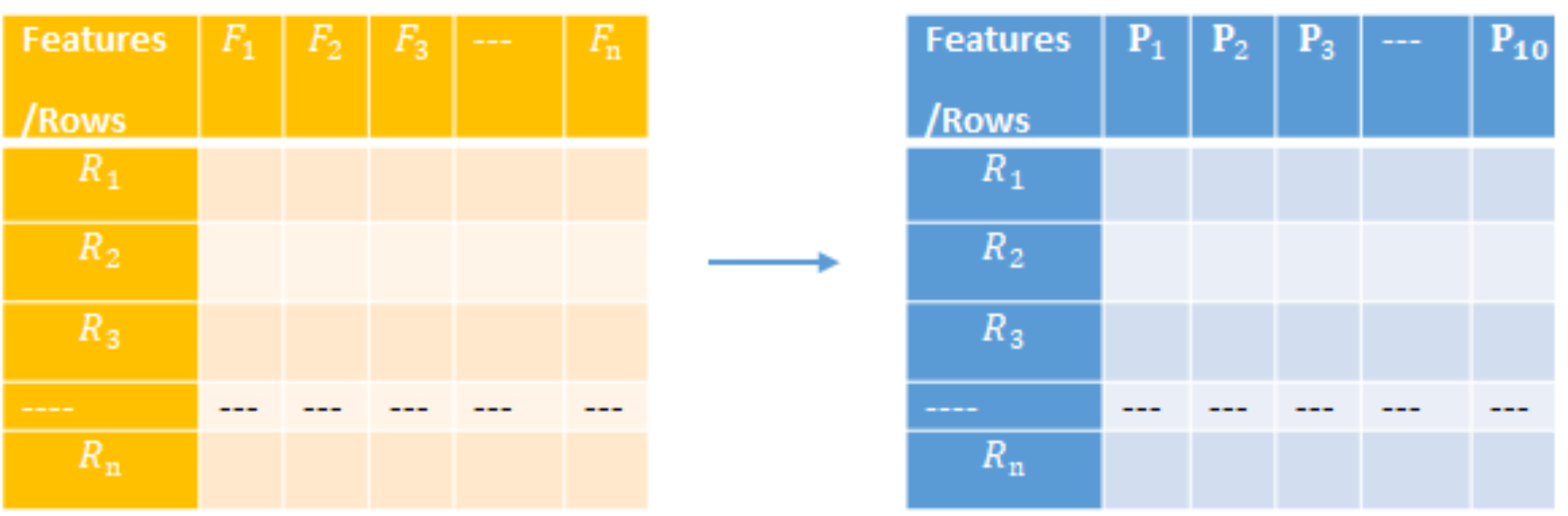}
\caption{Dimension reduction process using PCA}
\label{fig:pca}
\end{figure*}

In our proposed framework we adopted PCA to reduce the dimension of our datasets to get better performance with less number of features than the original. The reduced features contain the most important information of the datasets to produce the better performance to detect intrusion efficiently. During our proposed work, we considered the RR is 10:45 or 22.22\% for UNSW-NB15 and 10:79 or 12.65\% for the CIC-IDS2017 dataset, which is used to provide higher accuracy with a lower false rate. Several existing works took 13-22 or 28.88\%-48.88\% PCA for UNSW-NB15 \citep{kumar2020statistical, kasongo2020deep} and 22-52 or 27\%-65.82\% PCA for CIC-IDS2017 datasets \citep{al2021improved, stiawan2020cicids}. In this study we considered 10 PCA for both datasets to check the performance at these lower RR to prove the efficiency of our approach. The PCA process is depicted in Figure \ref{fig:pca}.
 
\subsection{ML Algorithms}

In this section, we have leveraged supervised ML algorithms to assess our performance in binary and multilabel classification tasks using our datasets. The detection of intrusion in the context of network security using ML is outlined as follows:

\subsubsection{Decision Tree (DT)}
A non-parametric supervised ML technique called the DT is used to solve problems with regression and classification. The prediction of the value of the output of a dataset is generated gripping decision rules from dataset features. It's easy to comprehend and interpret and it can be visualized. It can handle multi-output problems \citep{ahmim2019novel}. It is widely used in IDS. Decision node, having multiple branches and confirmed to make the decision and Leaf nodes, not contain any branches and the output of those decisions are the two nodes in DT. The starting decision node is called the root node. To build a decision tree, Attribute Selection Measure (ASM) is performed on information gain and the Gini index to select the feature \citep{sharmin2023hybrid}. The change of entropy based on a feature after splitting is called IG. Based on the value of IG, we have separated the node and constructed the decision tree based on the value of IG. The measure of purity or impurity creating a DT is called GI. To create binary splits, GI is used. The lower GI should be preferred as compared to the higher GI. Pruning is the practice of deleting nodes from a tree that is no longer needed to achieve the best decision tree possible, which is accomplished through Cost Complexity Pruning and Reduced Error Pruning. Equ. \ref{equ:gi} and \ref{equ:ig}show GI and IG respectively.

\begin{equation}
    Gini(D) = 1 - \sum\limits_{i = 1}^n {{{({p_i})}^2}}
    \label{equ:gi}
\end{equation}
\begin{equation}
    Gain(A) = Entropy(D) - Entropy{_A}(D)
    \label{equ:ig}
\end{equation}
where,
\begin{equation}
Entropy(D) =  - \sum\limits_{i = 1}^n {{p_i}{{\log }_2}({p_i})}
\end{equation}
\begin{equation}
Entrop{y_A}(D) = \sum\limits_{i = 1}^n {\frac{{{D_i}}}{{|D|}} \times Entropy({D_i})}
\end{equation}
and the probability of a data point in the subset of $D_i$ of a dataset $D$ is denoted by ($P_i$ ).

\subsubsection{Random Forest (RF)}
Random Forest (RF) is a renowned supervised ML technique rooted in the concept of ensemble learning which involves the amalgamation of multiple classifiers to tackle complex problems and enhance the overall performance of the model. RF serves as a meta predictor that leverages averaging to enhance predictive accuracy, all the while mitigating overfitting concerns by adapting various decision tree classifiers to diverse subsets of the dataset. A bootstrap randomized resampling method creates each decision tree \citep{breiman2001random}. It requires the least amount of training time compared to other algorithms and estimates output with high accuracy; it also operates efficiently on large datasets. It improves the model's accuracy and eliminates the problem of overfitting. The algorithm gathers the prediction results from each tree, sets up a voting mechanism and then performs a majority vote among the classifiers to make a classification decision. It builds a forecast using the results of many decision trees, which improves prediction accuracy \citep{uddin2023ensemble}.

\subsubsection{Extra Tree (ET)}
An ensemble ML technique and meta-estimator, Extra-Tree is also called Extremely Randomized Trees. In order to increase the model's prediction accuracy, it applies a series of randomized decision trees, referred to as extra-trees, to various sub-samples of datasets and averages them. This prevents over-fitting. It's an ensemble model, just like bagging and random forest in an ensemble decision tree. From the training datasets, it creates a huge number of unpruned decision trees in order to function. For regression and classification, respectively, the majority vote and average are used to predict the decision tree. It builds decision trees using the entire learning sample, and divides the nodes by randomly choosing all of the cut-points \citep{geurts2006louiswehenkel}.

\subsubsection{Extreme Gradient Boosting (XGB)}
% Extreme Gradient Boosting (XGB) is a supervised ML technique that improves speed and performance by using gradient boosted decision trees. It is fast as compared to other implementations of gradient boosting \citep{chen2015higgs}. The final prediction is made by a combination of the residuals of prior models which are predicted from the generated novel models. It uses a gradient descent strategy to reduce loss and to boost up the model for improved performance. It's a scalable end-to-end tree boosting method that data scientists utilize to obtain cutting-edge outcomes on a variety of ML problems \citep{chen2016xgboost}. The objective functions in XGB consist of two parts as training loss and regularization with $\theta$ the optimal settings for $x_i$ training data and $y_i$ labels. Equ. \ref{equ:loss} Shows objective functions of XGBboost.

A supervised ML method that uses gradient-boosted decision trees to improve speed and performance. XGB has remarkable speed as compared to other gradient boosting implementations \citep{chen2015higgs}. The combination of residuals from earlier models, which new models form—leads to its ultimate forecasts. This method employs a gradient descent to lessen loss and improve model performance. When looking for cutting-edge solutions for various ML problems, data scientists have come to appreciate it, as a scalable end-to-end tree-boosting technique.
\citep{chen2016xgboost}. Within XGB, the objective functions are composed of two key components: the training loss and regularization, with $\theta$ representing the optimal settings for the training data $x_i$ and the associated labels $y_i$. Equation \ref{equ:loss} illustrates the objective functions employed in XGBoost.

\begin{equation}
    O(\theta ) = L(\theta ) + \Omega (\theta )
    \label{equ:loss}
\end{equation}
 % Here, $L$ stands for the training loss function, which indicates how well our model predicts the training datasets. \\
 % A simple example of training loss that represents MSE:
In this context, $L$ represents the training loss function, which is a metric for evaluating the model's performance in predicting the training datasets.

For instance, a straightforward example of a training loss function that represents Mean Squared Error (MSE):

\begin{equation}
    L(\theta ) = \sum\limits_i {{{({y_i} - \mathop {{y_i}}\limits^ \wedge  )}^2}}
\end{equation}
The logistic loss function is a frequently used loss function in logistic regression:
\begin{equation}
    L(\theta ) = \sum\limits_i {[{y_i}\ln (1 + {e^{ - \mathop {{y_i}}\limits^ \wedge  }}) + (1 - {y_i})\ln (1 + {e^{ - \mathop {{y_i}}\limits^ \wedge  }})]}
\end{equation}
 $\Omega$ is the regularization that controls the complexity of the model, which helps us to avoid overfitting which is given by
\begin{equation}
    \Omega (f) = \gamma T + \frac{1}{2}\lambda \sum_{j = 1}^T \|w\|^2 
\end{equation}

In this case, $\gamma$ denotes encouraging pruning, $T$ denotes the number of terminal nodes, $w$ denotes the leaf weights, and $\lambda$ is expected to lower the outcome's sensitivity \citep{chen2015higgs}.

\section{Experimental setup and Evaluations}
\label{sec:experimental}
In this section, we have first covered the environment setup and performance evaluation measures in this section. Then, we have included descriptions for the CIC-IDS2017, CIC-IDS2018, and UNSW-NB15 benchmark datasets. We have employed four classification methods for our experiments: DT, RF, ET, and XGB. Data on binary and multilabel classification are used to examine the performance.

\subsection{Environment setup}

% The experiments are conducted on a computering environment consists of 2X-Large processer with Intel(R) 8 cores incorporating 64 GB RAM on 40GB Disk space. In Anaconda Navigator, the Jupyter notebook is used to perform experiments. The Python programming language and some popular libraries like pandas, NumPy, Matplotlib, Seaborn, TensorFlow, Keras, Scikit-learn, Imbalanced-learn, etc are used to implement the evaluation of the performance.

The experiments are conducted in a robust computing environment, utilizing a high-performance 2X-large virtual machine instance. This instance boasts 8 cores, allowing for efficient concurrent task handling and enhanced multi-threading capabilities. With 64 GB of RAM, the system is well-equipped to accommodate memory-intensive applications, and it offers a generous 40 GB of disk space for data storage. The experiments are seamlessly executed using the Jupyter notebook through Anaconda Navigator. To support our performance evaluation, we leverage the Python programming language and a suite of indispensable libraries, including TensorFlow, Keras, Pandas, Scikit-learn, NumPy, Seaborn, Matplotlib, Imbalanced-learn etc.

% \subsection{Performance Evaluation metrics:}
% Numerous performance metrics appraise our proposed model, including accuracy, RMSE, precision, recall, f1-score, ROC curve and confusion matrix. The matrices for appraising the performance are formulated in the following:
% \subsubsection{Confusion Matrix}
% The Confusion Matrix is a technique to assess ML classification efficiency. It is a table-like structure with four (TP, TN, FP, FN) combinations of predicted and actual values. Table \ref{table:confusion} shows a confusion matrix where the TP (True Positive) denotes a correctly anticipated positive value, TN (True Negative) denotes a correctly projected negative value, FN (False Positive) gives an inaccurately forecasted positive value and FN (False Negative) indicates an incorrectly predicted negative value. It is very useful in helping to determine Accuracy, Precision, Recall, F1-score and ROC Curve. 

\subsection{Performance Evaluation Metrics}

Several measures, such as accuracy, precision, recall, F1-score, ROC curve, and confusion matrix, are used to evaluate the performance of our proposed model. The following defines these performance matrices:

\subsubsection{Confusion Matrix}
The Confusion Matrix is a valuable tool for evaluating ML classification performance. It is a tabular representation containing four combinations of predicted and actual values: True Positive (TP), True Negative (TN), False Positive (FP), and False Negative (FN) \citep{talukder2023efficient}. Table \ref{table:confusion} illustrates a confusion matrix where TP represents correctly anticipated positive values, TN indicates accurately projected negative values, FP corresponds to incorrectly forecasted positive values, and FN signifies inaccurately predicted negative values. This matrix is essential for assessing  Recall, F1-score, Accuracy and Precision.

\begin{table}[]
    \centering
    \begin{tabular}{|l|l|l|}
    \hline
        & Actual   positive & Actual   negative \\ \hline
        Predicted   positive & TP & FP \\ \hline
        Predicted   negative & FN & TN \\ \hline
    \end{tabular}
    \caption{Confusion Matrix}
    \label{table:confusion}
\end{table}

\subsubsection{Accuracy}
Accuracy is a fundamental performance metric, representing the proportion of correctly predicted observations to the total observations. It is calculated as follows:
\begin{equation}
Accuracy = \frac{TP + TN}{TP + FP + FN + TN}
\end{equation}

\subsubsection{Precision}
Precision quantifies the ratio of correctly predicted positive values to the total number of predicted positive values:
\begin{equation}
Precision = \frac{TP}{TP + FP}
\end{equation}

\subsubsection{Recall}
Recall measures the ratio of correctly predicted positive values to all actual positive values:
\begin{equation}
Recall = \frac{TP}{TP + FN}
\end{equation}

\subsubsection{F1-Score}
The F1-score represents the harmonic mean of precision and recall for classification problems:
\begin{equation}
F1-Score = 2 \cdot \frac{(Precision \cdot Recall)}{(Precision + Recall)}
\end{equation}

% \subsubsection{RMSE (Root Mean Square Error)}
% RMSE is a widely used evaluation metric in regression problems. It quantifies the square root of the mean of the squared errors and is calculated as:
% \begin{equation}
% RMSE = \sqrt{\frac{\sum_{i = 1}^n (pred(i) - act(i))^2}{n}}
% \end{equation}
% where pred(i) represents the predicted value of I, act(i) is the actual value of I, and n is the total number of values.

\subsubsection{ROC Curve}
ROC curves are commonly employed two-dimensional plots for assessing the significance of classifiers \citep{rana2023robust}. These graphs provide a clear visualization of how a classifier's sensitivity and specificity trade-off at various classification thresholds. This feature is valuable for selecting classifiers that align with specific user requirements, often associated with variable error costs and accuracy expectations, as noted in studies by \citep{sameera2016binary, vergara2008star}. The Area Under the Curve (AUC) represents the degree of discrimination in the ROC curve, while the ROC curve itself is a probability curve that assesses the model's ability to distinguish between different categories. The true positive rate is plotted on the Y-axis, and the false positive rate is on the X-axis. An AUC value approaching 1 suggests that the model excels at distinguishing between class labels, while an AUC value approaching 0 indicates poor predictive performance, implying that the results mirror randomness. This method serves as a means to visualize the classification's efficiency, as emphasized by \citep{gorunescu2011data}. In essence, classifiers with higher ROC curves are considered superior, as supported by  \citep{yulianto2019improving}.

\subsection{K-fold Cross-Validation (CV)}
\begin{figure*}[!htbp]
\centering
  \includegraphics[width=0.65\textwidth]{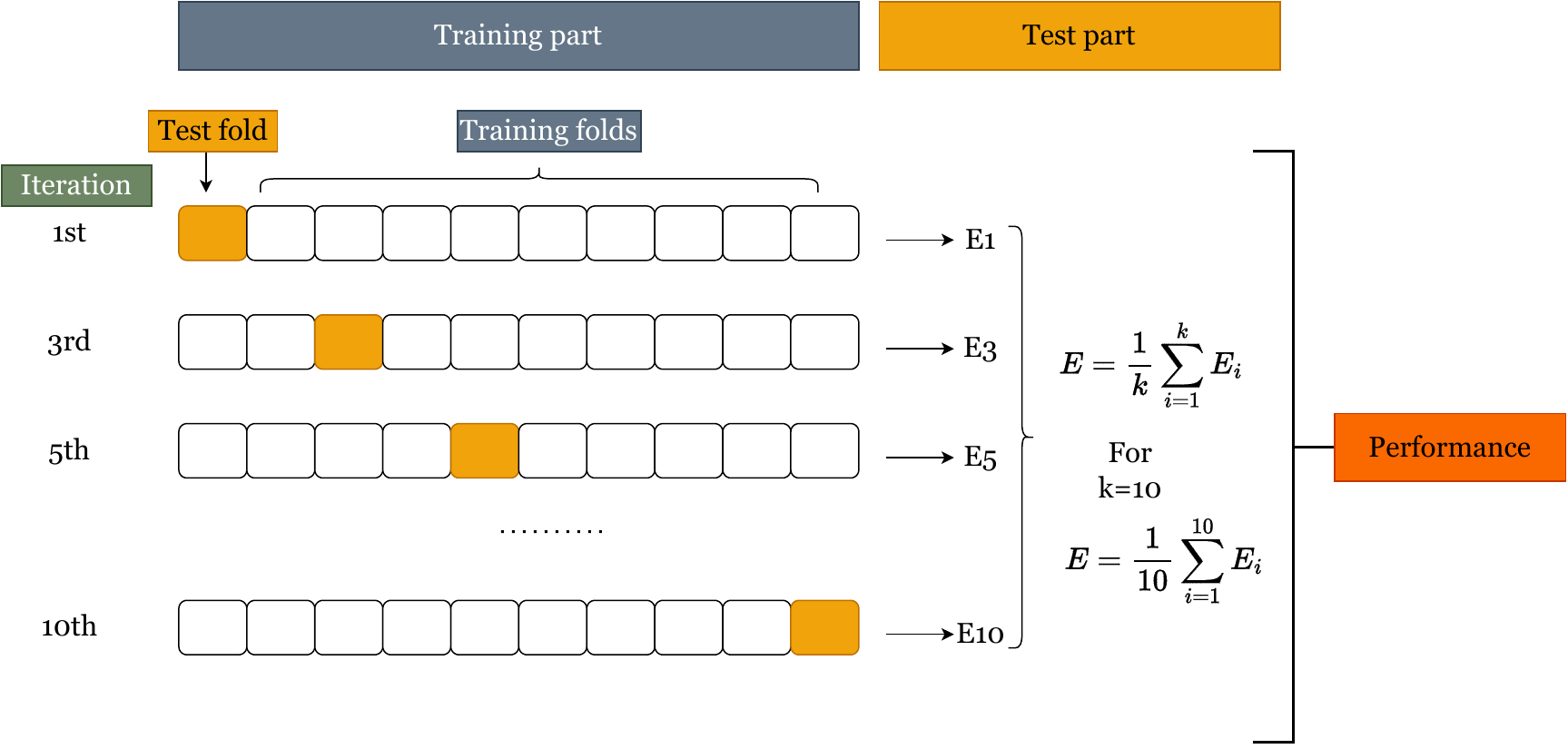}
\caption{K-fold cross-validation process}
\label{fig:kfld}
\end{figure*}

K-fold CV is a standard method that partitions the training set into k smaller sets or folds. The model is trained on each fold, and it is tested on the remaining data after that. Using k-fold CV, the performance measure is obtained as the average of these values. We use k-fold CV in our studies, where k is 10 and the dataset is split into 90\% training and 10\% testing for each fold. Figure \ref{fig:kfld} illustrates the k-fold CV process.

\section{Result Analysis}

In our analysis of the results, we evaluated the performance of four ML models for Intrusion IDS such as DT, RF, ET and XGB. Our focus was on assessing key performance metrics by considering "All Features" and a novel "Proposal" feature set in our evaluation.

\subsection{Results of UNSW-NB15 Dataset}

\begin{figure*}[!htbp]
	\centering
	\subfloat[Binary]{\includegraphics[scale=.420]{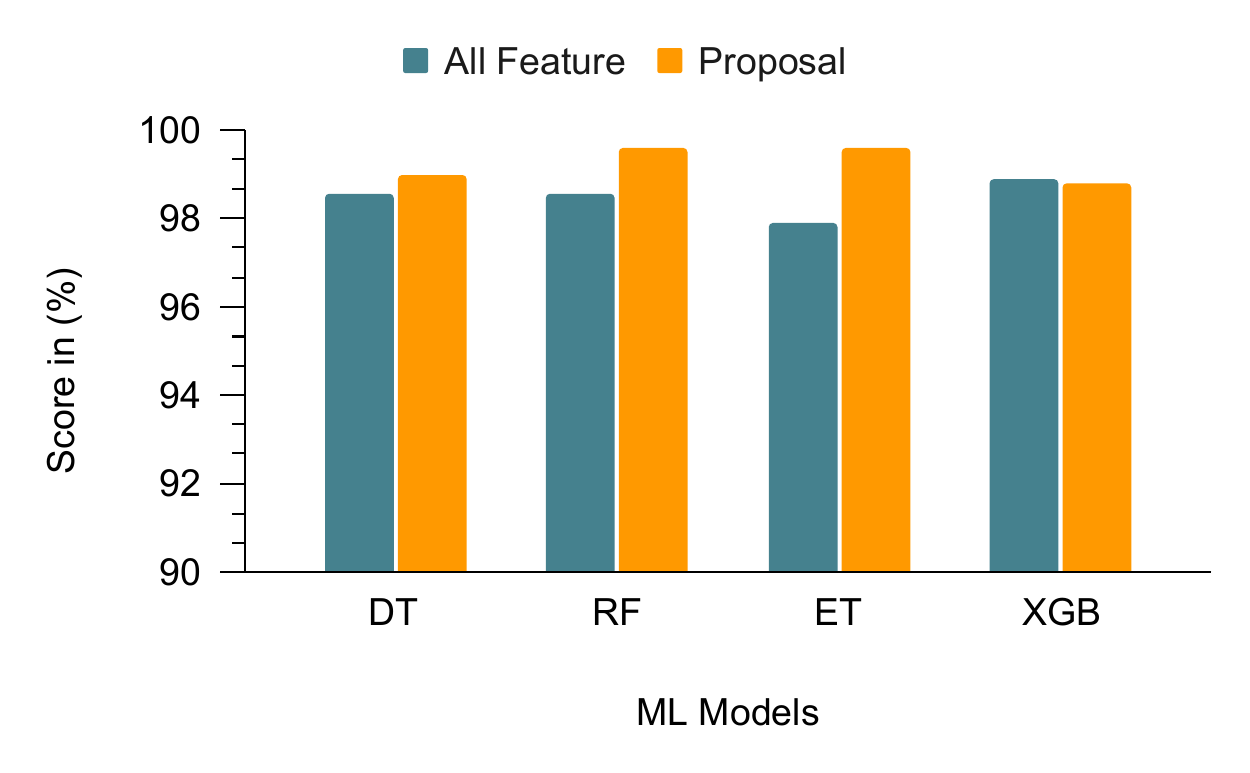}}\hspace{0.047cm}
	\subfloat[multilabel]{\includegraphics[scale=.420]{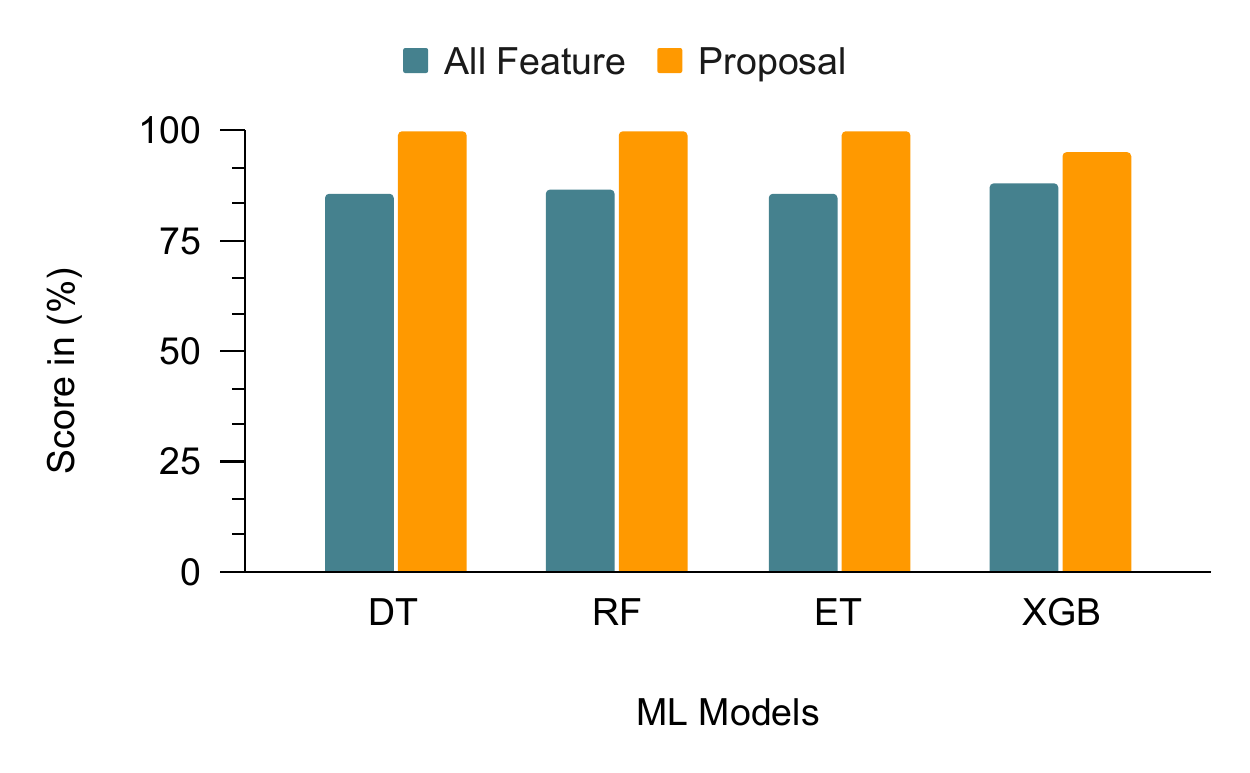}}
	\caption{Binary and multilabel Accuracy performance Bar Chart for All Features and Proposal Features for UNSW-NB15 Dataset}
	\label{fig:analysis_unswnb}
\end{figure*}

The performance results of binary and multilabel classification on the UNSW-NB15 dataset are presented in Figure \ref{fig:analysis_unswnb}, as well as in Table \ref{tab:binary_unswnb15} and Table \ref{tab:multi_unswnb15}. These figures and tables showcase the experimental results for two distinct scenarios: \textbf{All Features:} In this case, "All Features" represent the dataset where we did not oversample any features. We preprocessed, scaled, and applied these features directly to the machine learning models for training and performance evaluation. \textbf{Proposal Features:} Here, the "Proposal" refers to a methodology that encompasses various preprocessing steps that have been undertaken for evaluation. These results provide a comprehensive view of our model's performance and the impact of feature selection and preprocessing on IDS using the UNSW-NB15 dataset.

In the bar chart, it is evident that our proposed model exhibits a noteworthy increase in accuracy rates for binary and multilabel classification. Notably, the rate of accuracy improvement is more pronounced in the context of multilabel classification when compared to binary classification

% Dataset:  UNSW-NB15 [df shape (257673, 45)] 2lac 57thousand

\begin{table}[]
\centering
\resizebox{\textwidth}{!}{%
\begin{tabular}{lrrrrrrrr}
\hline
    & \multicolumn{2}{c}{Accuracy} & \multicolumn{2}{c}{Precision} & \multicolumn{2}{c}{Recall} & \multicolumn{2}{c}{F1-score} \\
ML &  \multicolumn{1}{l}{All Feature} &
  \multicolumn{1}{l}{Proposal} &
  \multicolumn{1}{l}{All Feature} &
  \multicolumn{1}{l}{Proposal} &
  \multicolumn{1}{l}{All Feature} &
  \multicolumn{1}{l}{Proposal} &
  \multicolumn{1}{l}{All Feature} &
  \multicolumn{1}{l}{Proposal} \\ \hline
DT  & 98.56         & 98.97        & 98.42         & 98.98         & 98.44        & 98.97       & 98.44         & 98.97        \\
RF  & 98.55         & 99.59        & 98.46         & 99.59         & 98.43        & 99.59       & 98.43         & 99.59        \\
ET  & 97.9          & 99.59        & 97.71         & 99.59         & 97.73        & 99.59       & 97.73         & 99.59        \\
XGB& 98.9          & 98.81        & 98.71         & 98.81         & 98.82        & 98.81       & 98.82         & 98.81     \\  \hline
\end{tabular}%
}
\caption{Performance Analysis of Binary Classification for UNSW-NB15 Dataset}
\label{tab:binary_unswnb15}
\end{table}

% The accuracy rate for binary classification is 98.97\%, 99.59\%, 99.59\% and 98.81\% for DT, RF, ET and XGB respectively on the proposed model.

The binary result analysis, as presented in Table \ref{tab:binary_unswnb15}, showcases the performance evaluation of various ML algorithms using both the "All Feature" and "Proposed" feature sets. Notably, among these algorithms, DT consistently emerges as the top performer in terms of accuracy, precision, recall, and F1-score. With the "Proposed" feature set, RF demonstrates outstanding performance, achieving an impressive accuracy of 99.59\%. This places RF at the forefront, surpassing other algorithms, including DT with an accuracy of 98.97\%, XGB with 98.81\%, and ET with 99.59\%. This remarkable increase in accuracy is consistently reflected in other metrics. The success of RF can be attributed to its ensemble learning approach, which harnesses the strengths of multiple decision trees to create a robust and highly accurate model.

\begin{table}[]
\centering
\resizebox{\textwidth}{!}{
\begin{tabular}{lrrrrrrrr}
\hline
    & \multicolumn{2}{c}{Accuracy} & \multicolumn{2}{c}{Precision} & \multicolumn{2}{c}{Recall} & \multicolumn{2}{c}{F1-score} \\
ML &
  \multicolumn{1}{l}{All Feature} &
  \multicolumn{1}{l}{Proposal} &
  \multicolumn{1}{l}{All Feature} &
  \multicolumn{1}{l}{Proposal} &
  \multicolumn{1}{l}{All Feature} &
  \multicolumn{1}{l}{Proposal} &
  \multicolumn{1}{l}{All Feature} &
  \multicolumn{1}{l}{Proposal} \\ \hline
DT  & 85.38         & 99.79        & 60.24         & 99.79         & 60.63        & 99.79       & 60.63         & 99.79        \\
RF  & 86.42         & 99.95        & 62.45         & 99.95         & 58.06        & 99.95       & 58.06         & 99.95        \\
ET  & 85.55         & 99.95        & 59.49         & 99.95         & 56.15        & 99.95       & 56.15         & 99.95        \\
XGB& 87.73         & 95.04        & 76.92         & 95.29         & 66.59        & 95.03       & 66.59         & 95.03       \\ \hline
\end{tabular}
}
\caption{Performance metrics of Multilabel Classification for UNSW-NB15 Dataset}
\label{tab:multi_unswnb15}
\end{table}

The multilabel performance analysis, as shown in Table \ref{tab:multi_unswnb15}, evaluates the performance of various ML algorithms on multiclass using both the "All Feature" and "Proposal" feature sets. Among these algorithms, RF consistently emerges as the top performer in terms of accuracy, precision, recall, and F1-score. Notably, with the "Proposal" feature set, RF and ET achieve an impressive 99.95\% accuracy, surpassing all other algorithms, including DT with an accuracy of 98.97\% and XGB with 95.04\%. This substantial accuracy enhancement extends to precision, recall, and F1-score metrics, underscoring RF's and ET's success attributed to their ensemble learning approach, which leverages the strengths of multiple decision trees to create a robust model

% 	      TP	TP (%)	 FP	    FP(%)	  FN	 FN(%)	TN	    TN(%)
% DT	1691600	49.15	24400	0.71	11000	0.32	1714800	49.82
% RF	1711000	49.71	5000	0.15	9100	0.26	1716700	49.88
% ET	1713100	49.77	2900	0.08	11100	0.32	1714700	49.82
% XGB	1703600	49.5	12400	0.36	28600	0.83	1697200	49.31

\begin{figure*}[]
	\centering
	\subfloat[DT]{\includegraphics[scale=.350]{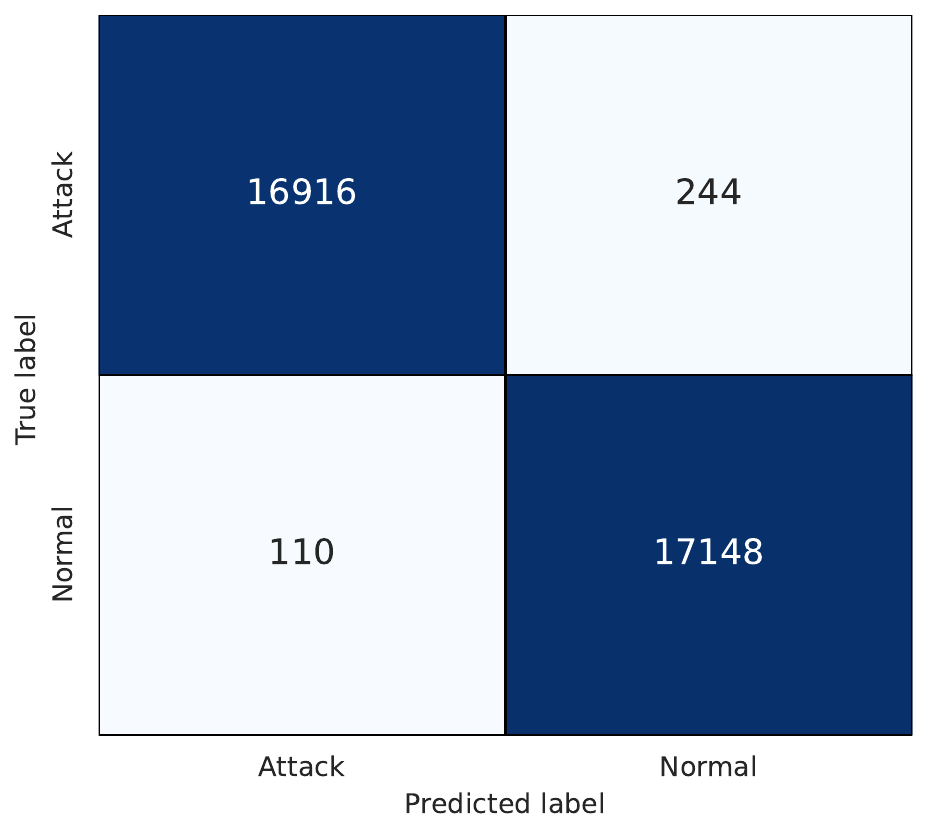}}
	\subfloat[RF]{\includegraphics[scale=.350]{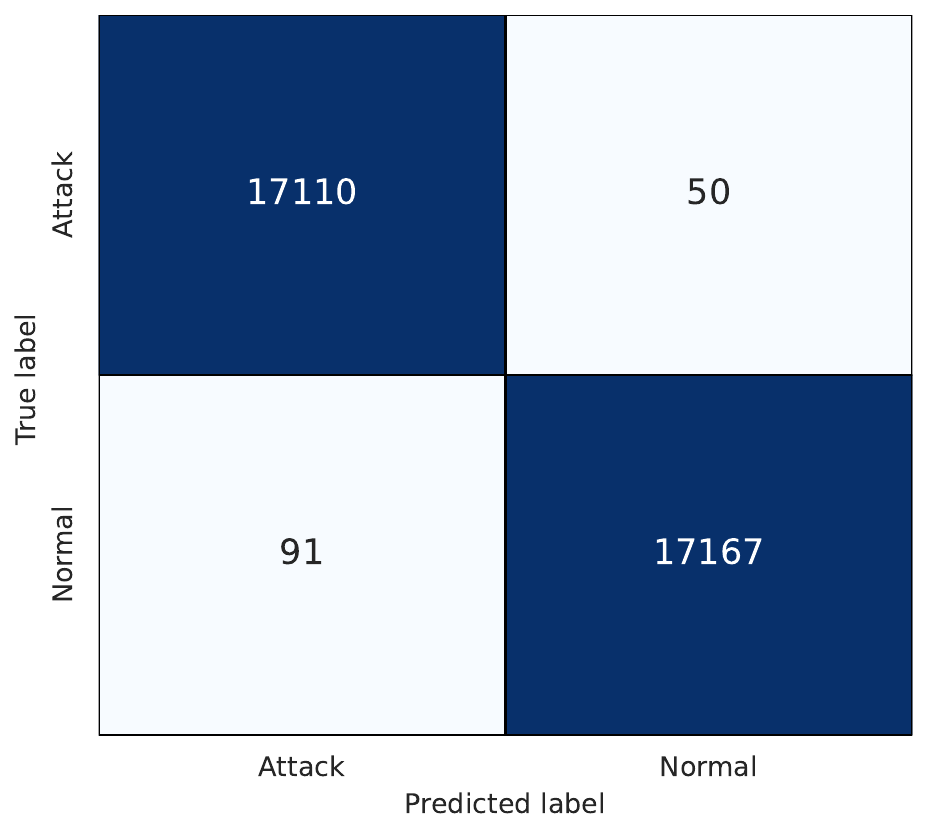}} \hspace{0.1cm}
	\subfloat[ET]{\includegraphics[scale=.350]{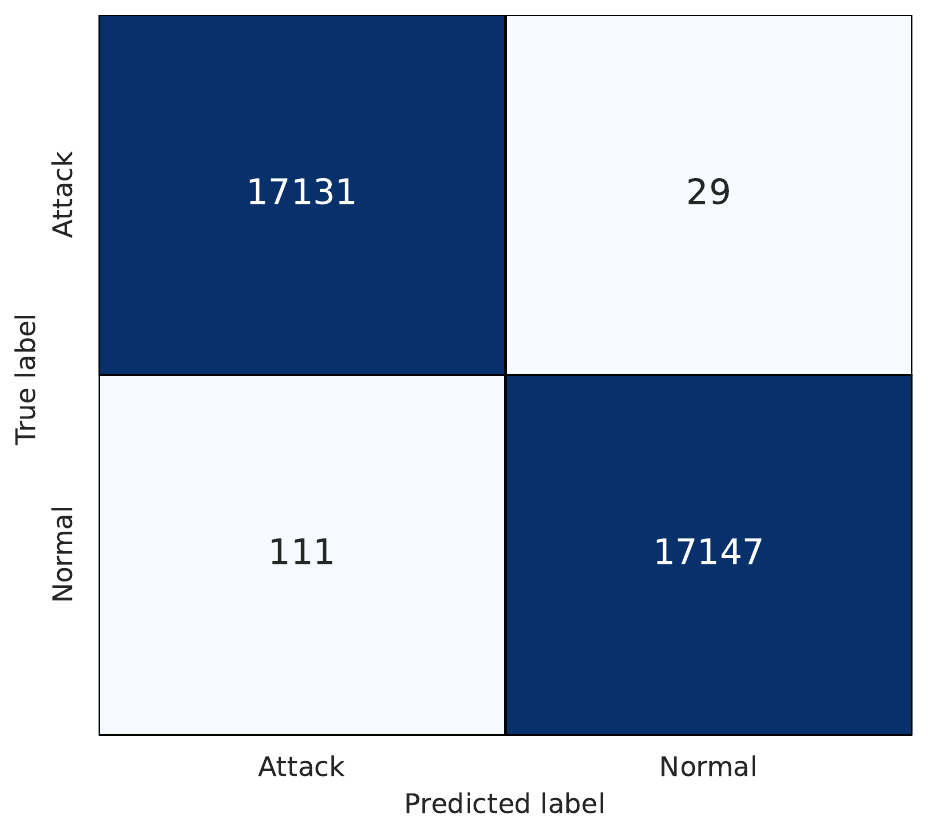}} \subfloat[XGB]{\includegraphics[scale=.350]{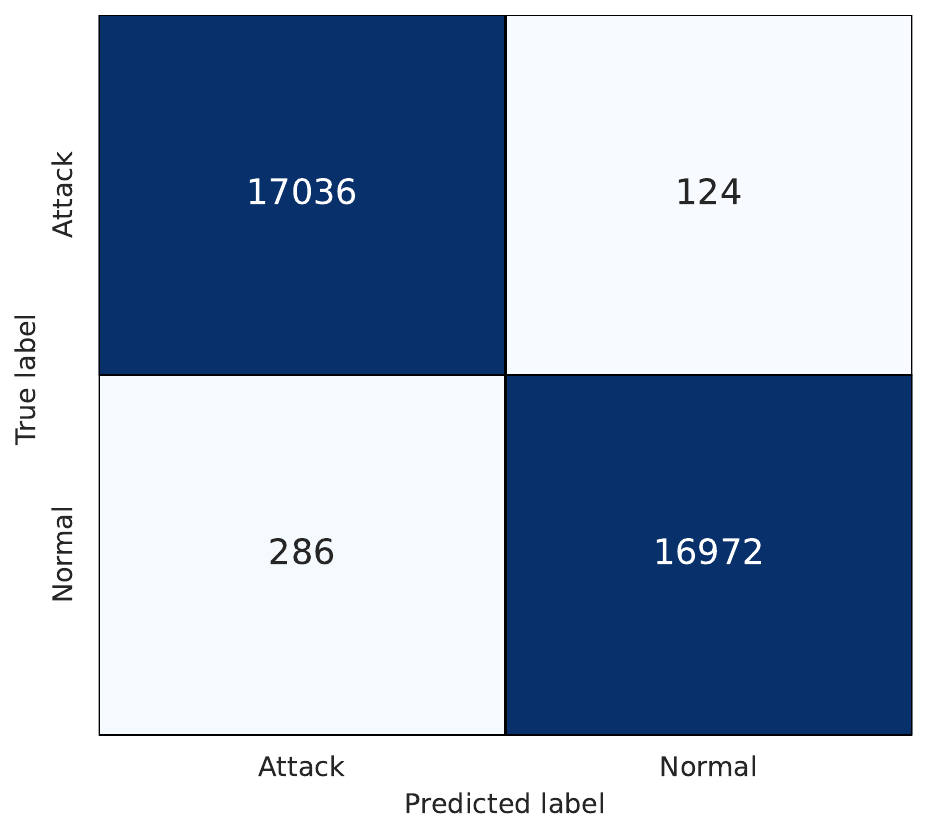}}
	
	\caption{Binary Confusion Matrix for UNSW-NB15 Dataset}
	\label{fig:bcon_unswnb}
\end{figure*}

The binary confusion matrix is displayed in Figure \ref{fig:bcon_unswnb}. A successful predictive model is characterized by a low number of Type 1 (FP) and Type 2 (FN) errors in the confusion matrix. For RF, the TP rate stands impressively at 49.71\%, while the TN rate is equally strong at 49.88\%. Additionally, the FP and FN rates are remarkably low, at 0.15\% and 0.26\%, respectively. These findings highlight RF's robust performance in accurately identifying positive cases (intrusions) and negative cases (non-intrusions), making it a compelling choice for intrusion detection. For ET, the TP rate is an outstanding 49.77\%, and the TN rate is equally impressive at 49.82\%. Furthermore, the FP and FN rates are notably low, at 0.08\% and 0.32\%, respectively. These results underscore the exceptional performance of ET in accurately identifying both positive cases (intrusions) and negative cases (non-intrusions), making it a compelling choice for intrusion detection.

Among all the evaluated models, it is evident that both RF and ET outperform the others, showcasing superior performance in terms of TP and TN rates for IDS. They consistently deliver higher TP and TN rates, demonstrating their effectiveness in accurately identifying intrusions while maintaining a low rate of false positives and false negatives.

\begin{figure*}[]
	\centering
	\subfloat[DT]{\includegraphics[scale=.250]{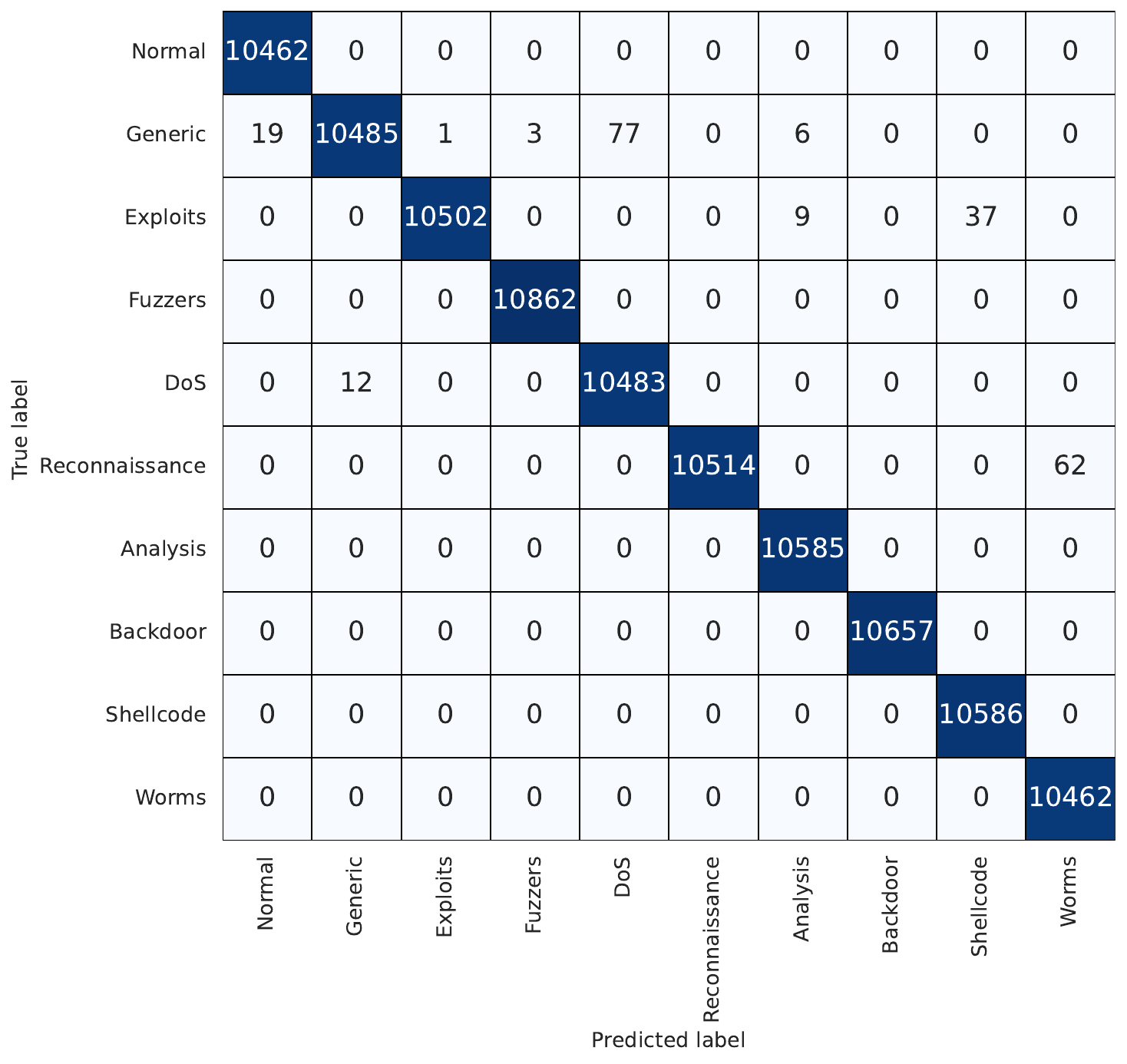}} 
 	\subfloat[RF]{\includegraphics[scale=.250]{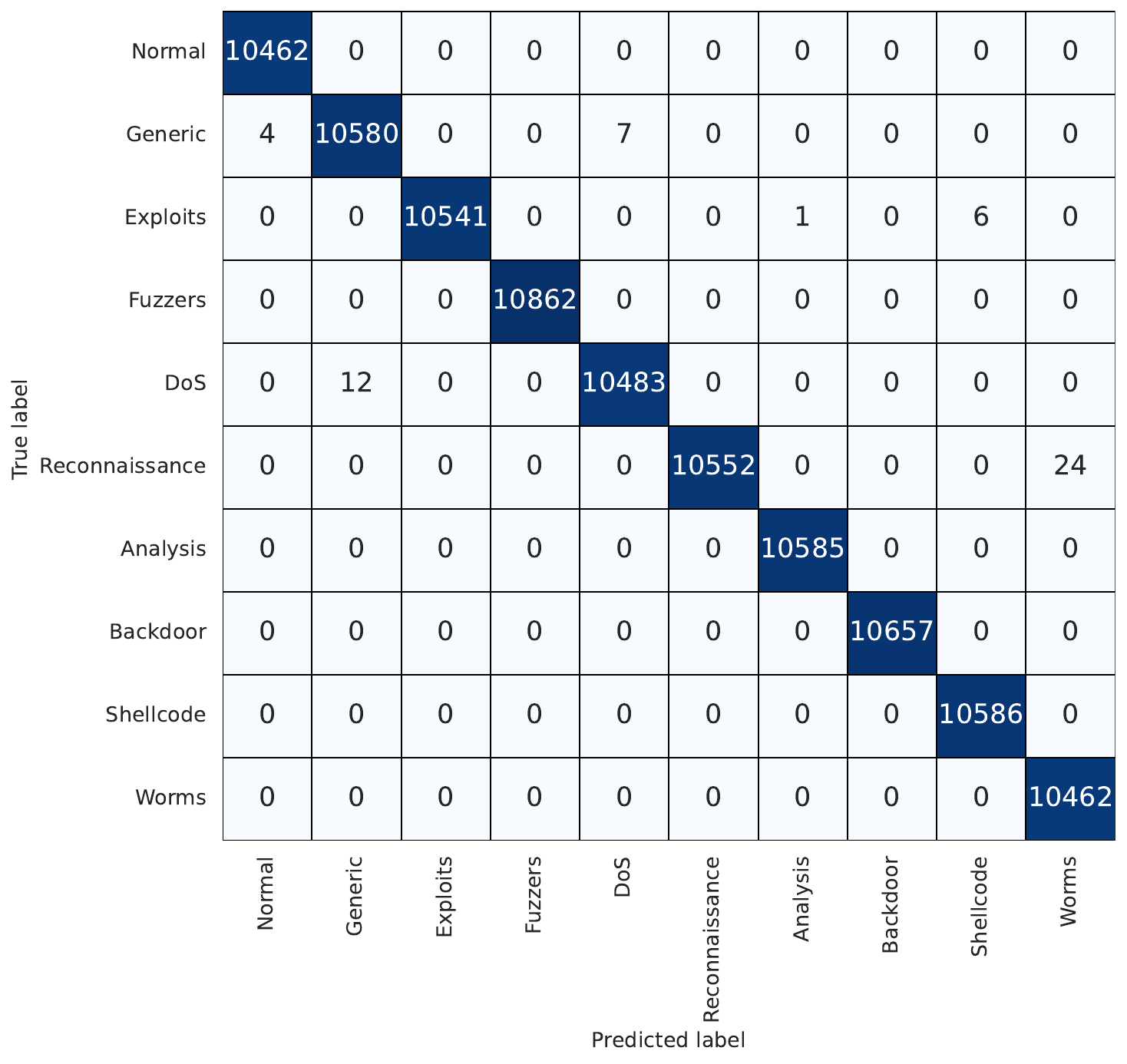}} 
  
	\subfloat[ET]{\includegraphics[scale=.250]{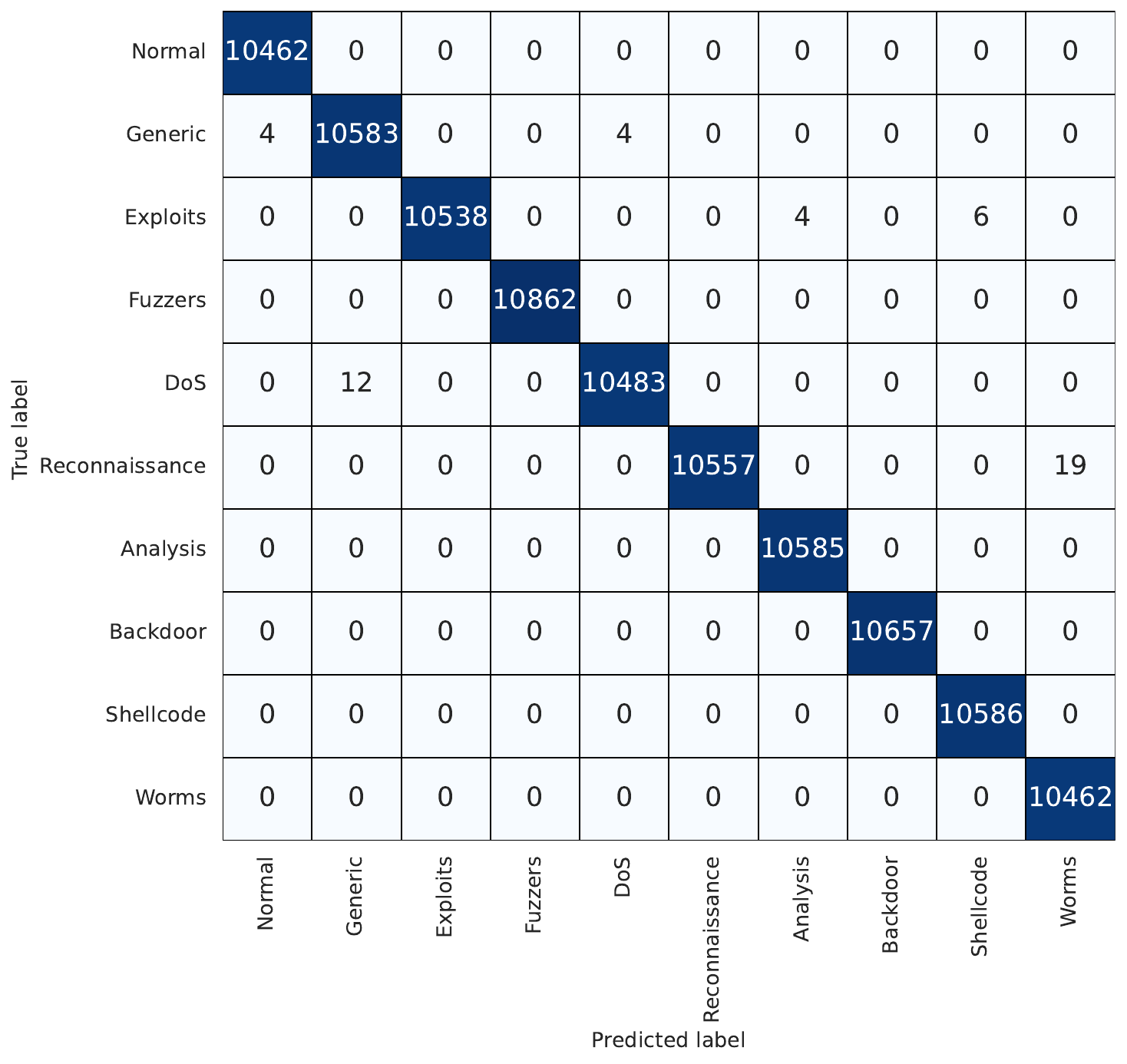}} 	
	\subfloat[XGB]{\includegraphics[scale=.250]{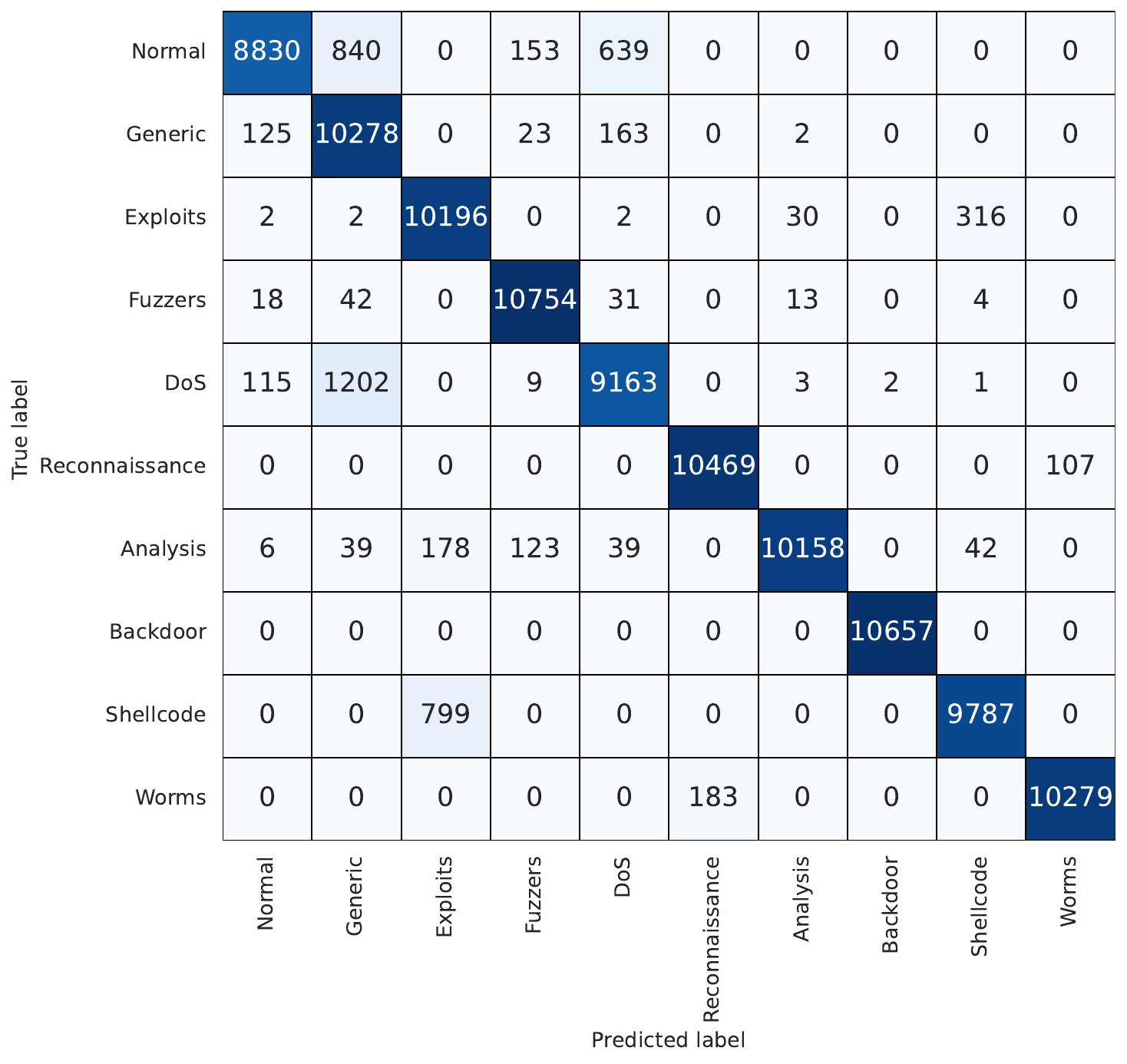}} 	

	\caption{Multilabel Confusion Matrix for UNSW-NB15 Dataset}
	\label{fig:mcon_unswnb}
\end{figure*}

\begin{figure*}[!htbptbp]
	\centering
	\subfloat[Binary]{\includegraphics[scale=.300]{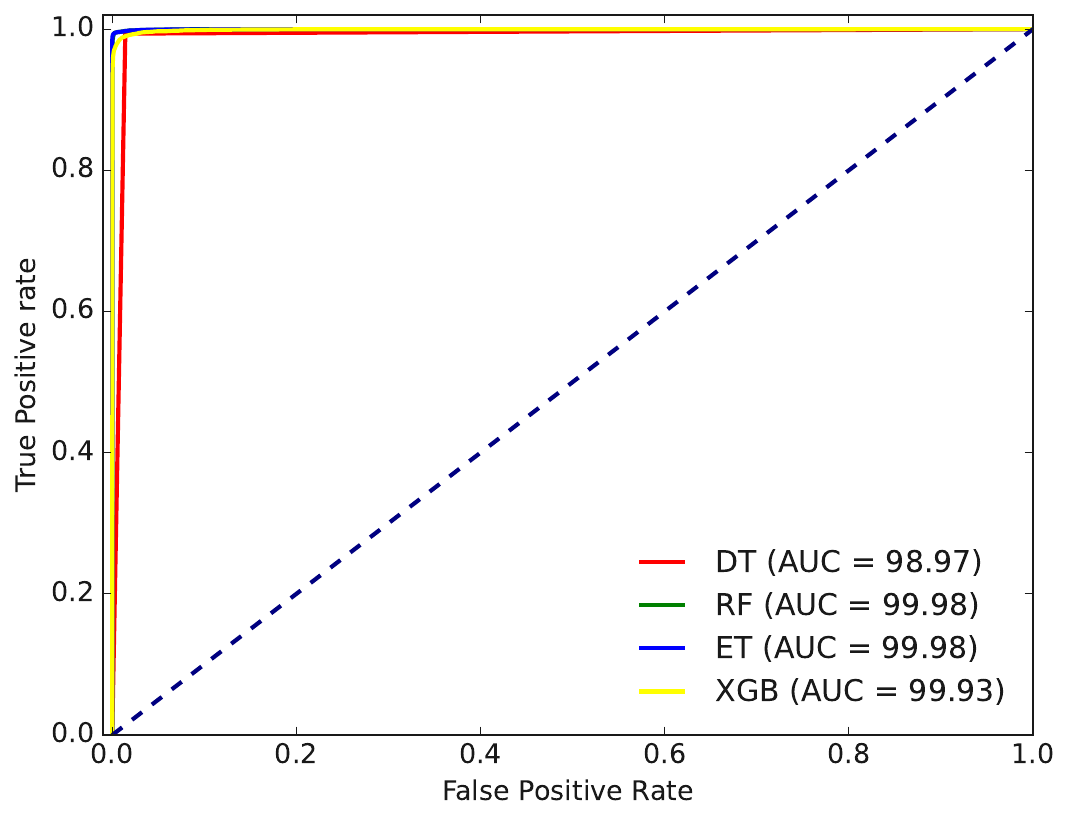}}\hspace{0.047cm}
	\subfloat[multilabel]{\includegraphics[scale=.300]{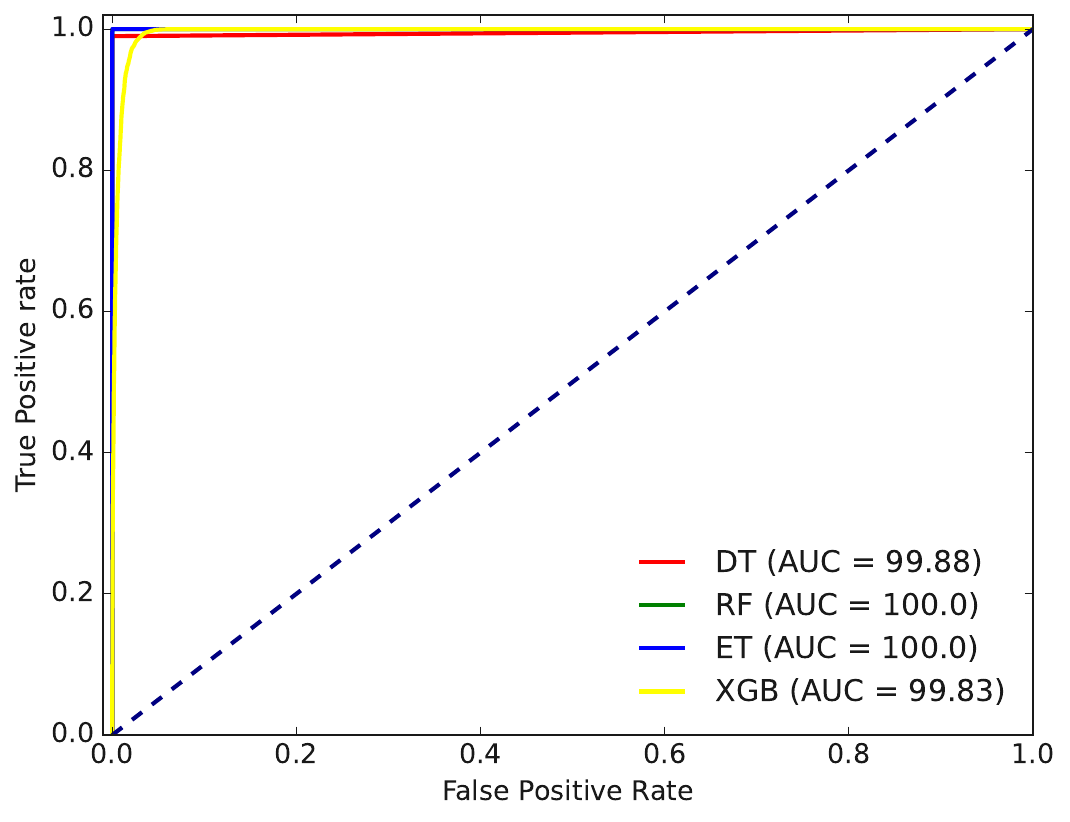}}
	\caption{Binary and Multilabel ROC Curve for UNSW-NB15 Dataset}
	\label{fig:roc_unswnb}
\end{figure*}

The ROC Curve is depicted in Figure \ref{fig:roc_unswnb} for binary and multilabel classification. The ROC Curves clearly illustrate that the AUC (Area Under the Curve) values are approaching 1, indicating a strong predictive model's ability to distinguish between classes. In the binary classification scenario, the AUC scores are 98.97\% for DT, 99.98\% for RF, 99.98\% for ET, and 99.93\% for XGB, with RF and ET algorithms leading in AUC score. In the multiclass classification, the AUC scores are 99.88\% for DT, 100\% for RF, 100\% for ET, and 99.83\% for XGB, with RF and ET again displaying superior AUC scores compared to the others. These high AUC scores, close to 1, indicate the strong predictive performance of the models on the UNSW-NB15 dataset, further validating their effectiveness.

\subsection{Results of CIC-IDS2017 Dataset}

\begin{figure*}[!htbp]
	\centering
	\subfloat[Binary]{\includegraphics[scale=.420]{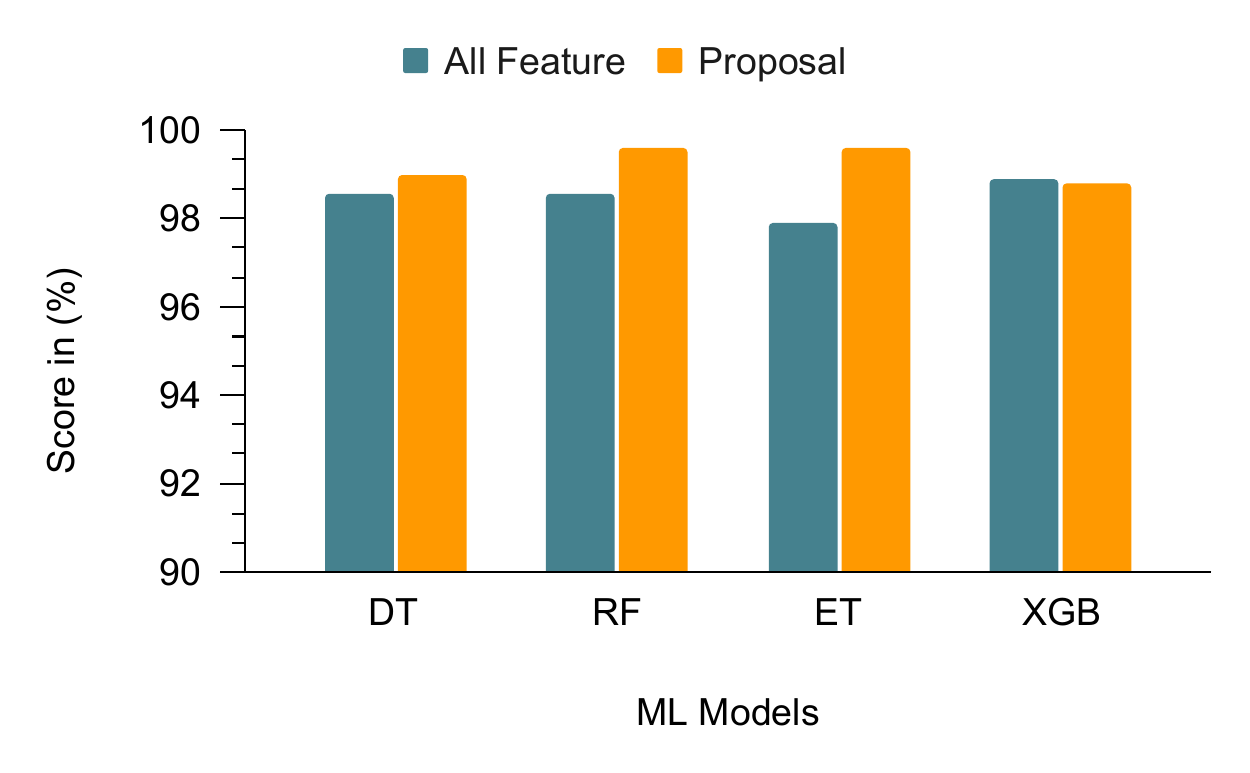}}
 
	\subfloat[multilabel]{\includegraphics[scale=.420]{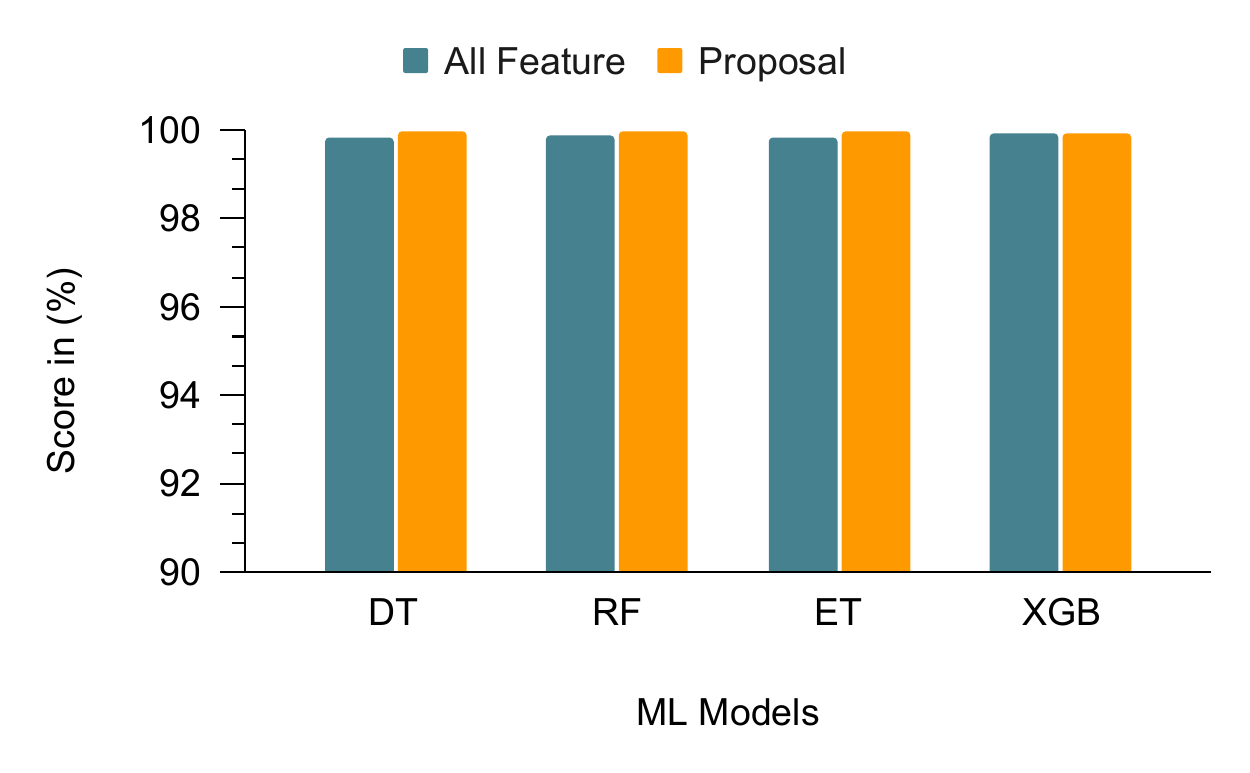}}
	\caption{Binary and Multilabel Accuracy Performance Bar Chart for All Features and Proposal Features for the CIC-IDS2017 Dataset}
	\label{fig:analysis_cicids}
\end{figure*}

The performance results for binary and multilabel classification on the CIC-IDS2017 dataset are presented in Figure \ref{fig:analysis_cicids}, along with detailed metrics in Table \ref{tab:binary_cicids2017} and Table \ref{tab:multi_cicids2017}. These figures and tables display the experimental results for two distinct scenarios: \textbf{All Features:} In this case, "All Features" represent the dataset where features are neither oversampled nor modified extensively. These features are preprocessed, scaled, and then used to train and evaluate the ML models. \textbf{Proposal Features:} The "Proposal" indicates a set of features that have undergone various preprocessing steps and modifications as part of our proposed methodology for evaluation. The inclusion of both scenarios allows for a comprehensive assessment of model performance on the CIC-IDS2017 dataset

In the bar chart, it's evident that our proposed model exhibits a substantial increase in accuracy for binary and multilabel classification. Interestingly, the rate of accuracy improvement is notably higher in multilabel classification when compared to binary classification.

% Dataset:  CIC-IDS2017 [df shape (2830743, 79)] 28 lac 30 thousand

\begin{table}[]
\centering
\resizebox{\textwidth}{!}{
\begin{tabular}{lrrrrrrrr}
\hline
    & \multicolumn{2}{c}{Accuracy} & \multicolumn{2}{c}{Precision} & \multicolumn{2}{c}{Recall} & \multicolumn{2}{c}{F1-score} \\ \hline
ML &
  \multicolumn{1}{l}{All Feature} &
  \multicolumn{1}{l}{Proposal} &
  \multicolumn{1}{l}{All Feature} &
  \multicolumn{1}{l}{Proposal} &
  \multicolumn{1}{l}{All Feature} &
  \multicolumn{1}{l}{Proposal} &
  \multicolumn{1}{l}{All Feature} &
  \multicolumn{1}{l}{Proposal} \\ \hline
DT  & 99.87         & 99.91        & 99.78         & 99.91         & 99.78        & 99.91       & 99.78         & 99.91        \\
RF  & 99.9          & 99.94        & 99.82         & 99.94         & 99.82        & 99.94       & 99.82         & 99.94        \\
ET  & 99.83         & 99.95        & 99.73         & 99.95         & 99.7         & 99.95       & 99.7          & 99.95        \\
XGB& 99.92         & 99.65        & 99.83         & 99.65         & 99.86        & 99.65       & 99.86         & 99.65       \\ \hline
\end{tabular}%
}
\caption{Performance metrics for Binary Classification for CIC-IDS2017 Dataset}
\label{tab:binary_cicids2017}
\end{table}

In the binary classification, the accuracy rates on the proposed model are as follows: 99.91\% for DT, 99.94\% for RF, 99.95\% for ET, and 99.65\% for XGB.

\begin{table}[]
\centering
\resizebox{\textwidth}{!}{%
\begin{tabular}{lrrrrrrrr}
\hline
    & \multicolumn{2}{c}{Accuracy} & \multicolumn{2}{c}{Precision} & \multicolumn{2}{c}{Recall} & \multicolumn{2}{c}{F1-score} \\ \hline
ML &
  \multicolumn{1}{l}{All Feature} &
  \multicolumn{1}{l}{Proposal} &
  \multicolumn{1}{l}{All Feature} &
  \multicolumn{1}{l}{Proposal} &
  \multicolumn{1}{l}{All Feature} &
  \multicolumn{1}{l}{Proposal} &
  \multicolumn{1}{l}{All Feature} &
  \multicolumn{1}{l}{Proposal} \\ \hline
DT  & 99.85         & 99.99        & 98.69         & 99.99         & 94.69        & 99.99       & 94.69         & 99.99        \\
RF  & 99.89         & 99.99        & 98.98         & 99.99         & 94.17        & 99.99       & 94.17         & 99.99        \\
ET  & 99.83         & 99.99        & 98.57         & 99.99         & 94.14        & 99.99       & 94.14         & 99.99        \\
XGB& 99.92         & 99.94        & 99.3          & 99.94         & 94.47        & 99.94       & 94.47         & 99.94       \\ \hline
\end{tabular}
}
\caption{Performance Analysis of Multilabel Classification for CIC-IDS2017 Dataset}
\label{tab:multi_cicids2017}
\end{table}

\begin{figure*}[]
	\centering
	\subfloat[DT]{\includegraphics[scale=.350]{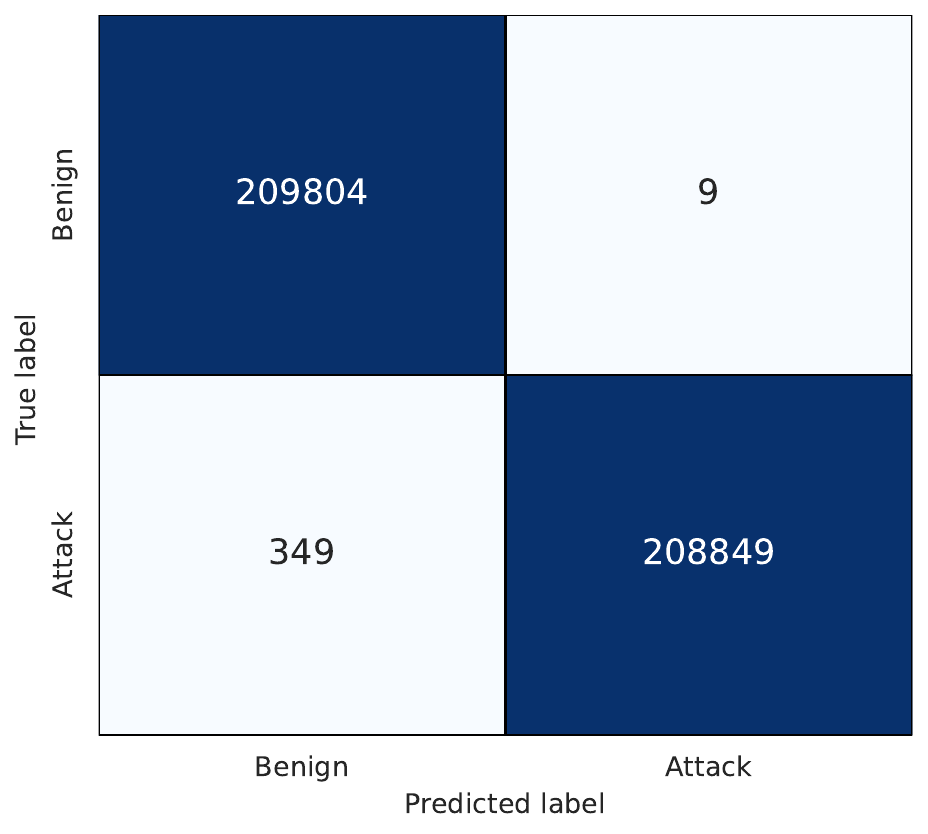}}
	\subfloat[RF]{\includegraphics[scale=.350]{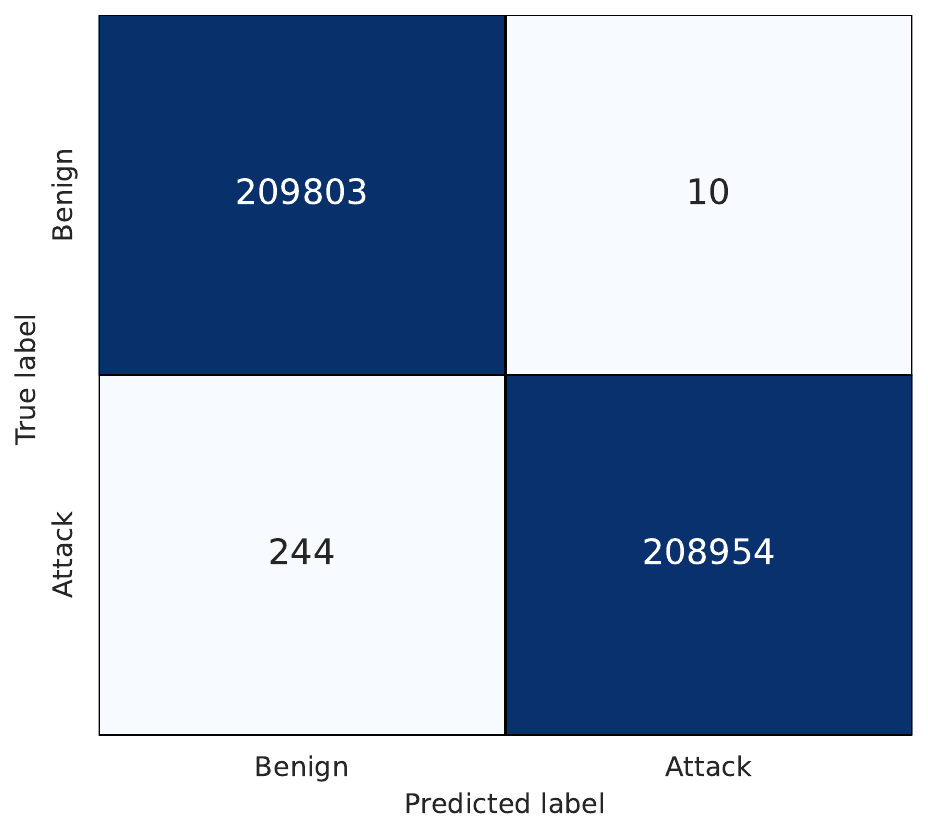}} \hspace{0.1cm}
	\subfloat[ET]{\includegraphics[scale=.350]{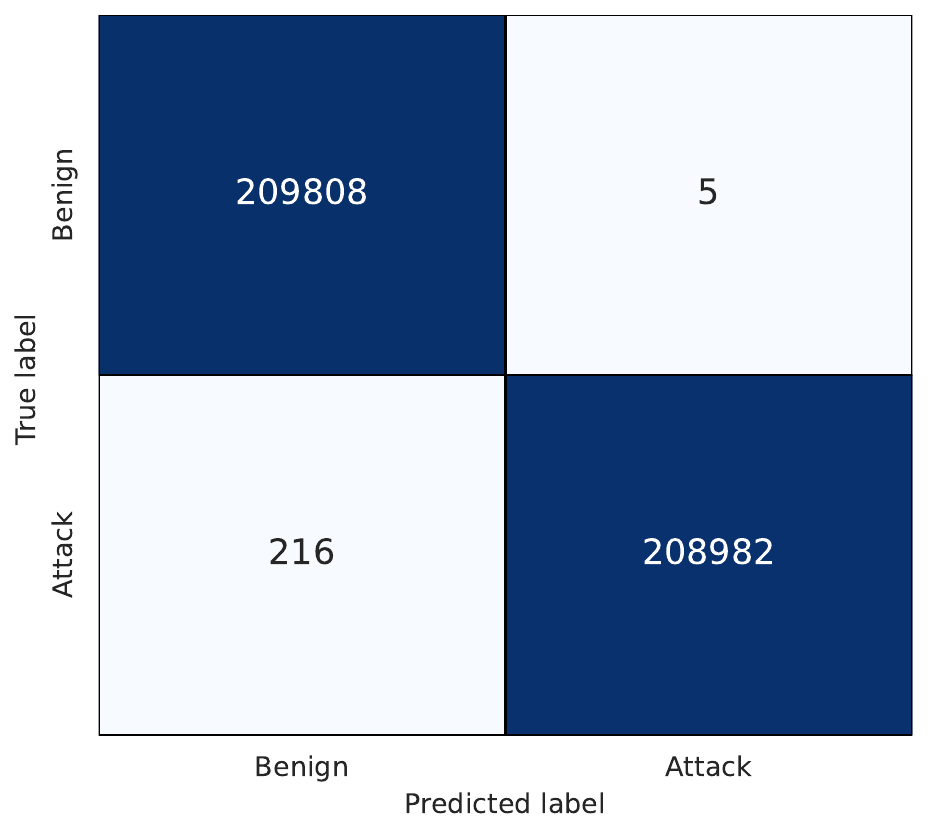}}  
	\subfloat[XGB]{\includegraphics[scale=.350]{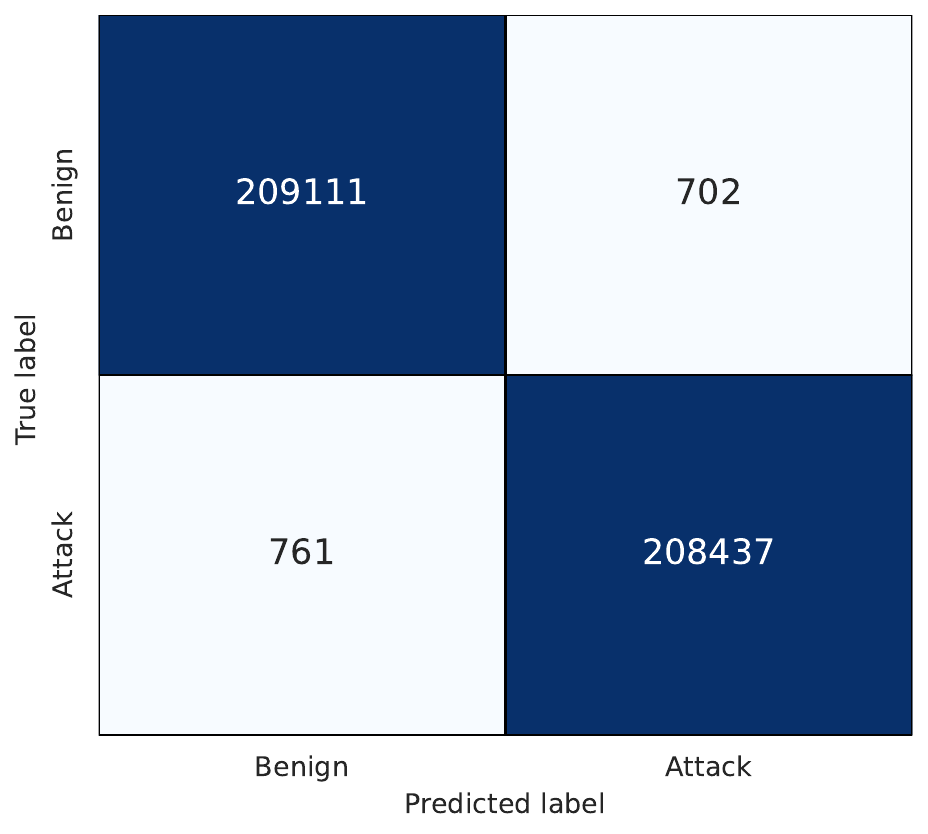}} 
	
	\caption{Binary Confusion Matrix for CIC-IDS2017 Dataset}
	\label{fig:bcon_CIC}
\end{figure*}

\begin{figure*}[]
	\centering
	\subfloat[DT]{\includegraphics[scale=.180]{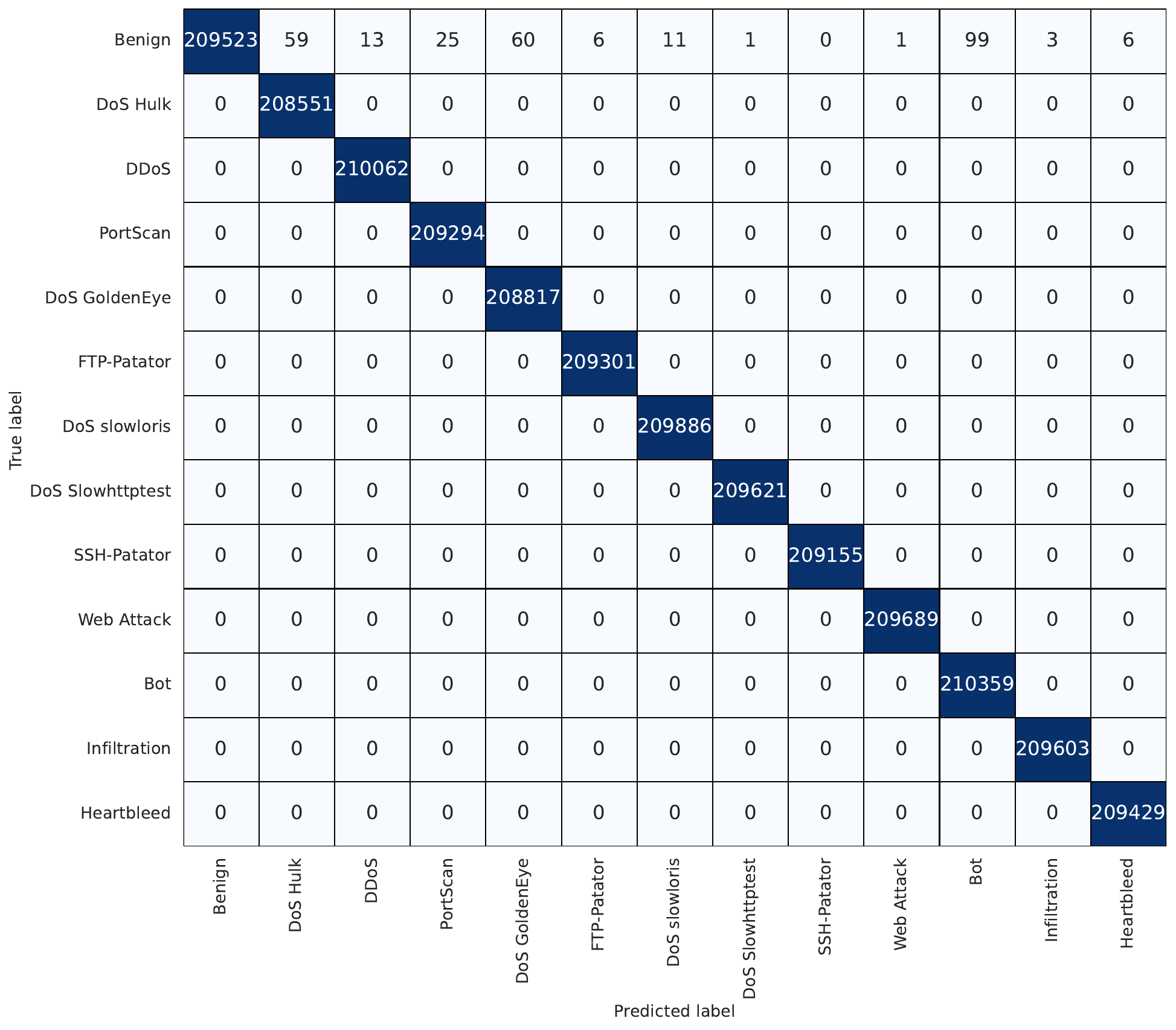}}
	\subfloat[RF]{\includegraphics[scale=.180]{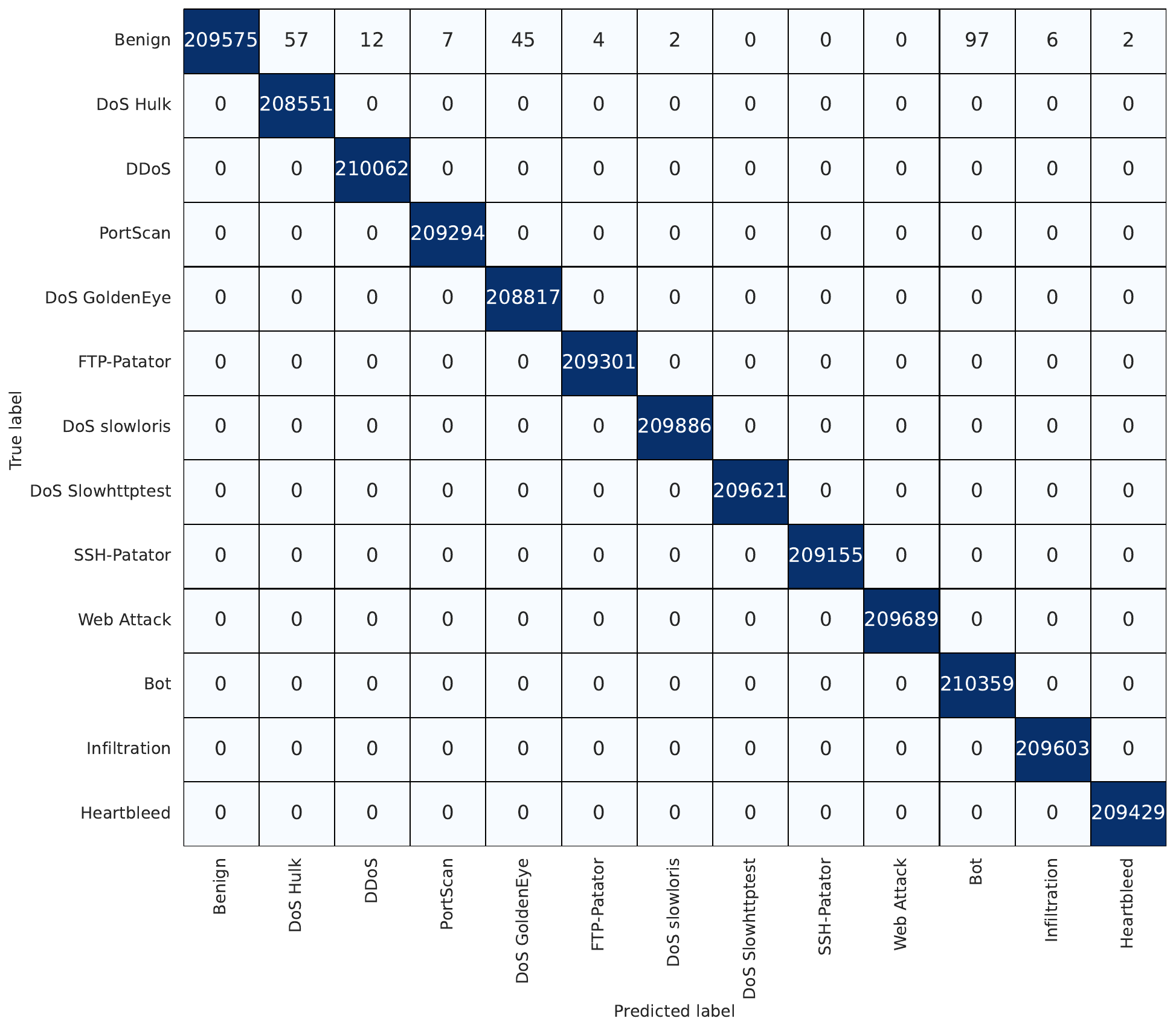}} \hspace{0.1cm}
	\subfloat[ET]{\includegraphics[scale=.180]{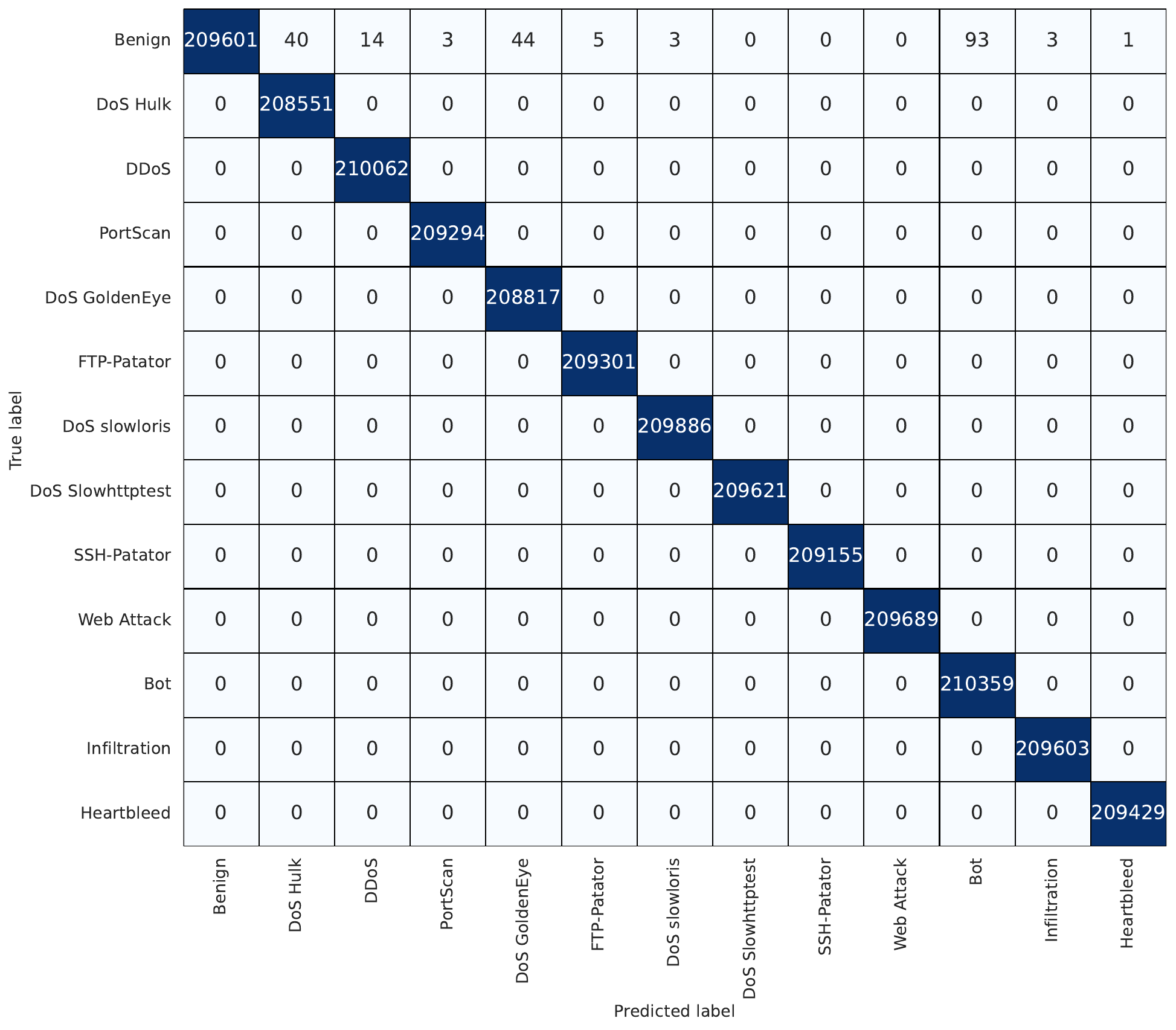}} 
	\subfloat[XGB]{\includegraphics[scale=.180]{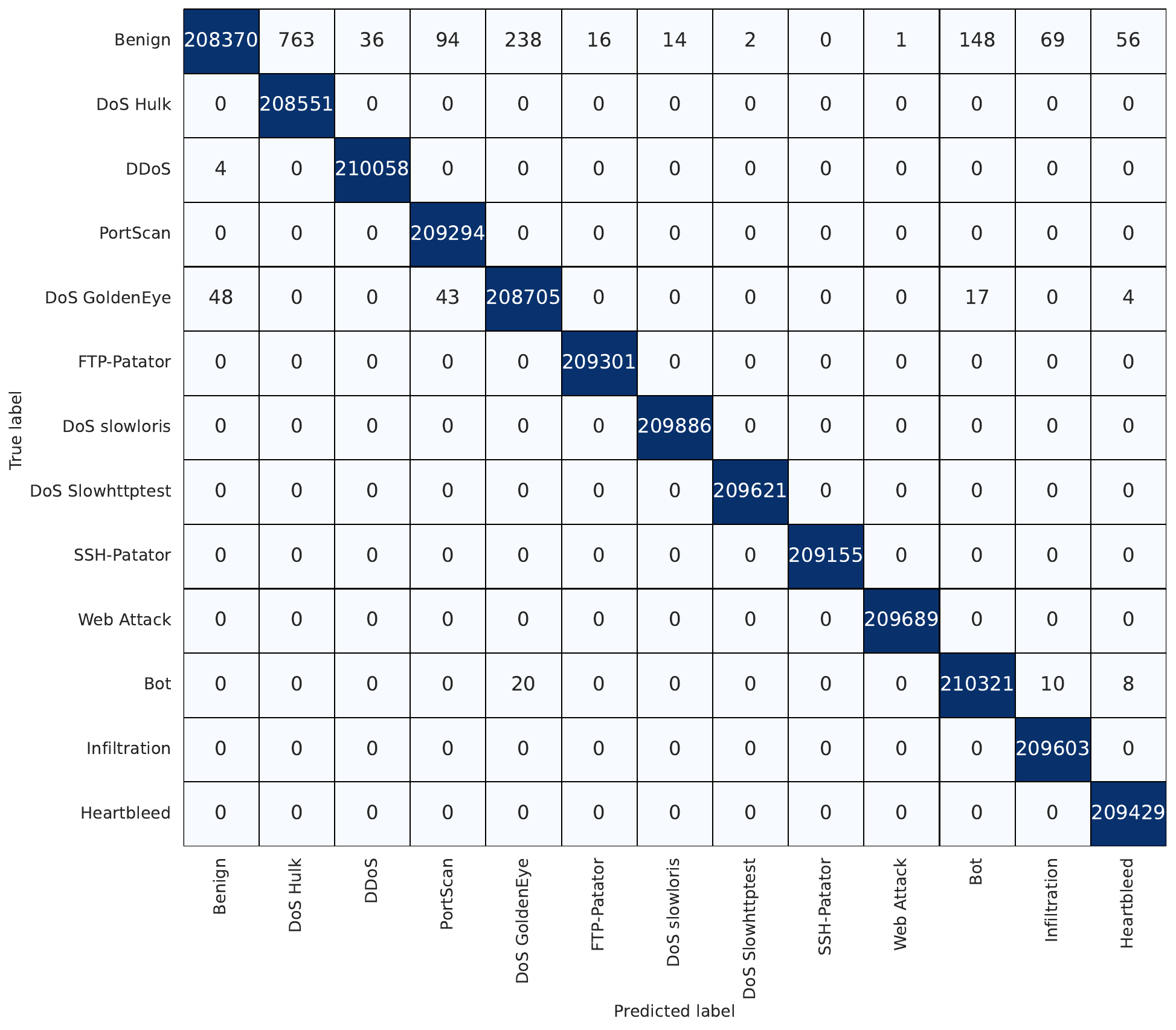}} 
	
	\caption{multilabel Confusion Matrix for CIC-IDS2017 Dataset}
	\label{fig:mcon_CIC}
\end{figure*}

The accuracy rates for multilabel classification are 99.91\% for DT, 99.94\% for RF, 99.95\% for ET, and 99.65\% for XGB on the proposed model. The confusion matrix is displayed in Figure \ref{fig:bcon_CIC} for binary classification and Figure \ref{fig:mcon_CIC} for multilabel classification. Upon examining the confusion matrix results for binary classification, it's observed that RF, ET, and XGB provide similar accuracy rates, with slight variations in their TP, FP, and FN rates.

%        TP	      TP (%)	FP	  FP (%)  FN	  FN (%)	  TN	TN (%)
% DT	20980400	50.07	900  	 0	    34900	0.08	20884900	49.84
% RF	20980300	50.07	1000	 0	    24400	0.06	 20895400	49.87
% ET	20980800	50.07	500	     0	    21600	0.05	20898200	49.88
% XGB	20911100	49.91	70200	0.17	76100	0.18	 20843700	49.74

In the binary confusion matrix, it's insightful to note that the TP rates are 50.07\% for DT, RF, ET, and 49.91\% for XGB. The TN rates are 49.84\% for DT, 49.87\% for RF, 49.88\% for ET, and 49.74\% for XGB. Additionally, the FP rates are 0.0\% for DT, RF, ET, and 0.17\% for XGB, while the FN rates are 0.08\% for DT, 0.06\% for RF, 0.05\% for ET, and 0.18\% for XGB. RF employs bootstrap repetitions and selects the best-split method, making it an effective ensemble of independent decision trees working together. ET, on the other hand, uses the entire original sample and selects the splitting operation at random, resulting in an ensemble of extra trees. This diversity in learning methods contributes to achieving the highest accuracy. For multilabel classification, DT, RF and ET exhibit similar and superior accuracy with minor variances in TP, FP, and FN rates. Notably, XGB demonstrates lower accuracy compared to the other algorithms in binary and multilabel classification scenarios.

\begin{figure*}[!htbptbp]
	\centering
	\subfloat[Binary]{\includegraphics[scale=.260]{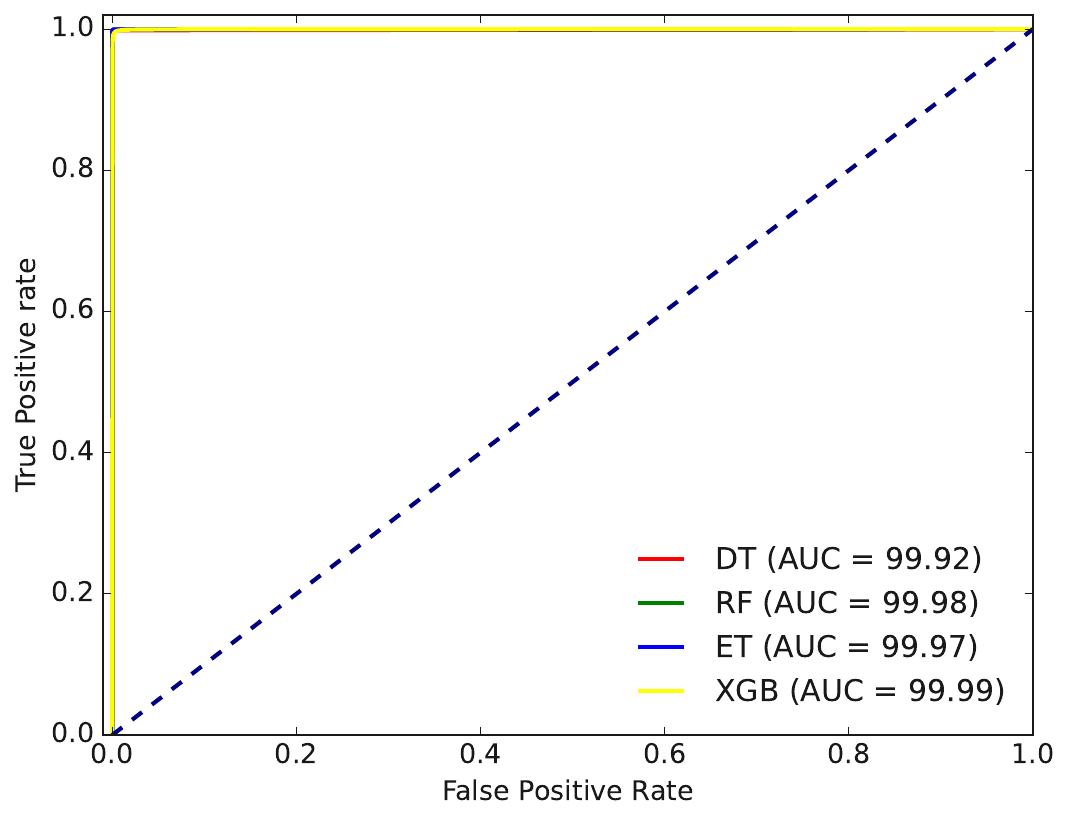}}\hspace{0.047cm}
	\subfloat[Multilabel]{\includegraphics[scale=.260]{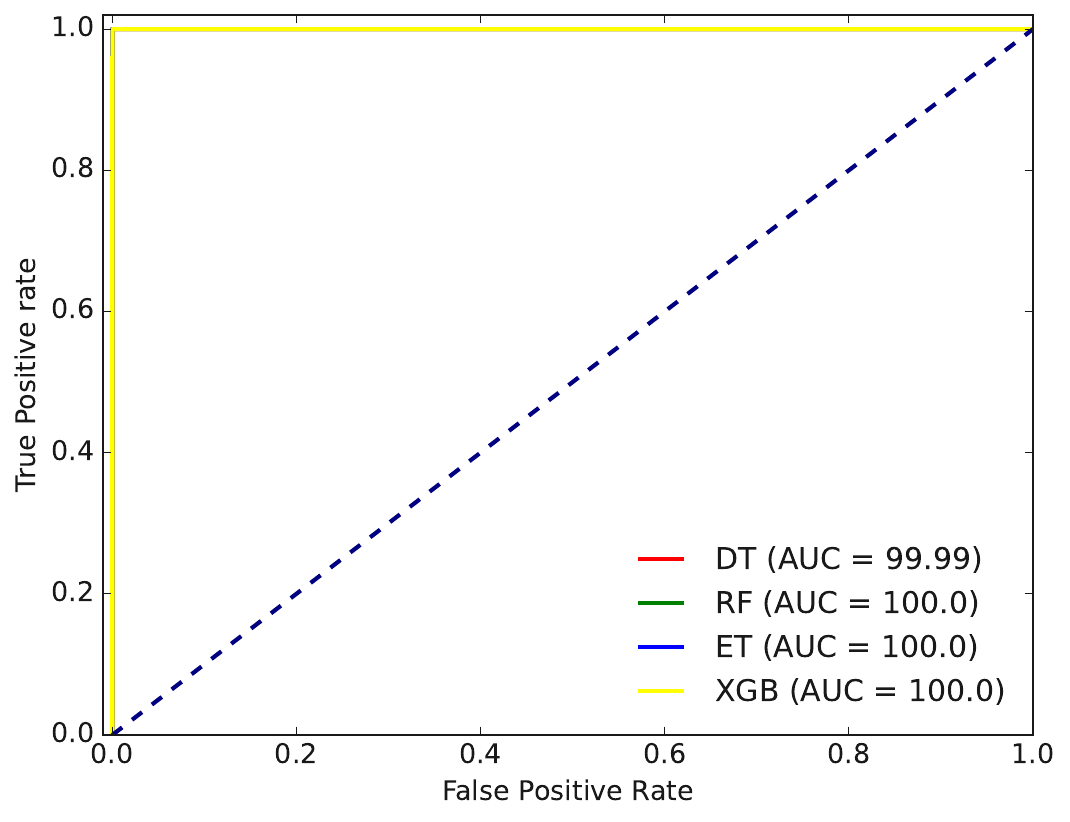}}
	\caption{Binary and multilabel ROC Curve for CIC-IDS2017 Dataset}
	\label{fig:roc_cicids}
\end{figure*}

The ROC Curve, as illustrated in Figure \ref{fig:roc_cicids}, provides a comprehensive view of the model's performance in binary and multilabel classification. The ROC Curves showcase that the Area Under the Curve (AUC) values are approaching the ideal value of 1, indicating a highly effective predictive model for distinguishing between classes. In the binary classification, the XGBoost (XGB) algorithm stands out with the highest AUC score. Specifically, the AUC scores are as follows: 99.92\% for Decision Trees (DT), 99.98\% for Random Forest (RF), 99.97\% for Extra Trees (ET), and an impressive 99.99\% for XGB. In multiclass classification, RF, ET, and XGB collectively achieve the highest AUC scores. Specifically, the AUC scores are 99.99\% for DT, a perfect 100\% for RF, ET, and XGB, emphasizing the exceptional predictive performance of these algorithms. These consistently high AUC scores, close to 1, underscore the strong predictive capabilities of the model on the CIC-IDS2017 dataset, indicating its suitability for the task.

\subsection{Results of CIC-IDS2018 Dataset}

\begin{figure}[!htbp]
	\centering
	\subfloat[Binary]{\includegraphics[scale=.420]{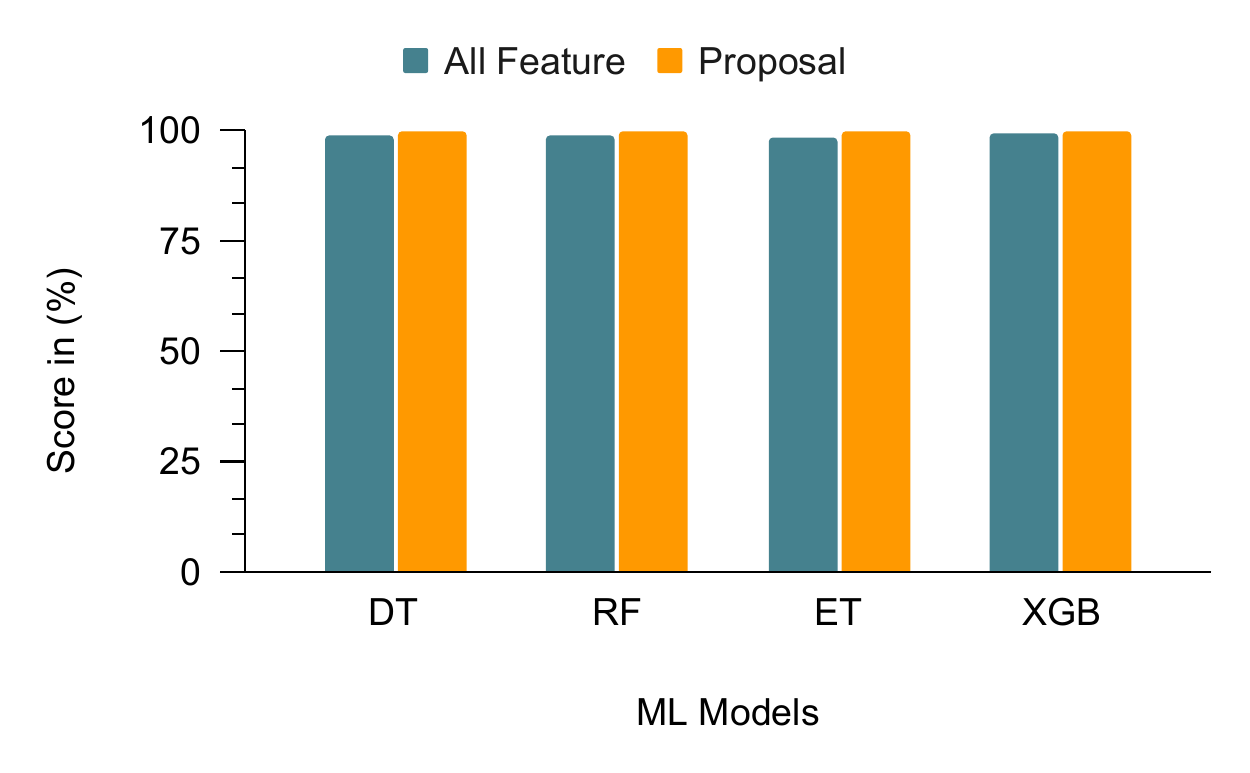}}
 
	\caption{Accuracy performance Bar Chart for All Features and Proposal Features for the CIC-IDS2018 Dataset}
	\label{fig:analysis_cicids2018}
\end{figure}

The performance results of the CIC-IDS2018 dataset are presented in Figure \ref{fig:analysis_cicids2018}, as well as in Table \ref{tab:analysis_cicids2018}. These  figure and table showcase the experimental results for two distinct scenarios: \textbf{All Features:} In this case, "All Features" represent the dataset where we did not oversample any features. We preprocessed, scaled, and applied these features directly to the machine learning models for training and performance evaluation. \textbf{Proposal Features:} Here, the "Proposal" refers to a methodology that encompasses various preprocessing steps that have been undertaken for evaluation. These results provide a comprehensive view of our model's performance and the impact of feature selection and preprocessing on IDS using the CIC-IDS2018 dataset.

In the bar chart, it is evident that our proposed model exhibits a noteworthy increase in accuracy rates for attack classification. Notably, the rate of accuracy improvement is more pronounced in the context of the proposed feature when compared to all feature.

\begin{table}[]
\centering
\resizebox{\textwidth}{!}{%
\begin{tabular}{lllllllll}
\hline
   & \multicolumn{2}{l}{Accuracy} & \multicolumn{2}{l}{Precision} & \multicolumn{2}{l}{Recall} & \multicolumn{2}{l}{F1-score} \\ \hline
ML & All Feature    & Proposal    & All Feature     & Proposal    & All Feature   & Proposal   & All Feature    & Proposal    \\ \hline
DT  & 98.88 & 99.94 & 95.25 & 99.94 & 95.21 & 99.94 & 95.21 & 99.94 \\
RF  & 98.74 & 99.93 & 98.25 & 99.93 & 91.55 & 99.93 & 91.55 & 99.93 \\
ET  & 98.37 & 99.94 & 96.81 & 99.94 & 90.38 & 99.94 & 90.38 & 99.94 \\
XGB & 99.11 & 99.87 & 98.51 & 99.87 & 96.51 & 99.87 & 96.51 & 99.87 \\ \hline
\end{tabular}%
}
\caption{Performance Analysis of CIC-IDS2018 Dataset
}
\label{tab:analysis_cicids2018}
\end{table}

The result analysis, as presented in Table \ref{tab:analysis_cicids2018}, highlights the performance assessment of various ML algorithms using both the "All Feature" and "Proposed" feature sets. Notably, among these algorithms, DT and ET consistently stand out as top performers in terms of accuracy, precision, recall, and F1-score. When considering the "Proposed" feature set, both DT and ET exhibit exceptional performance, achieving an impressive accuracy of 99.94\%, surpassing other algorithms, including RF with an accuracy of 99.93\% and XGB with 98.87\%. This remarkable increase in accuracy extends to precision, recall, and F1-score metrics, underscoring the effectiveness of DT and ET. Their success can be attributed to their decision tree-based learning approach, which allows them to effectively model complex relationships in the data. Additionally, they demonstrate robustness when applied to the "Proposed" feature set, which may involve a more intricate feature engineering process.

\begin{figure*}[!htbp]
	\centering
	\subfloat[DT]{\includegraphics[scale=.180]{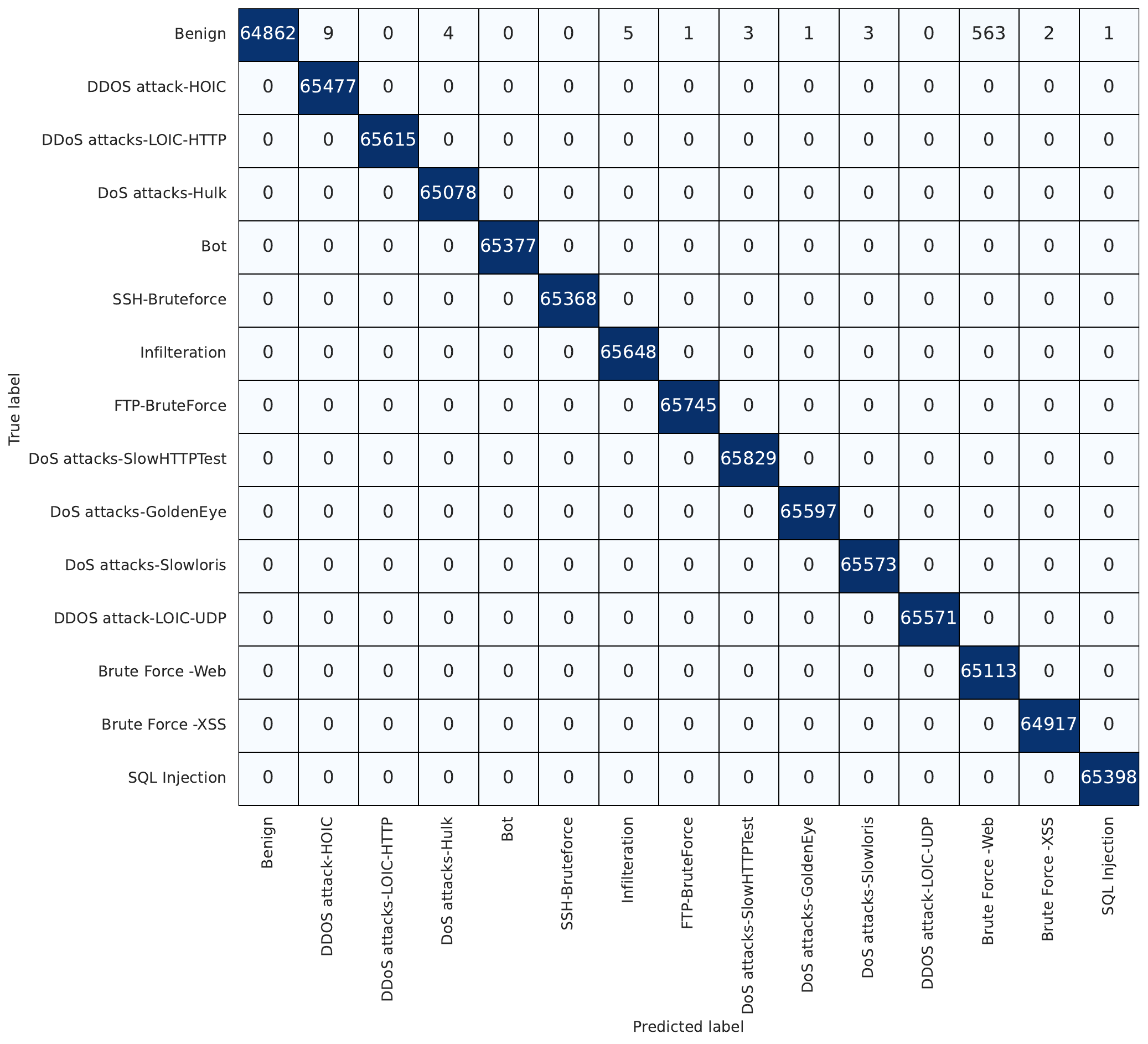}}
	\subfloat[RF]{\includegraphics[scale=.180]{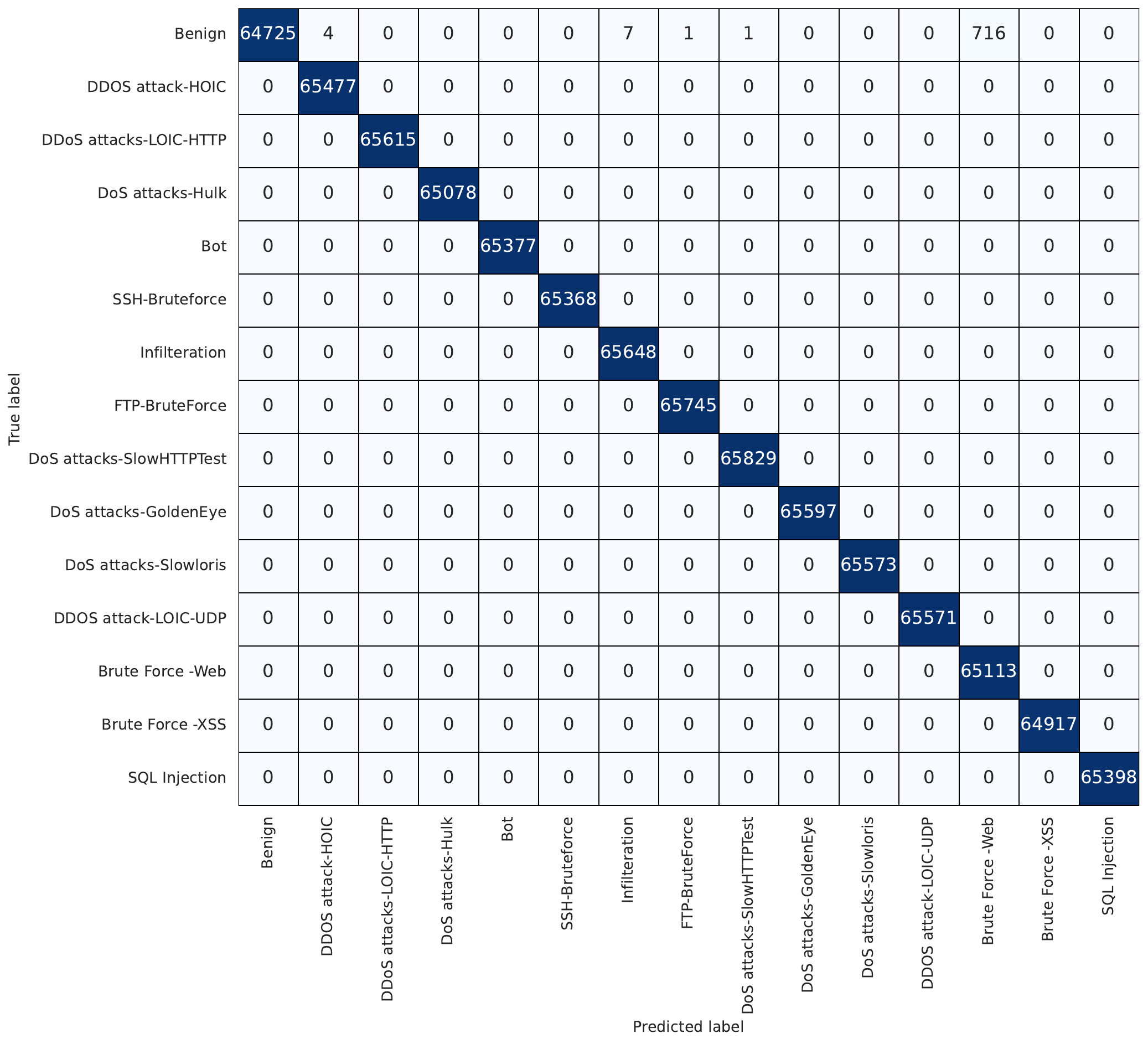}} \hspace{0.1cm}
	\subfloat[ET]{\includegraphics[scale=.180]{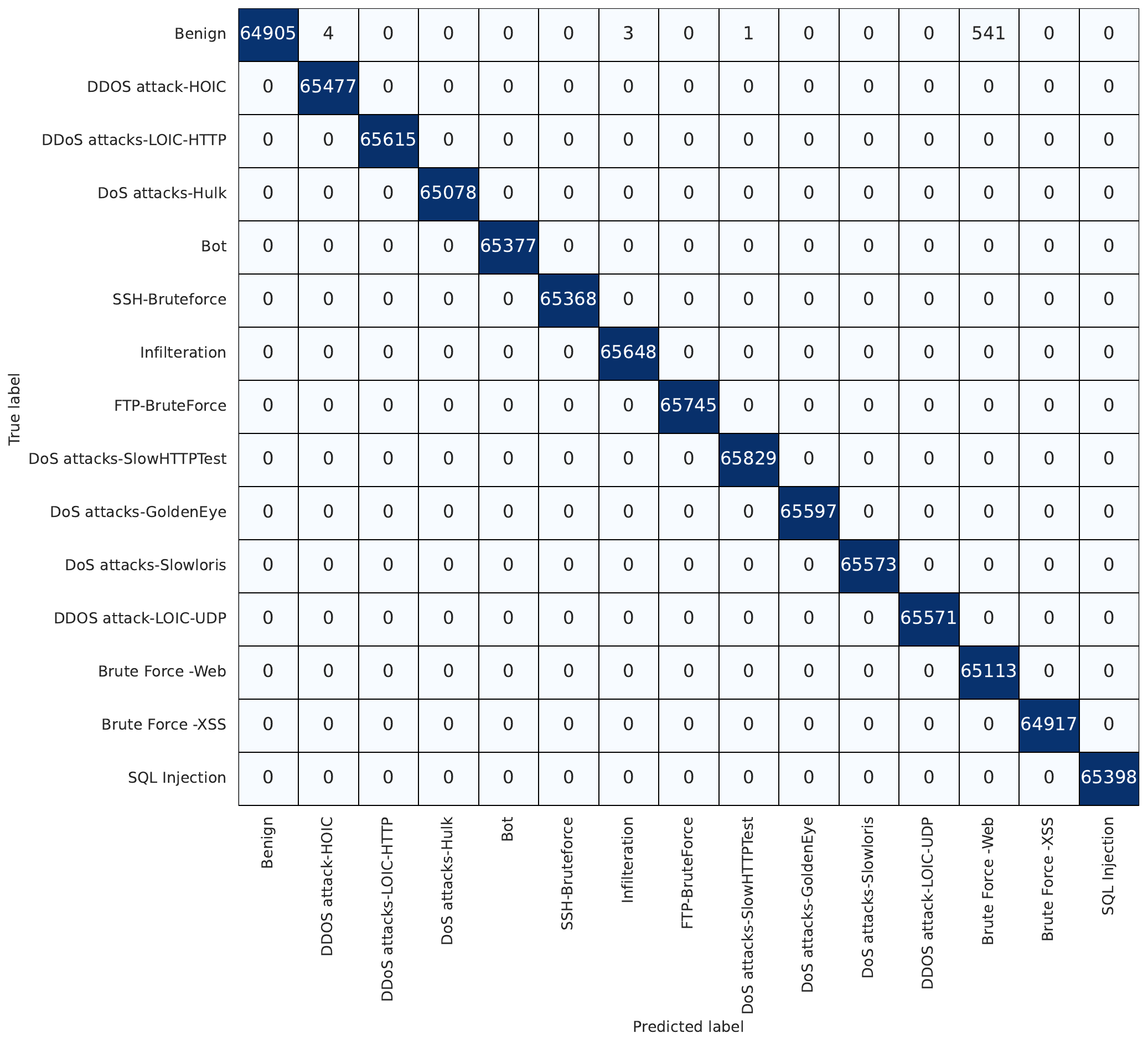}} 
	\subfloat[XGB]{\includegraphics[scale=.180]{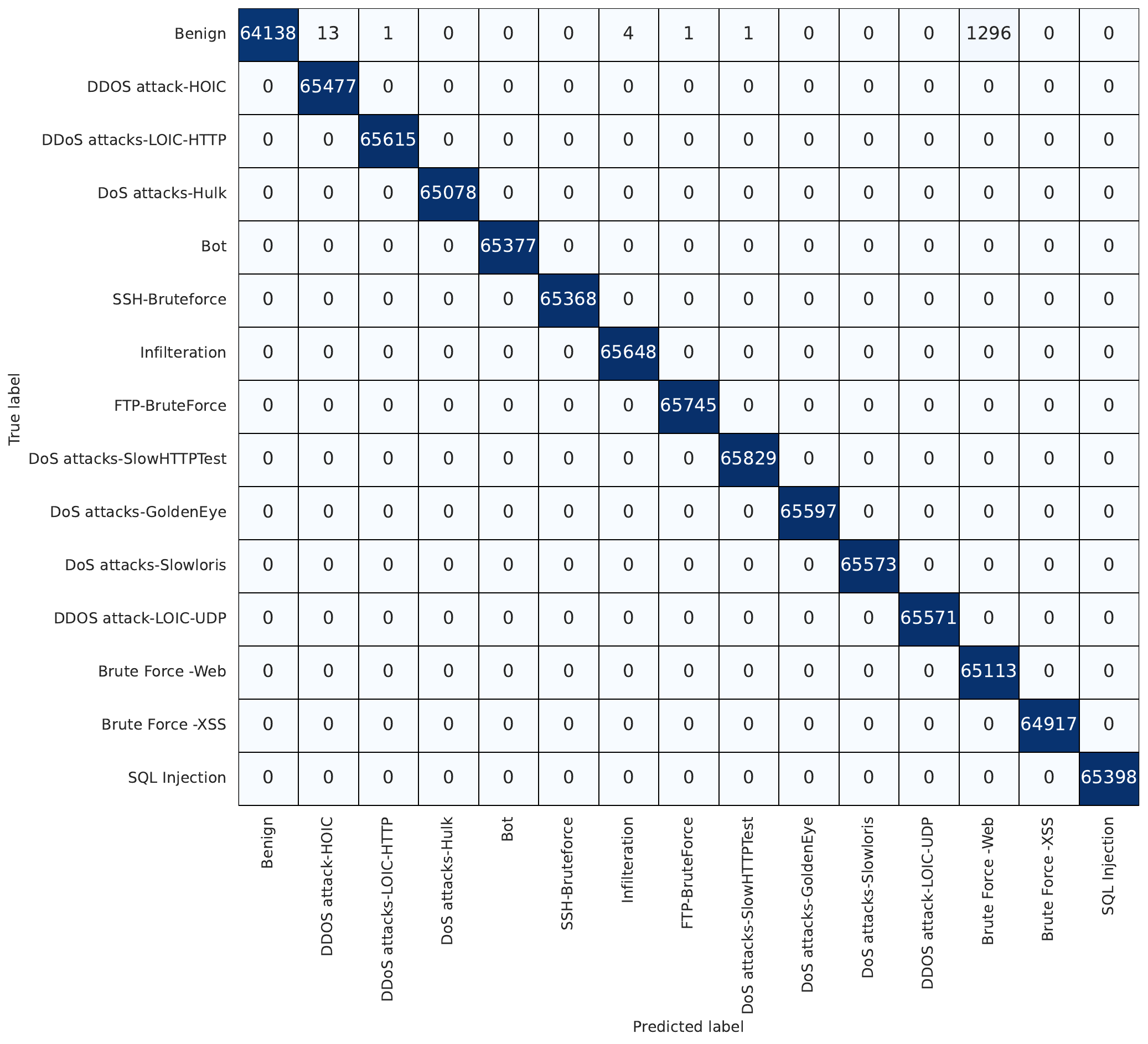}} 
	
	\caption{Confusion Matrix for CIC-IDS2018 Dataset}
	\label{fig:con_CIC2018}
\end{figure*}

The confusion matrix is displayed in Figure \ref{fig:con_CIC2018}. A successful predictive model is characterized by a low number of Type 1 (FP) and Type 2 (FN) errors in the confusion matrix. For DT, the TP rate stands impressively at 49.74\%, while the TN rate is equally strong at 49.76\%. Additionally, the FP and FN rates are remarkably low, at 0.16\% and 0.24\%, respectively. These findings highlight DT's robust performance in accurately identifying positive cases (intrusions) and negative cases (non-intrusions), making it a compelling choice for intrusion detection. For ET, the TP rate is an outstanding 49.78\%, and the TN rate is equally impressive at 49.82\%. Furthermore, the FP and FN rates are notably low, at 0.18\% and 0.28\%, respectively. These results underscore the exceptional performance of ET in accurately identifying both positive cases (intrusions) and negative cases (non-intrusions), making it a compelling choice for intrusion detection.

Among all the evaluated models, it is evident that both DT and ET outperform the others, showcasing superior performance in terms of TP and TN rates for IDS. They consistently deliver higher TP and TN rates, demonstrating their effectiveness in accurately identifying intrusions while maintaining a low rate of false positives and false negatives.

\begin{figure*}[!htbp]
	\centering
\includegraphics[scale=.260]{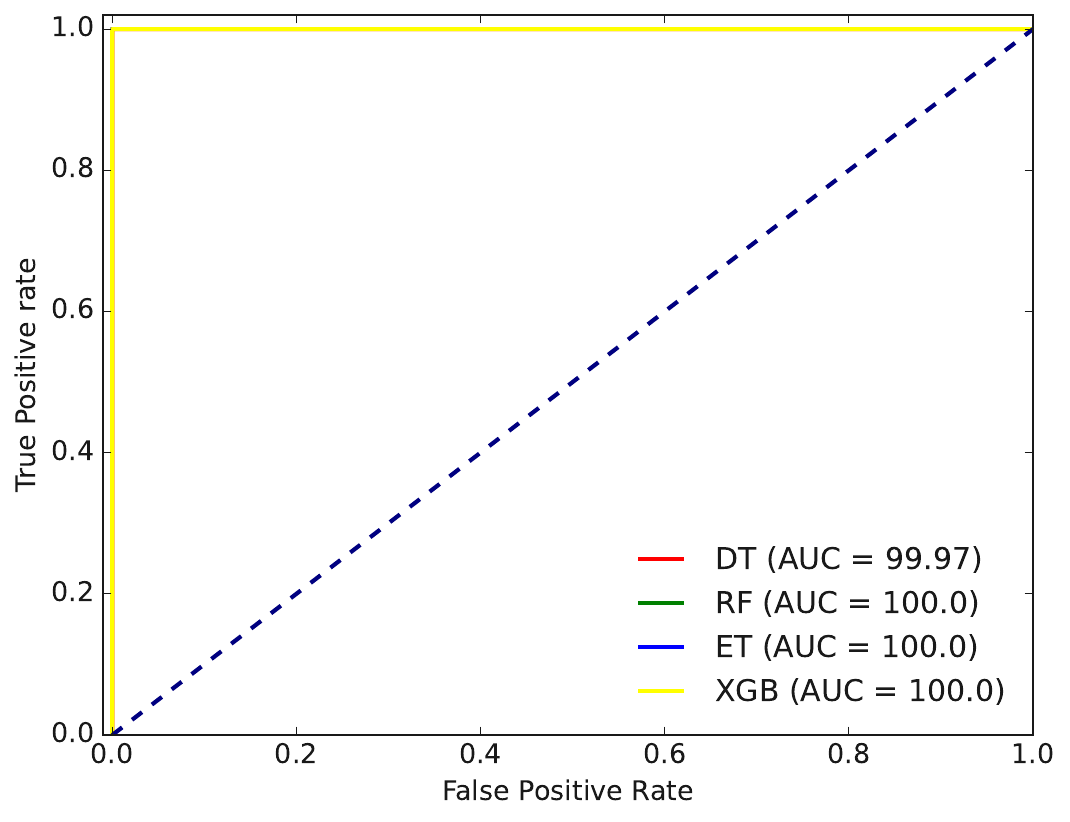}
	\caption{ROC Curve for CIC-IDS2018 Dataset}
	\label{fig:roc_cicids2018}
\end{figure*}

In Figure \ref{fig:roc_cicids2018}, the ROC Curve depicts the performance of machine learning models on the CIC-IDS2018 Dataset. The ROC Curves clearly demonstrate that the AUC values for DT, RF, ET, and XGB approach the desirable threshold of 1, indicative of an effective model for distinguishing between different classes. Notably, both RF and ET achieve the highest AUC scores. Specifically, the AUC scores are 99.99\% for DT, a perfect 100\% for RF and ET, and 84.85\% for XGB. These high AUC scores, approaching 1, signify the strong predictive performance of the model on the CIC-IDS2018 dataset, underlining its effectiveness in class differentiation.

\subsection{Discussion}

We have conducted a comprehensive comparative analysis of our proposed model in conjunction with other models. The comparative results are systematically presented in tabular format, as outlined in Table \ref{tab:com_UNSW} for the UNSW-NB15 dataset, Table \ref{tab:com_CIC2017} for the CIC-IDS2017 dataset, and Table \ref{tab:com_CIC2018} for the CIC-IDS2018 dataset. These tables offer valuable insights into the performance of our model relative to other models, facilitating a detailed examination of the results across different datasets.

% unswnb15>binary (multi+binary); cicids2017> multilabel (multi+binary(generated))

\begin{table}[]
\begin{tabular}{p{0.5cm}p{1.5cm}p{1cm}p{1.5cm}p{1cm}p{1cm}p{1cm}p{.8cm}p{.8cm}}
\hline
SI. NO. & Authors & Data Balancing & Dimension Reduction & Algorithm & Selected Feature & Binary Acc(\%) & Multilabel Acc(\%)\\ \hline
1 & \cite{moualla2021improving} & SMOTE & - & ELM & - & 98.43 & - \\ \hline
2 & \cite{kasongo2020performance} & - & XGB & DT, ANN & 19 & 90.85 (DT) & 77.51 (ANN) \\ \hline
3 & \cite{ahmad2021intrusion} & - & MQTT+TCP & RF & - & 98.67 & 97.37 \\ \hline
4 & \cite{kasongo2020deep} & - & WFEU & FFDNN & 22 & 87.10 & 77.16 \\ \hline
5 & \cite{choudhary2020analysis} & - & - & DNN & - & 91.50 & - \\ \hline
6 & \cite{aleesa2021deep} & - & - & ANN & - & - & 97.89 \\ \hline
7 & \cite{al2021stl} & STL & - & LSTM+CNN & - & - & - \\ \hline
8 & \cite{zhang2020effective} & SGM & - & CNN & - & - & 96.54 \\ \hline
9 & \cite{hassan2020hybrid} & - & - & CNN-WDLSTM & - & 97.17 & 98.43 \\ \hline
10 & Our Proposal & RO & SFE-PCA & RF & 10 & 99.59 & 99.95 \\ \hline
11 & Our Proposal & RO & SFE-PCA & ET & 10 & 99.59 & 99.95 \\ \hline

\end{tabular}

\caption{Comparison Analysis of UNSW-NB15 Dataset}
\label{tab:com_UNSW}
\end{table}

\begin{table}[]
\begin{tabular}{p{0.5cm}p{1.5cm}p{1cm}p{1.5cm}p{1cm}p{1cm}p{1cm}p{.8cm}p{.8cm}}

\hline
SI. NO. & Authors & Data Balancing & Dimension Reduction & Algorithm & Selected Feature & Accuracy(\%)\\ \hline
1 & \cite{kshirsagar2021efficient} & - & IGR+CR +ReF & PART & - & 99.95 (Binary) \\ \hline
2 & Our Proposal & RO & SFE-PCA & ET & 10 & 99.95 (Binary) \\ \hline \hline

3 & \cite{hammad2021t} & - & t-SNE & RF & - &  99.78 \\ \hline
4 & \cite{al2021improved} & - & NTLBO & LR & 22 & 97.00 \\ \hline
5 & \cite{guezzaz2021reliable} & - & EDFS & DT & - & 98.80 \\ \hline
6 & \cite{stiawan2020cicids} & - & IG+Ranking +Grouping & RF & 22 & 99.86 \\ \hline
7 & \cite{stiawan2020cicids} & - & IG+Ranking +Grouping & J48 & 52 & 99.87 \\ \hline
8 & \cite{bhardwaj2021hybrid} & - & - & DNN+ACO & - & 98.25 \\ \hline
9 & \cite{zhang2020effective} & SGM & - & CNN & - & 99.85 \\ \hline
10 & Our Proposal & RO & SFE-PCA & DT & 10 &  99.99  \\ \hline
11 & Our Proposal & RO & SFE-PCA & RF & 10 &  99.99  \\ \hline
12 & Our Proposal & RO & SFE-PCA & ET & 10 &  99.99  \\ \hline

\end{tabular}

\caption{Comparison Analysis of CIC-IDS2017 Dataset}
\label{tab:com_CIC2017}
\end{table}

\begin{table}[]
\begin{tabular}{p{0.5cm}p{1.5cm}p{1cm}p{1.5cm}p{1cm}p{1cm}p{1cm}p{.8cm}p{.8cm}}

\hline
SI. NO. & Authors & Data Balancing & Dimension Reduction & Algorithm & Selected Feature &  Accuracy(\%)\\ \hline
1 & \cite{seth2021novel} & - & HFS &  LightGBM & 24 & 97.73 \\ \hline
2 & \cite{khan2021hcrnnids} & - & - & CNN +RNN & - & 97.75 \\ \hline
3 & \cite{kim2020cnn} & - & - & CNN & - & 91.50 \\ \hline

4 & Our Proposal & RO & SFE-PCA & DT & 10 & 99.94  \\ \hline
5 & Our Proposal & RO & SFE-PCA & RF & 10 & 99.94  \\ \hline

\end{tabular}

\caption{Comparison Analysis of CIC-IDS2018 Dataset}
\label{tab:com_CIC2018}
\end{table}

In our research, we address the challenge of imbalanced datasets by employing Random Oversampling (RO). Additionally, we apply the Stacking Feature Embeded (SFE) technique to augment feature sets and create metadata. Subsequently, we reduce the dimensionality to 10 features using Principal Component Analysis (PCA). Our model is then trained using popular machine learning algorithms, including DT, RF, ET, and XGB. The performance of our model is evaluated on two prominent datasets: UNSW-NB15 and CIC-IDS2017.
For the UNSW-NB15 dataset, we achieve noteworthy accuracy scores. In binary classification, our model attains accuracy rates of 98.97\% (DT), 99.59\% (RF), 99.59\% (ET), and 98.81\% (XGB). In multilabel classification, the accuracy scores reach 99.79\% (DT), 99.95\% (RF), 99.95\% (ET), and 95.04\% (XGB).
For the CIC-IDS2017 dataset, our model continues to excel. In binary classification, we obtain accuracy rates of 99.91\% (DT), 99.94\% (RF), 99.95\% (ET), and 99.65\% (XGB). In multilabel classification, the accuracy scores are impressive, with 99.99\% for DT, RF, and ET, and 99.94\% for XGB.
Furthermore, in the evaluation on the CIC-IDS2018 Dataset, our model maintains high accuracy rates. We achieve accuracy scores of 99.94\% (DT), 99.93\% (RF), 99.94\% (ET), and 99.87\% (XGB).
The results analysis reveals that the highest accuracy rates for binary and multilabel classification are 99.59\% and 99.95\%, achieved with RF and ET algorithms on the UNSW-NB15 dataset. In contrast, for the CIC-IDS2017 dataset, the highest accuracy rate is 99.99\% for binary classification using the ET model and 99.99\% for multilabel classification using DT, RF, and ET models.

In summary, our proposed model consistently outperforms existing methods in binary and multilabel classification scenarios with reduced features. The 10-feature dimension reduction (RR is 10:N, where N represents the input features) significantly enhances intrusion detection accuracy, minimizing false positive and negative rates. These results emphasize the importance of considering at least 10 features for optimal intrusion detection accuracy across various dimensional datasets.

\section{Time Complexity}

Time complexity, which denotes the time required for executing an operation, plays a crucial role in assessing the efficiency of algorithms \citep{talukder2022machine, talukder2024empowering}. In the context of IDS, evaluating the time complexity of ML models is paramount for efficient operation. We analyze the time complexity of key algorithms such as DT, RF, MLP, KNN, LGB and XGB models as follows.

\begin{itemize}
    \item \textbf{DT}: The time complexity is typically \( O(n \cdot m \cdot \log(m)) \), with \( n \) representing data points and \( m \) as features. It constructs a tree by recursively partitioning data.
    
    \item \textbf{RF}: Comprising multiple DTs, its time complexity is \( O(t \cdot n \cdot m \cdot \log(m)) \), where \( t \) is the number of trees.
    
    \item \textbf{MLP (Multi-Layer Perceptron)}: With multiple layers and neurons, its time complexity is \( O(w \cdot e \cdot n) \), where \( w \) is the number of weights, \( e \) the number of epochs, and \( n \) the number of data points.
    
    \item \textbf{KNN (K-Nearest Neighbors)}: This non-parametric method has a time complexity of \( O(n \cdot m) \) for training, where \( n \) is the number of data points and \( m \) is the number of features. The prediction phase can be more computationally intensive.
    
    \item \textbf{LGB (Light Gradient Boosting Machine)}: Known for efficiency and low memory usage, LGB's time complexity is \( O(t \cdot n \cdot m) \), where \( t \) is the number of trees.
    
    \item \textbf{XGB (XGBoost)}: This model's time complexity varies but is generally \( O(t \cdot d) \), where \( t \) is the number of trees and \( d \) the depth of the trees.
\end{itemize}

\begin{table}[!h]
\centering 
\begin{tabular}{|c|c|c|}
\hline
\textbf{SI. No.} & \textbf{ML Model} & \textbf{Time Complexity} \\
\hline
1 & DT (Decision Trees) & $O(n \cdot m \cdot \log(m))$ \\
2 & RF (Random Forests) & $O(t \cdot n \cdot m \cdot log(m))$ \\
3 & ET (Extra Trees) & $O(t \cdot n \cdot m \cdot log(m))$ \\
4 & XGB (XGBoost) & $O(t \cdot d)$ \\
\hline
\end{tabular}
\caption{Time Complexity of ML Models in IDS}
\label{tab:time_complexity}
\end{table}

\section{Conclusion}
\label{sec:conclusion}

In conclusion, our research has introduced a novel approach to network intrusion detection by combining various techniques to address the challenges of imbalanced data, feature embedding, and dimension reduction. Our model leverages the Random Oversampling (RO) method to tackle data imbalance, utilizes feature embedding through Kmeans and GM clustering results, and employs Principal Component Analysis (PCA) for dimension reduction. We have evaluated our model's performance with four prominent ML algorithms, DT, RF, ET and XGB for binary and multilabel classification studies using three benchmark datasets: UNSW-NB15, CIC-IDS2017 and CIC-IDS2018. Our experimental results have demonstrated exceptional accuracy rates, with RF and ET achieving 99.59\% and 99.95\% accuracy, respectively, on the UNSW-NB15 dataset. Besides, our model has achieved remarkable accuracy rates, with DT, RF, and ET reaching 99.99\% on the CIC-IDS2017 dataset and for CIC-IDS2018 we achieved 99.94\% accuracy rate using DT and RF models. These performance scores surpass those of existing methods, indicating the effectiveness of our approach in enhancing network intrusion detection. Our proposed model brings about substantial improvements across these benchmark datasets. Specifically, we observed a significant increase in accuracy, with enhancements ranging from 1.52\% to 22.19\% for UNSW-NB15, 0.12\% to 2.99\% for CIC-IDS2017, and 1.99\% to 8.44\% for CIC-IDS2018 when compared to prior research. These results highlight the remarkable advancements our model introduces in the field of intrusion detection.

Our contributions to the field of network intrusion detection are substantial. We have addressed the persistent challenge of imbalanced data, ensuring that our model can handle real-world scenarios where intrusion instances are often rare compared to benign network traffic. The incorporation of feature embedding techniques has allowed us to capture more nuanced patterns and anomalies within the data, thus improving detection accuracy. Additionally, the application of PCA for dimension reduction has not only reduced computational complexity but also enhanced the interpretability of the model. The benefits of our new model extend beyond accuracy improvements. It offers a more robust and adaptable solution for intrusion detection, capable of handling varying data distributions and network environments. By combining multiple machine learning algorithms, our model harnesses the strengths of each, providing a versatile tool for network security professionals. Furthermore, its enhanced accuracy and lower false positive rates can significantly reduce the burden of false alarms in intrusion detection systems, allowing security teams to focus on the most critical threats.

In practical terms, our model can be invaluable for IDS in safeguarding network infrastructure. Its high accuracy rates and adaptability make it well-suited for identifying both known and novel threats, enhancing the overall security posture of organizations. The reduced false positive rates contribute to a more efficient use of resources, as security teams can concentrate their efforts on genuine security incidents. Ultimately, our model represents a significant advancement in the field of network intrusion detection, offering a more reliable and efficient solution for protecting critical network assets

The limitation of our research is that we did not employ deep learning models along with optimization techniques. While our current approach has demonstrated remarkable results, there remains untapped potential for further improving the performance of intrusion detection systems. In the future, we envision expanding our work to incorporate deep learning models, which have shown great promise in various fields, including intrusion detection. DL algorithms, such as DNN, RNN or Hybrid are capable of capturing intricate patterns and representations in complex data, which can be particularly advantageous in the realm of network security.

\section*{Declarations}

\subsection*{Conflict of interest}
The authors have no conflicts of interest to declare that they are relevant to the content of this article.

\subsection*{Ethics approval}  Not applicable
\subsection*{Consent to participate} Not applicable
\subsection*{Consent to Publish} Not applicable

\subsection*{Availability of data and materials}
The selected datasets are sourced from free and open-access sources such as 
UNSW-NB15: \url{https://research.unsw.edu.au/projects/unsw-nb15-dataset},
CIC-IDS2017: \url{https://www.unb.ca/cic/datasets/ids-2017.html} and
CIC-IDS2018: \url{https://www.unb.ca/cic/datasets/ids-2018.html}.

\subsection*{Authors’ contributions}

Md. Alamin Talukder: Conceptualization, Data curation, Methodology, Software, Formal analysis, Visualization, Writing – original draft, Writing – review \& editing. 
Md. Manowarul Islam: Methodology, Investigation, Validation, Project administration. 
Md. Ashraf Uddin: Supervision, Investigation, Resources, Validation. 
Khondokar Fida Hasan: Investigation, Validation, Visualization
Selina Sharmin: Software, Validation. 
Salem A. Alyami: Investigation, Validation.  
Mohammad Ali Moni: Investigation, Validation, Visualization, Writing – review \& editing.

\section*{Acknowledgments}
This research was supported by the Deanship of Scientific Research Large Groups at King Khalid University, Kingdom of Saudi Arabia (RGP.2/219/43)

\bibliography{mbibs}

\end{document}